# Alternative Methylated Biosignatures. I. Methyl Bromide, a Capstone Biosignature

Michaela Leung[1,2,3], Edward W. Schwieterman[1,2,3,4], Mary N. Parenteau[2,3,5], and Thomas J. Fauchez[3,6,7,8]
[1] Department of Earth and Planetary Sciences, University of California, Riverside, CA 92521, USA; michaela.leung@email.ucr.edu
[2] NASA Alternative Earths Team, Riverside, CA, USA
[3] NASA Nexus for Exoplanet System Science, Virtual Planetary Laboratory Team, Box 351580, University of Washington, Seattle, WA 98195, USA
[4] Blue Marble Space Institute of Science, Seattle, WA, USA
[5] NASA Ames Research Center, Moffett Field, CA, USA
[6] NASA Goddard Space Flight Center, 8800 Greenbelt Road, Greenbelt, MD 20771, USA
[7] Sellers Exoplanet Environment Collaboration (SEEC), NASA Goddard Space Flight Center, Greenbelt, MD, USA
[8] American University, Washington DC, USA

## Abstract

The first potential exoplanetary biosignature detections are likely to be ambiguous due to the potential for false positives: abiotic planetary processes that produce observables similar to those anticipated from a global biosphere. Here we propose a class of methylated gases as corroborative "capstone" biosignatures. Capstone biosignatures are metabolic products that may be less immediately detectable, but have substantially lower false-positive potential, and can thus serve as confirmation for a primary biosignature such as $O_2$. $CH_3Cl$ has previously been established as a biosignature candidate, and other halomethane gases such as $CH_3Br$ and $CH_3I$ have similar potential. These gases absorb in the mid-infrared at wavelengths that are likely to be captured while observing primary biosignatures such as $O_3$ or $CH_4$. We quantitatively explore $CH_3Br$ as a new capstone biosignature through photochemical and spectral modeling of Earthlike planets orbiting FGKM stellar hosts. We also reexamine the biosignature potential of $CH_3Cl$ over the same set of parameters using our updated model. We show that $CH_3Cl$ and $CH_3Br$ can build up to relatively high levels in M dwarf environments and analyze synthetic spectra of TRAPPIST-1e. Our results suggest that there is a coadditive spectral effect from multiple $CH_3X$ gases in an atmosphere, leading to an increased signal-to-noise and greater ability to detect a methylated gas feature. These capstone biosignatures are plausibly detectable in exoplanetary atmospheres, have low false-positive potential, and would provide strong evidence for life in conjunction with other well-established biosignature candidates.

## 1. Introduction

The recent launch of the James Webb Space Telescope (JWST) and construction of the ground-based extremely large telescopes (ELTs) will allow for tremendous advances in the quality of possible spectral measurements of exoplanets (Fujii et al. 2018; López-Morales et al. 2019; Lustig-Yaeger et al. 2019). These instruments will yield new insight into exoplanetary atmospheres and provide the first opportunities to characterize terrestrial exoplanets and search for biosignatures. To prepare for this opportunity, it is necessary to robustly understand the relationship between life and the gases that can be spectroscopically detected in an exoplanetary atmosphere. Preparing a toolkit of potential biosignature candidates viable across a variety of planetary and stellar environments will streamline future analysis and interpretation of terrestrial planetary spectra. It is especially critical to simulate spectra of favorable targets that can be observed with current and near future observatories in the 2020 and 2030s.

Most well-known biosignatures are single or paired gases produced by photosynthetic and chemosynthetic life on the Earth, such as oxygen (Meadows 2017; Meadows et al. 2018a), methane, nitrous oxide, sulfur gases (Domagal-Goldman et al. 2011), or combinations of the prior listed gases (Grenfell 2017; Kaltenegger 2017; Schwieterman et al. 2018). However, many of these gases have potential false positives from abiotic sources; therefore, it is essential to interpret them within the context of the environment. Confirming the biogenicity of putative planetary biosignatures will be an ongoing process in practice, requiring verification through less easily detected corroborative evidence beyond the primary biosignature feature. Positive affirmation of the presence of life may be realized by detecting "capstone" biosignatures. We define a capstone biosignature as one with high biological specificity and low false-positive potential, which may be detected alongside a primary biosignature such as $O_2$ or $CH_4$, potentially through intensive follow-up observations. Previous work has considered the detectability of trace gases with high biological specificity such as $CH_3Cl$, $(CH_3)_2S$, and $C_5H_8$ (Pilcher 2003; Segura et al. 2005; Domagal-Goldman et al. 2011; Zhan et al. 2021). In particular, the promising nature of the methylated gases $CH_3Cl$ and $(CH_3)_2S$ establishes the first members of a process-based biosignature class, where members are defined by their generation from a common microbial metabolism.

Methylated gases are known to be produced in a variety of environments, both terrestrial and marine, by both microbes (Meyer et al. 2008; Fujimori et al. 2012; Shibazaki et al. 2016) and higher organisms including macroalgae, plants, and fungi (Tait & Moore 1995; Rhew et al. 2000, 2001; St et al. 2016). Ongoing marine and terrestrial studies for atmospheric surveillance, marine cycling, and environmental toxicity management have revealed plentiful fluxes of methylated

**Table 1**
Methylated Halogen (Cl, Br, I) and Chalcogen (S, Se, Te) Gases Are Produced by a Variety of Organisms across a Range of Environments

| Chemical Species | Example Environments | Example Producers | Reference(s) |
|---|---|---|---|
| $CH_3Cl$ | Marine, Terrestrial | Algae, Bacteria, Fungi, Plants | Tait & Moore (1995), Farhan Ul Haque et al. (2017) |
| $CH_2Cl_2$ | Industrial/Potential Marine | Industrial/Unknown | Ooki & Yokouchi (2011) |
| $CHCl_3$ | Marine, Terrestrial | Algae | Harper (1995), Macdonald et al. (2020) |
| $CCl_4$ | Marine | Algae | Harper (1995) |
| $CH_3Br$ | Marine | Algae, Bacteria | Paul & Pohnert (2011), Fujimori et al. (2012) |
| $CH_2Br_2$ | Marine, Terrestrial | Algae | Montzka et al. (2011), Macdonald et al. (2020) |
| $CHBr_3$ | Marine, Terrestrial | Algae, Bacteria | Montzka et al. (2011), Macdonald et al. (2020) |
| $CBr_4$ | Marine | Algae | Paul & Pohnert (2011) |
| $CH_2BrCl$ | Marine | Algae | Carpenter et al. (2003), Yokouchi et al. (2005) |
| $CHBr_2Cl$ | Marine | Algae | Yokouchi et al. (2005), Shibazaki et al. (2016) |
| $CHBrCl_2$ | Marine | Algae | Schall et al. (1994), Yokouchi et al. (2005) |
| $CH_3I$ | Marine, Terrestrial | Algae, Bacteria | Manley et al. (1992), Manley et al. (2006) |
| $CH_2I_2$ | Marine | Algae | Schall et al. (1994), Carpenter et al. (2003) |
| $CHI_3$ | Marine | Algae, Bacteria | Carpenter et al. (2003), Fujimori et al. (2012) |
| $(CH_3)_2CHI$ | Marine | Algae | Schall et al. (1997), Carpenter et al. (2003) |
| $CH_2ClI$ | Marine | Algae | Klick & Abrahamsson (1992), Carpenter et al. (2003) |
| $CH_2IBr$ | Marine | Algae | Carpenter et al. (2003) |
| $CHIBr_2$ | Marine | Algae | Carpenter et al. (2003) |
| $(CH_3)_2S$ | Marine, Terrestrial | Algae, Bacteria | Stefels et al. (2007), Carrión et al. (2015) |
| $(CH_3)_2S_2$ | Lacustrine, Marine, Terrestrial | Algae, Bacteria | Visscher et al. (2003), Hu et al. (2007) |
| $CH_3SH$ | Marine | Algae, Bacteria | Visscher et al. (2003) |
| $(CH_3)_2Se$ | Lacustrine, Terrestrial | Bacteria, Fungi, Plants | Chau et al. (1976), Bañuelos et al. (2017) |
| $(CH_3)_2Se_2$ | Lacustrine, Terrestrial | Bacteria, Fungi, Plants | Chau et al. (1976), Bañuelos et al. (2017) |
| $CH_3SeS$ | Marine | Algae | Dungan et al. (2003) |
| $(CH_3)_2SeS$ | Lacustrine | Algae | Fan et al. (1997) |
| $CH_3SeH$ | Marine | Algae | Amouroux & Donard (1996) |
| $(CH_3)_2Te$ | Laboratory/Potential Terrestrial | Bacteria, Fungi | Basnayake et al. (2001), Chasteen & Bentley (2003) |
| $(CH_3)_2Te_2$ | Laboratory/Potential Terrestrial | Fungi | Chasteen & Bentley (2003) |

**Note.** Examples of terrestrial environments are salt marshes and flooded fields (e.g., rice paddies). Marine production occurs in both coastal and open-water regimes. Terrestrial plant producers include *Brassica* species (Bañuelos et al. 2017). Example producers are not comprehensive (i.e., not all organisms that produce the given gas are necessarily listed or known).

halogen (Cl, Br, I) and chalcogen (S, Se, Te) gases across a wide variety of environments (Chasteen & Bentley 2003; Rhew et al. 2010; Simmonds et al. 2010; Macdonald et al. 2020). The major producers of these gases are marine micro- and macroalgae as well as bacteria and fungi in terrestrial soils. Some of these oxygenic phototrophs such as cyanobacteria produce both methylated gases and $O_2$, generating a biogenic gas pair that may help discriminate false positives for $O_2$ (Shibazaki et al. 2016). See Table 1 for a more extensive accounting of methylation substrates, environments, and source organisms.

Methylated gases are especially compelling biosignature candidates because (1) their production is widespread across all domains of life on Earth, and (2) they represent a basic metabolic need to detoxify environmental metals and halides (via methylation and volatilization) that can inhibit growth. Therefore, it is reasonable to speculate that this basic metabolic process could evolve on habitable exoplanets. The biological processes that generate methylated forms of halogens, chalcogens, and other elemental substrates likely evolved as a byproduct of the primordial methanogenesis metabolism (Manley 2002), which was then adapted to the detoxification of metal- and halide-rich environments, among other purposes (Jia et al. 2013), and subsequently radiated across the domains of life. Halogenated compounds can also be produced as metabolic byproducts, another close connection between the production of methylated halogenated compounds (Neilson 2003; van Pée & Unversucht 2003). Additionally, the production of methylated compounds is catalyzed by enzymes, which are a fundamental trait of life on Earth, and the ultimate source of atmospheric chemical disequilibrium. For Earth organisms, the specific methylation pathway is dependent on the environmental context (Manley 2002; Chasteen & Bentley 2003; Paul & Pohnert 2011). Because metabolism and the need to transfer information are universal features of life, parallels to enzymes, proteins, and genetic coding may evolve on an exoplanet and produce analogous methylated gases. Additionally, methylation uses abundant atom types (C and H) and has been shown to utilize a broad variety of substrates including As, Bi, S, Se, Te, Cl, Br, and I (Thayer 2002; Chasteen & Bentley 2003; Paul & Pohnert 2011; Macdonald et al. 2020). Since there are finite chemical possibilities for life to use and methylated gases are not produced by equilibrium reactions, their kinetic signature is deeply tied to the chemical disequilibrium that underpins life. If the biological production of methane and the presence of these elements is common in the universe, it is reasonable to likewise anticipate methylation of diverse elemental substrates to be similarly widespread. Importantly, because methylation is a metabolic process used for environmental detoxification, among other adaptations, fluxes are not directly related to crustal abundances. Instead, the process represents detoxification and volatilization of local concentrations of metals and halides in the immediate growth environment. Therefore, the biologically mediated volatilization of the element substrates (e.g., Cl, Br, and I) will not necessarily be in proportion to their global crustal abundance. Importantly, to assess the potential of various methylated gases as remote biosignatures, we must also determine their survivability to atmospheric photolysis and chemistry as well as their detectability given their unique absorption properties.

### *1.1. Past Analysis of Methylated Halogen Biosignatures: $CH_3Cl$ and $CH_3Br$*

Since first described as a potential biosignature by Segura et al. (2005), $CH_3Cl$ has been considered a strong biosignature candidate, particularly around M dwarfs where high atmospheric mixing ratios can build up due to the photochemical stellar environment (tens of parts per million). This analysis, and subsequent studies, of the photochemical environment generated by M dwarf stars, have relied on the well-characterized nature of photochemical OH radical cycling on the Earth, which has been established for decades (Comes 1994; Jacob 1999). Recent studies such as those by Scheucher et al. (2018) have confirmed the potential for atmospheric buildup of $CH_3Cl$ in these systems.

Terrestrial planets orbiting late M dwarfs such as TRAPPIST-1e (Gillon et al. 2017) will be among the first targets observed by JWST (Morley et al. 2017). Several recent studies have evaluated the potential for characterizing the atmosphere of TRAPPIST-1e and have extensively examined its potential for biosignature detection and analysis (Lustig-Yaeger et al. 2019; Pidhorodetska et al. 2020; Gialluca et al. 2021). Conveniently, M dwarfs are both the most favorable stellar environment for atmospheric buildup of methylated biosignatures and some of the best targets for transmission spectroscopy in the near to intermediate future. In the immediate future, the results of Gialluca et al. (2021) suggest that $CH_3Cl$ is most favorable to detect using the NIRSpec instrument on JWST due to the lower noise floor at shorter infrared wavelengths. It must be noted that this feature overlaps strongly with $CH_4$ absorption and may be challenging to uniquely resolve using JWST. As previous studies have noted, multiple $CH_3Cl$ spectral bands heavily overlap those of other gases, with the best possible detection location at 13.7 $\mu$m, outside of the range of the JWST MIRI-LRS instrument (Rugheimer et al. 2015).

While TRAPPIST-1e is a popular target for atmospheric modeling, other favorable stellar environments have also been studied for potential near term biosignature characterization. Recently, Kaltenegger et al. (2020) simulated the detectability of atmospheric biosignatures, including $CH_3Cl$, for Earthlike planets orbiting white dwarfs, which have an observational advantage due to the large relative size of the planet to the star and much higher likelihood of transit. Recent discoveries of (nonterrestrial) white dwarf planets (Vanderburg et al. 2020; Blackman et al. 2021) have further motivated the study of potentially habitable white dwarf planetary atmospheres, to utilize their unique observational advantages (Lin et al. 2022). If an appropriate transiting white dwarf planet is found, such a promising target could be well-characterized by JWST, though substantial uncertainties remain regarding the likelihood of generating or maintaining habitability given the post-main-sequence evolution of the host star and corresponding consequences for the planet.

Hypothetical Earthlike or Earth-similar planets are a common proving ground for simulating the detectability of biosignatures since they require the least speculative divergence from our only validated example of an inhabited world. However, the relatively small size and high molecular weight atmosphere of Earth impede the detectability of atmospheric gases in the transmission spectroscopy observing mode relative to larger and/or lower molecular weight atmospheres. Some recent studies have evaluated "super-Earth" (1–1.5 $R_{\mathrm{Earth}}$) planets for detection of biosignatures using JWST and other current or future instrumentation. Wunderlich et al. (2021) considered the potential for detection of $CH_3Cl$ in the atmosphere of LHS 1140b, a super-Earth, finding $CH_3Cl$ to be a leading biosignature candidate in $CO_2$- and $H_2$-dominated atmospheres. Madhusudhan et al. (2021) also analyzed $CH_3Cl$ for a "Hycean" world, a super-Earth with an $H_2$-dominated atmosphere over an ocean. They found that $CH_3Cl$ at 1 ppmv abundance could be detected with $3\sigma$ confidence on K2-18b, a JWST GO Cycle 1 target. Interpretation of observations of these super-Earth targets will require an abundance of planetary and stellar context, and caution given the unique properties and the needed extrapolation beyond the validated Earth example. However, their observational favorability makes them among the first targets for atmospheric characterization in the coming years.

### *1.2. Additional Methylated Biosignatures*

$CH_3Cl$ is the most established, but not the only, methylated exoplanetary biosignature gas candidate described in the literature. Domagal-Goldman et al. (2011) showed that another methylated gas, $(CH_3)_2S$, can have a detectable presence in anoxic atmospheres similar to the early Earth (Pilcher 2003), both as a primary biosignature with low near-UV (NUV) flux stars such as M dwarfs, and also as a secondary biosignature by changing the ratio of $CH_4$ and $C_2H_6$. This results from the photochemical conversion of methyl radicals from $(CH_3)_2S$ or $(CH_3)_2S_2$. The suggestion of $(CH_3)_2S$ as a biosignature in anoxic environments further motivates the exploration of additional methylated biosignature gases.

Given the demonstrated biosignature potential of previously established methylated gases, here we analyze a novel methylated biosignature gas, $CH_3Br$, and present a roadmap for continuing evaluation of future methylated biosignature gas candidates. $CH_3Br$, like $CH_3Cl$, is a methyl halide with the general formula $CH_3X$ (X = F, Cl, Br, I;). Note that fluorinated organics are rare, and no volatile methylated F gases have been detected from direct biogenic production (Carvalho & Oliveira 2017).

$CH_3Br$ is a trace constituent of Earth's atmosphere (Schaefer & Fegley 2008) and has been briefly described as a biosignature in past studies (Seager et al. 2016). Messenger (2013) examined the signal-to-noise ratio (S/N) needed to detect biosignature gases with large ground-based telescopes using transmission spectroscopy and determined that $CH_3Br$ could be detected with $3.5\sigma$–$6.6\sigma$ confidence using a 35 m telescope with varying atmospheric abundances from 1 ppm to 100%. However, this study did not consider the self-consistent conditions necessary to maintain high levels of atmospheric $CH_3Br$ nor simulate its absorption signature in emitted light. The application of current and next-generation space telescopes was also not considered.

Our objective includes self-consistently modeling the photochemistry of $CH_3Br$ in terrestrial exoplanetary atmospheres across a range of stellar hosts amenable to either direct imaging in emitted light or transmission spectroscopy in the near- to mid-infrared. We consider a broader parameter space than previous studies spanning F4V to M8V host stars, a range of methylated gas surface fluxes informed by empirical measurements of Earth life, and an additional methylated gas, $CH_3Br$. We also include a reanalysis of $CH_3Cl$, a previously studied biosignature gas, as a benchmark to contextualize our results. The self-consistent, vertically integrated photochemical and spectral modeling of $CH_3Br$ is the first step to expanding the evaluation of halomethanes and other alternative methylated gases as a general class of biosignatures. Figure 1 shows an artistic impression of an exoplanet with possible biogenic methylated gas production, buildup, and detection.

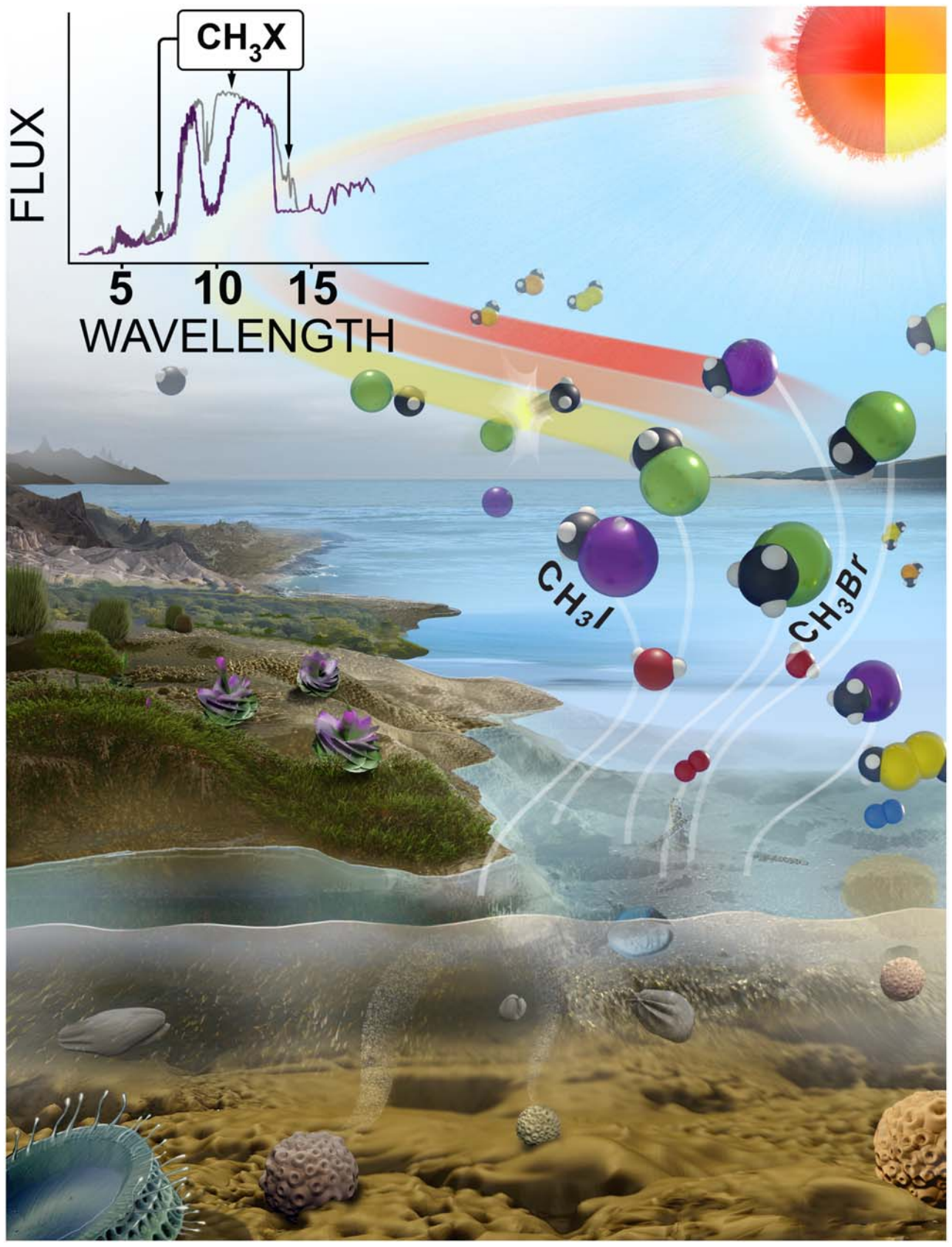

**Figure 1.** Concept figure showing the production, reaction, and absorption by $CH_3Br$ in a theoretical exoplanet. Methylated chalcogens such as $(CH_3)_2Se$ are shown in orange and yellow.

In Section 2, we describe the photochemical and spectral models used in this study and their assumed boundary conditions. In Section 3, we quantify the photochemical buildup of $CH_3Br$ and $CH_3Cl$ under a variety of conditions including variations in flux and the spectral type of the host star. Section 4 presents synthetic emission and transmission spectra of $CH_3Br$ and $CH_3Cl$ for a range of fluxes examined in Section 3 and provides a proof-of-concept test case for detecting these gases on TRAPPIST-1e in transmission for global environments similar to the most productive seen on Earth. In Section 5, we discuss methylated biosignature gases as a general class of biosignature, consider potential false positives, and outline avenues for future work. We conclude in Section 6.

## 2. Methods

To explore atmospheric buildup of novel methylated biosignatures, we used a number of models, beginning with the atmos 1D photochemical model, originally developed by Kasting et al. (1979), expanded and modified by Zahnle et al. (2006), and most recently updated by Arney et al. (2016) and Lincowski et al. (2018). Atmos is often used to simulate exoplanetary atmospheres and biosignatures (e.g., Domagal-Goldman et al. 2011; Arney et al. 2016; Schwieterman et al. 2019; Teal et al. 2022). For this work, we modified atmos to include bromine (Br) chemistry by incorporating Br species and expanding the chemical reaction network.

### 2.1. Atmos Modifications

To modify the atmos photochemical model to include bromine gases, we started from the ModernEarth+Cl reaction scheme template (Catling et al. 2010). We added Br reactions sourced from Burkholder et al. (2015) and Burkholder et al. (2020). The newly added reactions create closed-loop cycles for Br, BrO, HOBr, $CH_3Br$, $CH_2Br$, and $BrONO_2$. We generated a new reaction scheme template that can be easily run in atmos. Additional photochemical cross sections for $CH_3Cl$ and $CH_3Br$ were added from the Max Planck Institute MAINZ cross section database (Keller-Rudek et al. 2013; Burkholder et al. 2015). Our modified version of atmos contains 89 unique photochemically and kinetically active species and 413 reactions.

### 2.2. Atmos Inputs and Boundary Conditions

For all cases presented here, we assumed an Earthlike bulk atmosphere with 21% $O_2$ and 78% $N_2$. Likewise, we assumed the Earth's temperature–pressure profile with a globally averaged surface temperature of 288 K. We fixed these parameters to ensure that our results were dependent only as a function of changing surface molecular fluxes for $CH_3X$ gases. Our base surface molecular fluxes of Br-containing species are sourced from Yang et al. (2005). We also adopt rainout parameters for HOBr, HBr, and $Br_2$ from Yang et al. (2005) to account for wet deposition of these gases. When using globally averaged $CH_3Br$ flux values from Yang et al. (2005), equivalent to $5.13 \times 10^6$ molecules $cm^{-2}$ $s^{-1}$, our model predicts an atmospheric mixing ratio of 11 ppt $CH_3Br$, which is about 122% of the modern-day concentration of 9 ppt (Carpenter et al. 2014). We chose to replicate the modern system because available information about gas sources and sinks is based in the contemporary system. The modern value is less than a factor-of-two greater than the preindustrial value of 5.5 ppt. This is a reasonable approximation because the atmospheric mixing ratio is known to be highly spatially and temporally variable in the Earth's atmosphere, so a calculation that agrees within an order of magnitude can be considered an accurate model output. There is a known discrepancy in the $CH_3Br$ sources and sinks on the Earth, and error introduced by this is an additional explanation for our output value (Carpenter et al. 2014; Seinfeld & Pandis 2016; Engel et al. 2019).

To produce the 9 ppt modern-day atmospheric abundance of $CH_3Br$, a surface flux of $4.19 \times 10^6$ molecules $cm^{-2}$ $s^{-1}$ is necessary. This is the flux level we use for the Earthlike cases in this paper. The corresponding $CH_3Cl$ flux necessary to produce an atmospheric abundance of 0.5 ppb $CH_3Cl$ is $2.25 \times 10^8$ molecules $cm^{-2}$ $s^{-1}$ (Seinfeld & Pandis 2016). We adopt this flux value for $CH_3Cl$ as our base case throughout the paper.

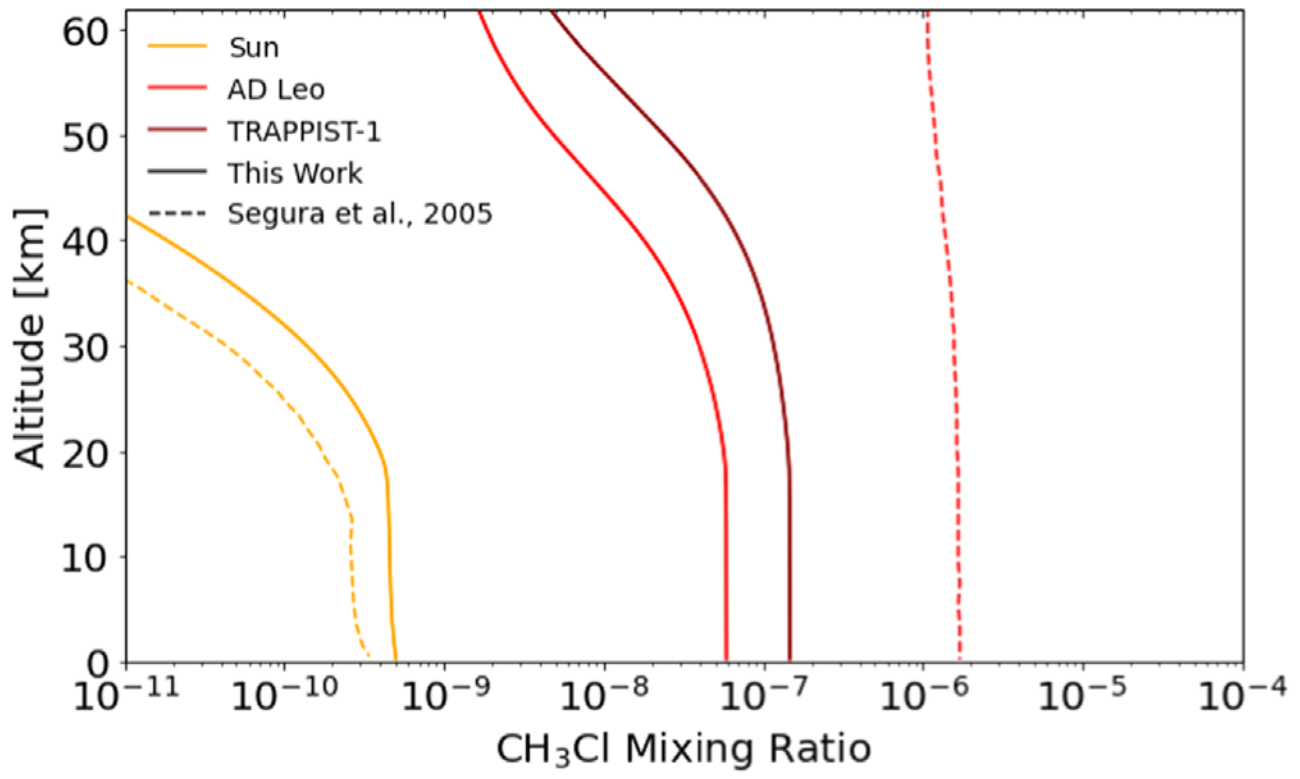

**Figure 2.** Comparison of mixing ratios from Segura et al. (2005) and this work for the Sun and AD Leo. Our work also shows the photochemical buildup around TRAPPIST-1, which is the latest-type M dwarf typically evaluated for exoplanet studies. We construct a $CH_3Cl$ profile with surface mixing ratio of 0.5 ppb based on best available data for the Earth (Seinfeld & Pandis 2016), which is slightly higher than the profile used by Segura et al. (2005). We also generate a mixing ratio profile for AD Leo that is lower than than the previous results by more than an order of magnitude. These changes can be attributed to significant overhauls of the atmos model in the intervening years (e.g., Arney et al. 2016; Lincowski et al. 2018).

For the TRAPPIST-1 system, we used the planetary parameters reported by Agol et al. (2021). We added spectra for the K type stars from the MUSCLES program (version 2.2, Loyd et al. 2016), and used a median averaged TRAPPIST-1 spectrum from Peacock et al. (2019). Other stars, reactions, and chemical species included were previously described in Arney et al. (2016) and Arney (2019). For a complete list of our model boundary conditions, see Table 2 in Appendix A. The complete reaction list can be found in Table 3 in Appendix B. The spectra used for Proxima Centauri and TRAPPIST-1 are scaled to the flux received by the habitable zone planets Proxima Centauri b and TRAPPIST-1e, respectively, resulting in minor deviations from overall trends with stellar type. The planetary size and gravity for these planets are also self-consistently modeled throughout the atmos runs shown here.

### *2.3. Radiative Transfer Model*

To generate synthetic spectra based on the atmospheric profiles from atmos, we used the Spectral Mapping and Radiative Transfer (SMART) model (Meadows & Crisp 1996; Crisp 1997). SMART has been repeatedly validated against solar system objects (Tinetti et al. 2005; Robinson et al. 2011; Arney et al. 2014; Robinson et al. 2014; Schwieterman et al. 2015) and is frequently used to model terrestrial exoplanetary atmospheres (e.g., Meadows et al. 2018b; Lincowski et al. 2018; Lustig-Yaeger et al. 2019). The model is well suited to this task as it is both highly flexible, with the addition of novel gases into the atmosphere easily handled, and has very high resolution by default. As input, SMART uses absorption profiles generated by the Line-by-Line Absorption Coefficient model using the HITRAN line lists (Gordon et al. 2022). SMART also uses separate cross section files to account for opacities not contained within HITRAN. For $CH_3Br$, we incorporated supplementary cross sections from the NIST WebBook (National Institute of Standards & Technology 2018), parsed using the JCAMP Python package.[9] Specifically, for $CH_3Br$, the HITRAN line lists cover from 5.86–12.59 $\mu$m, with supplementary cross sections covering the remaining wavelengths. For $CH_3Cl$, the HITRAN region is from 3.13–15.4 $\mu$m, which spans the same region as available cross sections.

### *2.4. Instrument Models*

We used the Planetary Spectrum Generator (PSG) to simulate transmission spectra and feature detectability for TRAPPIST-1e. PSG is a public access radiative transfer suite that is capable of simulating observables for a variety of objects, compositions, and instruments (Villanueva et al. 2018). For this paper, we primarily used the ability of PSG to simulate realistic noise for JWST and future concept missions to enhance the realism of our simulated data. PSG has been used extensively to quantitatively simulate exoplanet observables (Parmentier et al. 2018; Fauchez et al. 2020; Pidhorodetska et al. 2020; Suissa et al. 2020) and for solar system simulations, such as to retrieve Martian surface features (Liuzzi et al. 2020).

## 3. Photochemical Buildup of $CH_3X$ Gases

### *3.1. $CH_3Cl$ Revisited and Flux–Abundance Relationships*

First we try to reproduce the $CH_3Cl$ concentrations predicted by Segura et al. (2005) for the Sun and AD Leo. For the Earth–Sun case, we produce a $CH_3Cl$ profile with a surface mixing ratio of 0.5 ppb (Seinfeld & Pandis 2016), which is slightly offset at high concentrations from simulations by Segura et al. (2005). However, while Segura et al. (2005) found a surface-level $CH_3Cl$ concentration of ∼1 ppm for an Earthlike planet orbiting AD Leonis, an M3.5V star, we calculate a concentration of only 0.07 ppm for Earthlike surface fluxes, which is more than an order of magnitude lower. This is likely due to intervening atmos model updates (e.g., Domagal-Goldman et al. 2011; Arney et al. 2016; Lincowski et al. 2018; Ranjan et al. 2020; Teal et al. 2022). Specific intervening improvements to the model include expansion of the reaction list and updates to the reaction rates, extensions of the $H_2O$ cross sections (Ranjan et al. 2020), and enhanced resolution of the wavelength grid used for the input stellar spectra and molecular cross sections. For a complete list of changes to the reaction list made since Segura et al. (2005), see Table 3 in Appendix B. We find a steady-state $CH_3Cl$ mixing ratio of ∼0.2 ppm from Earthlike flux levels for the TRAPPIST-1 case, which represents the largest enhancement of the atmospheric mixing ratio of any stellar host investigated. A comparison of the mixing ratio profiles from our simulations and Segura et al. (2005) can be seen in Figure 2. Our results here suggest that future studies incorporating $CH_3Cl$ should carefully check and update their reaction schemes.

Figure 3 shows the atmospheric mixing ratios of methyl chloride for seven different stellar types and five flux scenarios from our photochemical calculations. The thick and thin lines show the mixing ratio for 10 and 1000 times the Earth's average surface flux, respectively, while the central white circle shows the atmospheric abundance maintained for Earth flux levels. For the F, G, and early K types, there is little additional buildup in the $CH_3Cl$ level, shown at the bottom of the plot. The late K and M types show significant enhancement in the atmospheric abundance with an increase by a factor of 1000 for the latest-type star (TRAPPIST-1, M8V). For three cases, M3.5e, M5V, and M8V, at least 1 ppm atmospheric concentrations can be reached with 10× the Earth's surface flux. For the F4, G2, and K cases, a much higher surface flux level than considered here is necessary to reach parts per million levels of abundance (1000× Earthlike or greater). Our incident stellar spectra for Proxima Centauri and TRAPPIST-1 are scaled to their habitable zone planets, which results in the Proxima Centauri planet building up slightly higher $CH_3Cl$ fluxes than the TRAPPIST-1 case due to lower incident total and UV fluxes.

[9] https://pypi.org/project/jcamp/

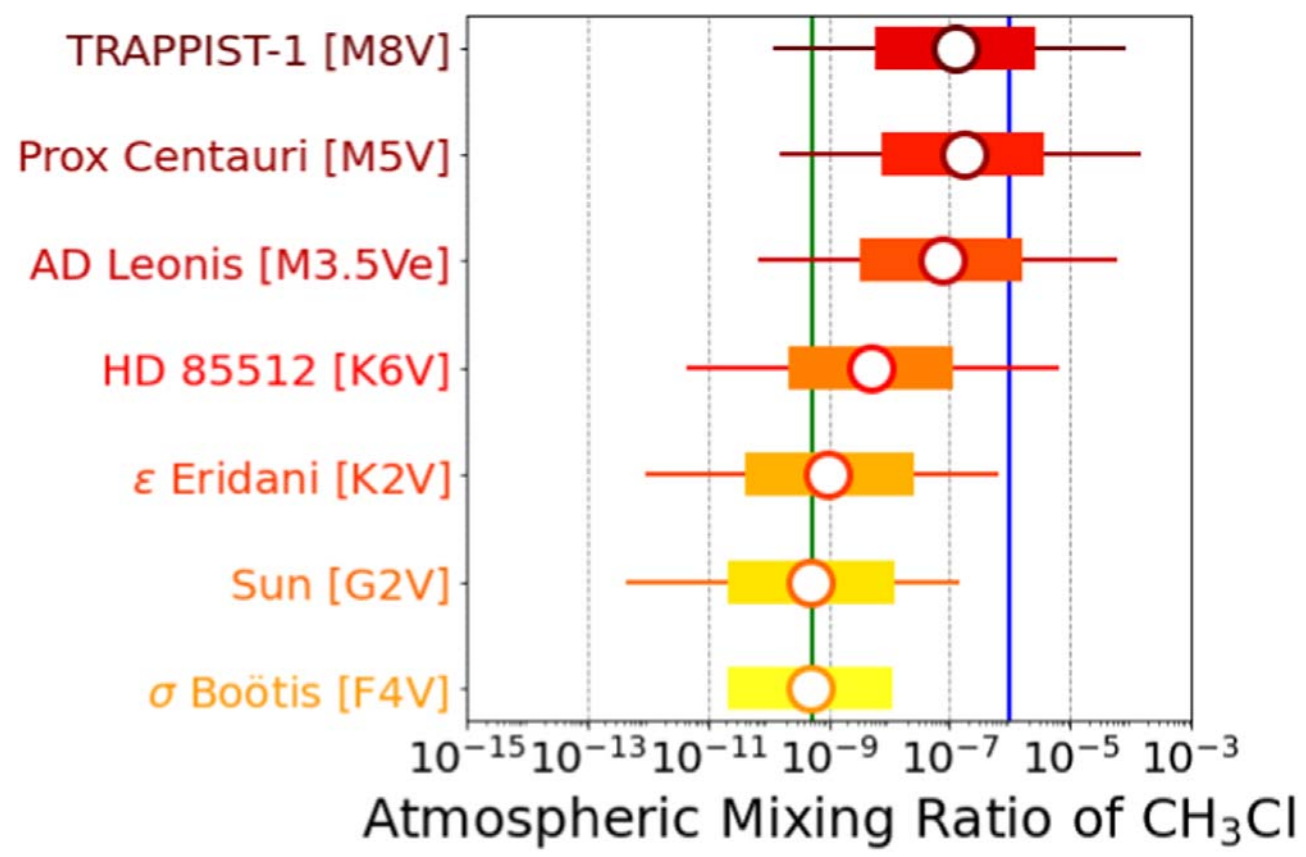

**Figure 3.** Buildup of $CH_3Cl$ in planetary atmospheres around different stellar types. The white circle shows the Earth's flux levels, the thick bar shows an order of magnitude above and below, and the thinner line represents a 1000× enhancement or depletion of the flux vs. Earth's global average. The green line shows the Earth's global average mixing ratio, and the blue line indicates 1 ppm.

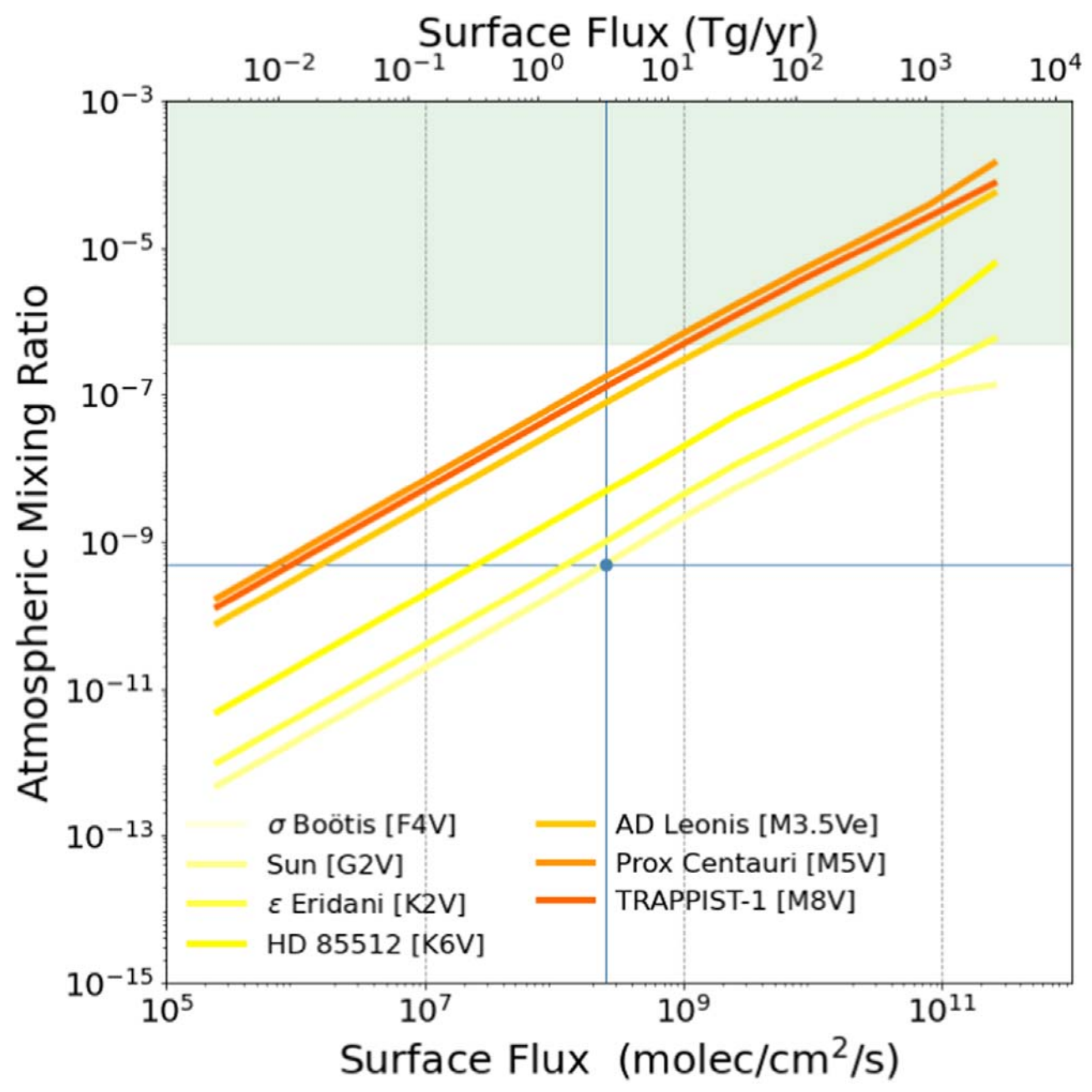

**Figure 4.** Atmospheric buildup of $CH_3Cl$. Crosshairs show the Earth flux and atmospheric abundance. The region highlighted in green is comparable to preindustrial methane concentrations on Earth.

Figure 4 shows the similar flux–abundance results for $CH_3Cl$ flux scenarios between a factor of 1000 higher or lower than the modern Earth's globally average surface flux. For context, salt marshes on the Earth, a highly productive ecosystem for methylated gases, locally produce 1000× the globally averaged surface flux of $CH_3Cl$ and $CH_3Br$ (Rhew et al. 2000). A 1000× flux scenario therefore simulates a planet where the majority of the surface is covered in highly productive environments, which include oceans, swamps, and terrestrial ecosystems. The green area highlights the region where the mixing ratio reaches 0.5 ppm, a concentration comparable to preindustrial methane on the Earth. As shown in the previous plot, the later star types consistently have higher atmospheric mixing ratios for the same surface flux values with the enhancement becoming correspondingly greater at higher flux values. The predicted abundance at these high levels reaches into the hundreds of parts per million, levels comparable to $CO_2$ on the modern Earth.

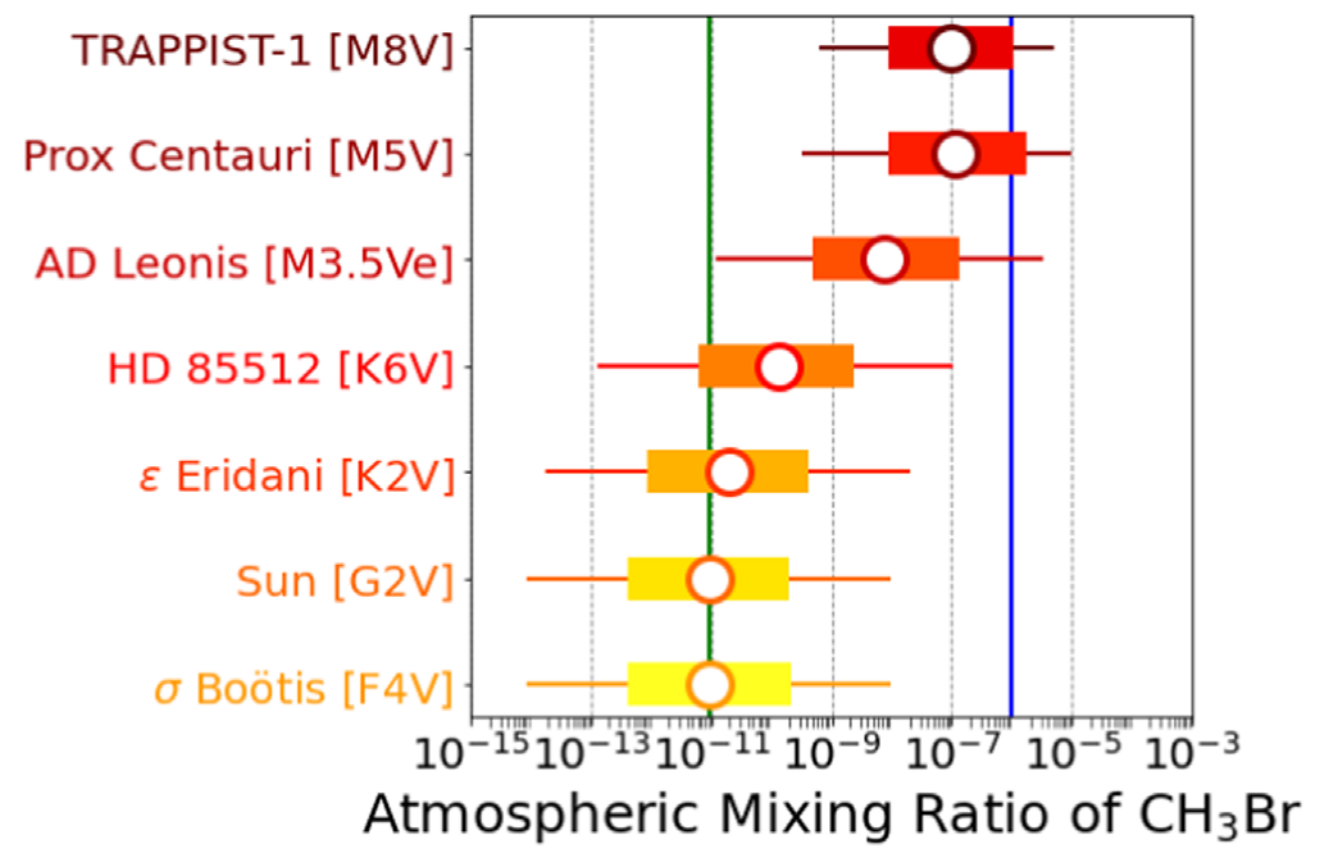

**Figure 5.** Atmospheric buildup of $CH_3Br$ in planetary atmospheres around different stellar types. The white circle shows the Earth's flux levels, the thick bar shows an order of magnitude above and below, and the thinner line represents a 1000× extension of the flux. The green line shows the Earth's mixing ratio, and the blue line indicates 1 ppm. Compare to Figure 3.

The $CH_3Cl$ levels are a strong function of stellar type because the OH radical is the dominant photochemical sink for $CH_3Cl$. As originally described in Segura et al. (2005) and elaborated by numerous workers thereafter, OH production is strongly sensitive to the NUV photon flux from the star, which is sourced from the stellar photosphere and not from magnetic activity, and is thus a strong function of stellar effective temperature. Tropospheric $O_3$ is photolyzed by NUV photons ($\lambda < 340$ nm) to yield $O^1D$ ($O_3 + h\nu \rightarrow O_2 + O^1D$). This $O^1D$ subsequently reacts with water to yield hydroxl radicals ($O^1D + H_2O \rightarrow 2$ OH). These OH radicals then destroy $CH_3Cl$ ($OH + CH_3Cl \rightarrow CH_2Cl + H_2O$). This process is highly dependent on NUV photons in a narrow wavelength range that are not shielded (absorbed) by the overlying $O_2$ and $O_3$ in the atmosphere, but nonetheless possess the energy to photolyze $O_3$, and results in a strong dependence on the Wien tail of the star's photospheric emission. Hence, there is a very strong inverse correlation between the abundance of a trace gas whose major sink is OH, and the effective temperature of the host star. This is the correlation we see for $CH_3Cl$ in Figures 3 and 4 and will be demonstrated for $CH_3Br$ below.

### 3.2. $CH_3Br$ Flux–Abundance Relationships

Figure 5 shows $CH_3Br$ mixing ratios for selected FGKM stars (the selected stars are the same ones considered for $CH_3Cl$). The general trends show the same qualitative relationship as demonstrated for $CH_3Cl$ above. For the F through K2 stars, there is a consistent trend with small increases in the central (average Earth-flux equivalent) value as well as the endpoints with decreasing stellar photospheric temperature. Beginning with the K6V case, an inflection point is reached. The gas concentration enhancement becomes increasingly larger, with the late M dwarfs reaching or exceeding 0.1 ppm of atmospheric buildup with the globally averaged Earth's surface flux. For the M dwarf cases considered, greater than $\sim 20\times$ the Earth's average flux will lead to 1 ppm levels of atmospheric buildup or more. The same deviations from the overall trend due to planetary distance scaling of TRAPPIST-1e and Proxima Centauri b shown for $CH_3Cl$ are seen here as well.

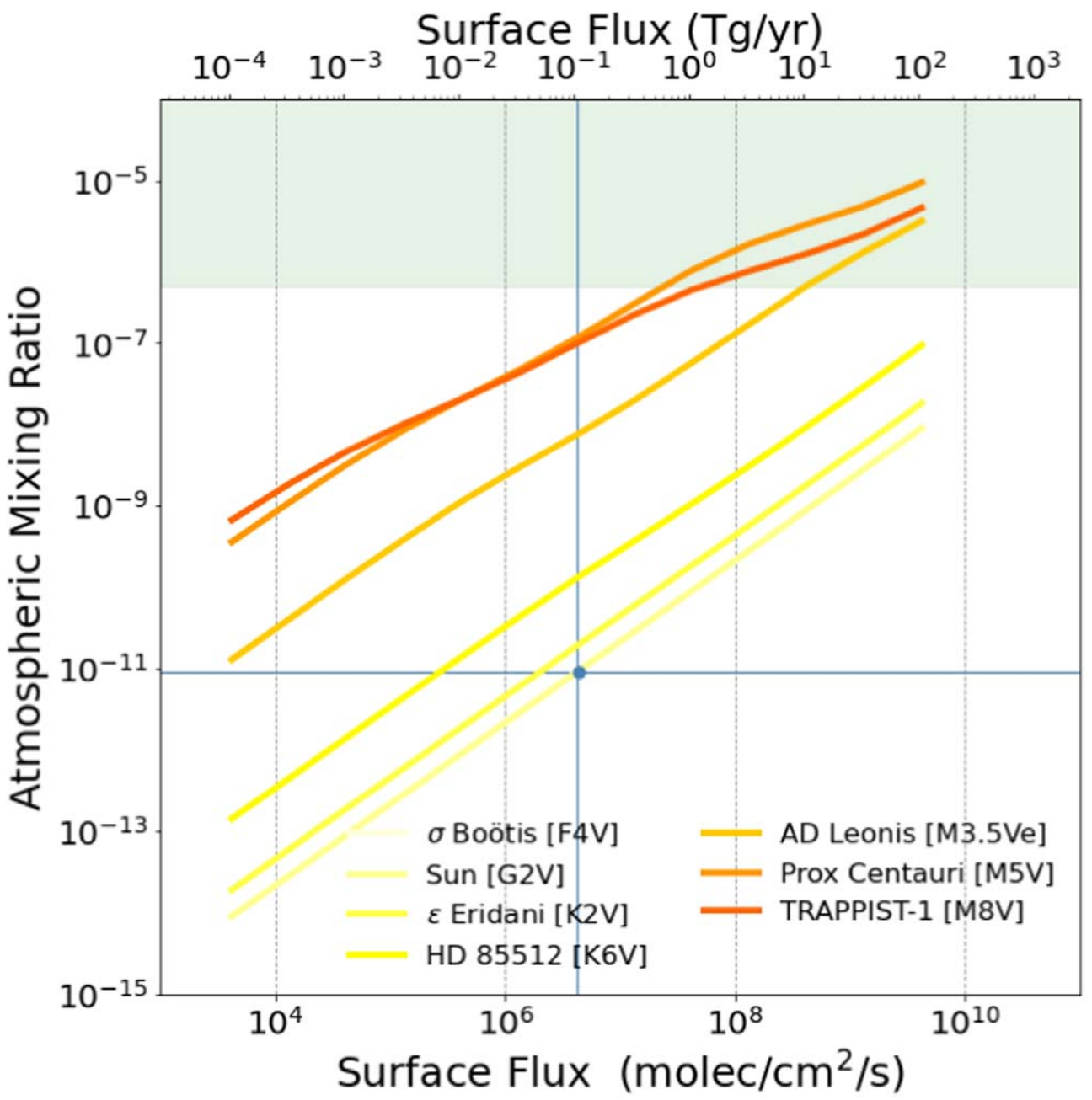

**Figure 6.** Atmospheric buildup of $CH_3Br$ in planetary atmospheres around different stellar types for a variety of surface flux conditions based on those seen on the Earth. Compare to Figure 4. Mixing ratios in the green region are comparable to Earth's preindustrial methane abundance.

Figure 6 shows the $CH_3Br$ mixing ratio for each star as the surface flux increases. This figure shows the difference in outcomes between the early- and late-type stars. The changes in slope and structure of the lines at high fluxes are due to different changing proportional contributions from photochemical sinks, such as radicals generated from photolysis of $CH_3Br$ and downstream impacts on the atmospheric chemistry. The F4V and G2V stars show the same flux–abundance relationship except at the highest fluxes. We do not see indications of photochemical runaway in the flux–abundance relationship at the highest fluxes examined here. This is a logical result since the fluxes explored here are approximately an order of magnitude less than those in Figure 4.

While both $CH_3Cl$ and $CH_3Br$ have similar flux–abundance relationships, the relative enhancement from early-to-late-type stars is substantially greater for $CH_3Br$ versus $CH_3Cl$. Specifically, the enhancement in atmospheric concentration is 4 orders of magnitude for $CH_3Br$ when comparing a Sunlike (G2V) host and TRAPPIST-1 (M8V), while $CH_3Cl$ shows an enhancement of just over 2 orders of magnitude for the same scenario.

We also tested the sensitivity of our results to changes in $CO_2$. Increasing the $CO_2$ mixing ratio from 400 ppm to 5% $CO_2$ results in minor changes in the $CH_3Cl$ and $CH_3Br$ concentrations of at most 3%. This result matches that found by Wunderlich et al. (2021), who also reported additional buildup of $CH_3Cl$ in a high $CO_2$ atmosphere. However, we find changes to $CO_2$ abundances do not significantly impact the photochemical outcomes of methylated gases in our study.

Additionally, we explored the sensitivity of our results to surface temperature and temperature profiles. Higher temperatures result in greater $H_2O$ mixing ratios due to higher vapor pressures, which in turn supports the production of more OH radicals, which is the major sink for $CH_3Cl$ and $CH_3Br$. The methylated gas buildup is therefore sensitive to changes in temperature. Specifically, we find a 50% concentration decrease for the K6V stellar type when using a surface temperature of 300 K. For a surface temperature of 275 K, we find a maximum of sixfold increase in concentration for Earthlike flux levels for the same stellar type. However, because the production of OH decreases at later-type stars where the highest buildup is otherwise seen, the K6V simulations show the largest projected change. For planets orbiting the Sun, the increase is threefold for cooler (275 K) planets and the decrease is 25% for the warmer (300 K) scenario. We note that on a low-temperature planet considered here, biological productivity would likely be limited via midlatitude glaciation. While these gases are sensitive to temperature, the impact is asymmetrical, as higher temperatures result in small decreases while lower temperatures result in substantially higher concentrations. Since we assume a surface temperature of 288 K, our simulations are conservative with respect to temperature effects.

Since both $CH_3Cl$ and $CH_3Br$ are produced on Earth, we considered the impact of both gases at high levels in a planetary atmospheres. We find that the introduction of additional Cl radicals from $CH_3Cl$ pushes $CH_3Br$ into a runaway state at $1000\times$ the Earth's flux level of both gases. This coadditive flux results in hundreds of parts per million of $CH_3Br$. In the next section, we consider the observing potential of both pre- and postrunaway conditions.

## 4. Simulated Spectra

To understand the spectral impact of our $CH_3Br$ and $CH_3Cl$ results, we first used SMART to generate moderate resolution spectra ($R = 1000$) for a broad wavelength range in the mid-infrared. Figure 7 shows the result of a grid of moderate resolution SMART emission spectra for a parameter space including several surface flux scenarios of (from top to bottom) $CH_3Cl$, $CH_3Br$ and an atmosphere with both gases. We simulated emission spectra for Earth around a late K (HD 85512), early-mid M (AD Leo), and mid-late-M (Proxima Centauri) dwarf star because those will be favorable targets for ground-based ELTs and future space-based mid-infrared interferometers (Defrère et al. 2018; Fujii et al. 2018; Quanz et al. 2018, 2022; López-Morales et al. 2019).

The absorption features in the emission spectra show increasing width and depth as a function of the molecular flux levels. AD Leo and Proxima Centauri show significantly larger atmospheric features than those modeled for HD 85512, consistent with our photochemical modeling showing higher $CH_3Cl$ and $CH_3Br$ buildup for planets orbiting M dwarf stars. The strongest emission features for $CH_3Cl$ can be seen at 7, 9.8, and 13.7 $\mu$m. The combined absorption core of the 9.8 $\mu$m feature becomes less deep in the intermediate-flux cases resulting from $O_3$ depletion at 9.65 $\mu$m. The increase in absorption centered around 9.65 $\mu$m is much larger in the simulations of AD Leo and Proxima Centauri. There are $CH_3Cl$ features located at 3.4, 4.1, and 13.7 $\mu$m as well, with the near-infrared features being confounded by other absorbers. For $CH_3Br$, the main emission feature can be seen between 9.3 and 11.75 $\mu$m with two additional features at 7.7 $\mu$m and 17.5 $\mu$m.

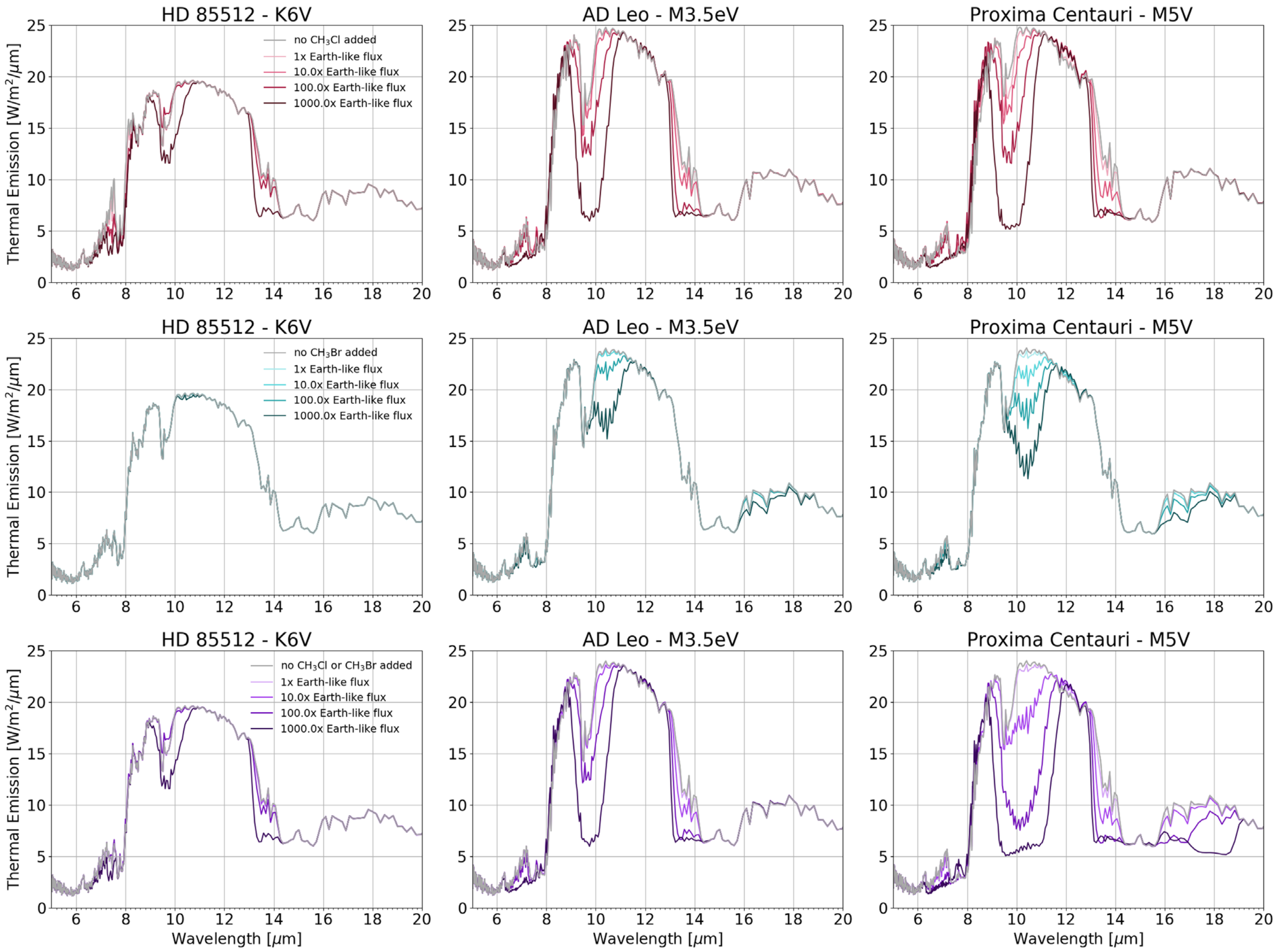

**Figure 7.** Comparison of emission features for atmospheres with methylated biogenic gases around a K6V, M3.5Ve, and M5V star. From top to bottom: $CH_3Cl$, $CH_3Br$, and combination of $CH_3Cl$ and $CH_3Br$ for a range of fluxes ranging from globally averaged modern-Earth fluxes to most productive local terrestrial and marine environments. Spectra are shown for a 50% clear sky, 50% cloudy with equal parts cirrus and stratocumulus clouds.

The atmospheric scenarios with both gases (bottom panel in Figure 7) shows a combination of these two features with increased absorption at ∼7 $\mu$m, 10 $\mu$m, 13.7 $\mu$m, and 17.5 $\mu$m. There is a minor decrease in absorption at 9.65 $\mu$m resulting from the destruction of $O_3$, primarily by $CH_3Cl$. At fluxes between 10× and 100× the globally averaged flux, the absorption features of these methylated gases become comparable to other major atmospheric absorption features.

We consider the potential application of high-resolution ground-based spectroscopy to image the absorption of these gases around 10 $\mu$m, a wavelength regime where both $CH_3Cl$ and $CH_3Br$ absorb (with absorption peaks at 9.8 and 10.5 $\mu$m, respectively). Figure 8 compares high-resolution ($R = 100{,}000$) transmission spectroscopy of Proxima Centauri b for four different simulated atmospheres (productive $CH_3Cl$, productive $CH_3Br$, and both gases). The spectra have been separated so the band structure is clear. The two methylated gases, $CH_3Cl$ and $CH_3Br$, have distinct absorption bands with $CH_3Br$ having more dense lines over the same region, compared to $CH_3Cl$. The atmosphere with both gases shows features of each methylated gas absorber. Any of the added-gas atmospheres are clearly distinct from the base atmosphere case with no $CH_3Cl$ or $CH_3Br$, shown at the bottom of the plot in gray.

### 4.1. Detectability Estimates

Using the PSG (Villanueva et al. 2018, 2022), we simulated observations using current and proposed space-based telescope concepts. Following the methods of Pidhorodetska et al. (2020), we calculated S/Ns and determined the number of transits necessary to detect atmospheric features. We also add out-of-transit noise using a transit to out-of-transit time ratio of 1:3, which lowers the S/N by a factor of $\sqrt{4/3} \sim 1.155$. This ratio was selected to match planned observations of TRAPPIST-1e with JWST (program 1331, PI Nikole Lewis; for more information see Lewis et al. 2017). See Section 4.1.1 for further discussion of JWST and its ability to detect methylated biosignature gases.

With the goal of these gases being corroborative "capstone" biosignatures, we performed detectability analyses for next-generation instruments that might be capable of the extensive follow-up investigations necessary to confirm a primary biosignature detection.

The Origins Space Telescope is a next-generation mid-to-far-IR 5.9 m telescope concept that was submitted to the 2020 Decadal Survey (Battersby et al. 2018; Meixner et al. 2019). While it was ultimately not selected for further development at the time of writing, its potential capabilities have been thoroughly described, and provide a roadmap for the capabilities of a next-generation IR telescope capable of characterizing exoplanetary atmospheres in transit. We used the Origins concept parameters to explore the detectability of $CH_3Cl$ and $CH_3Br$ features on TRAPPIST-1e. The Origins estimated noise floor of ∼5 ppm is significantly more favorable in the mid IR than the projected MIRI noise floor, and could enable greater analysis of mid IR features.

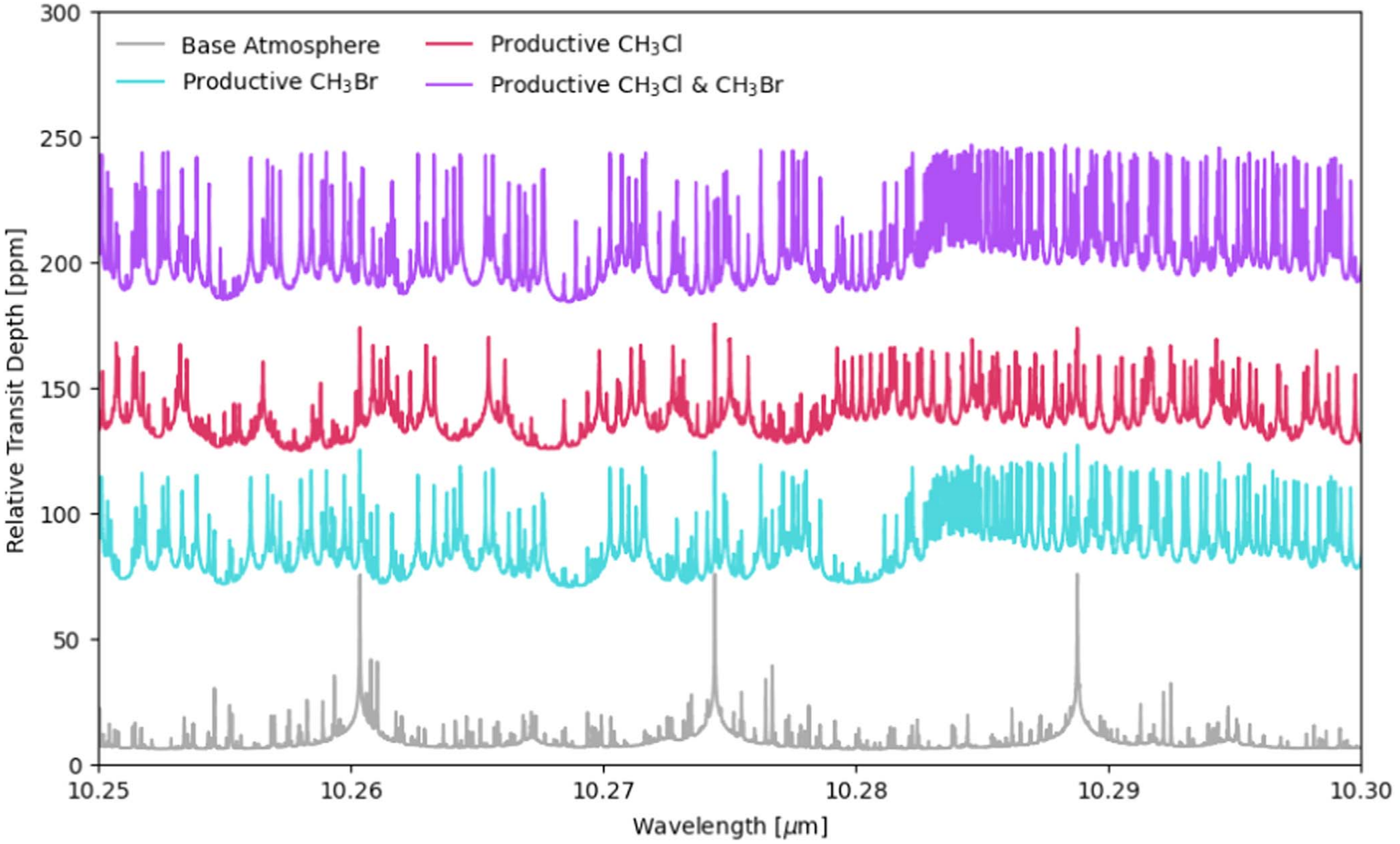

**Figure 8.** Simulated high-resolution ($R = 100{,}000$) transmission spectroscopy of Proxima Centauri b. For maximally productive environments of $CH_3Cl$ (red) and $CH_3Br$ (blue), clearly distinct absorption lines are produced, which can be distinguished from each other, and from the base atmosphere, which is shown in gray. Also shown is an atmosphere with both $CH_3Cl$ and $CH_3Br$, which shows both components (purple). The spectra are vertically offset to better show the line structure and are simulated for a clear sky atmosphere.

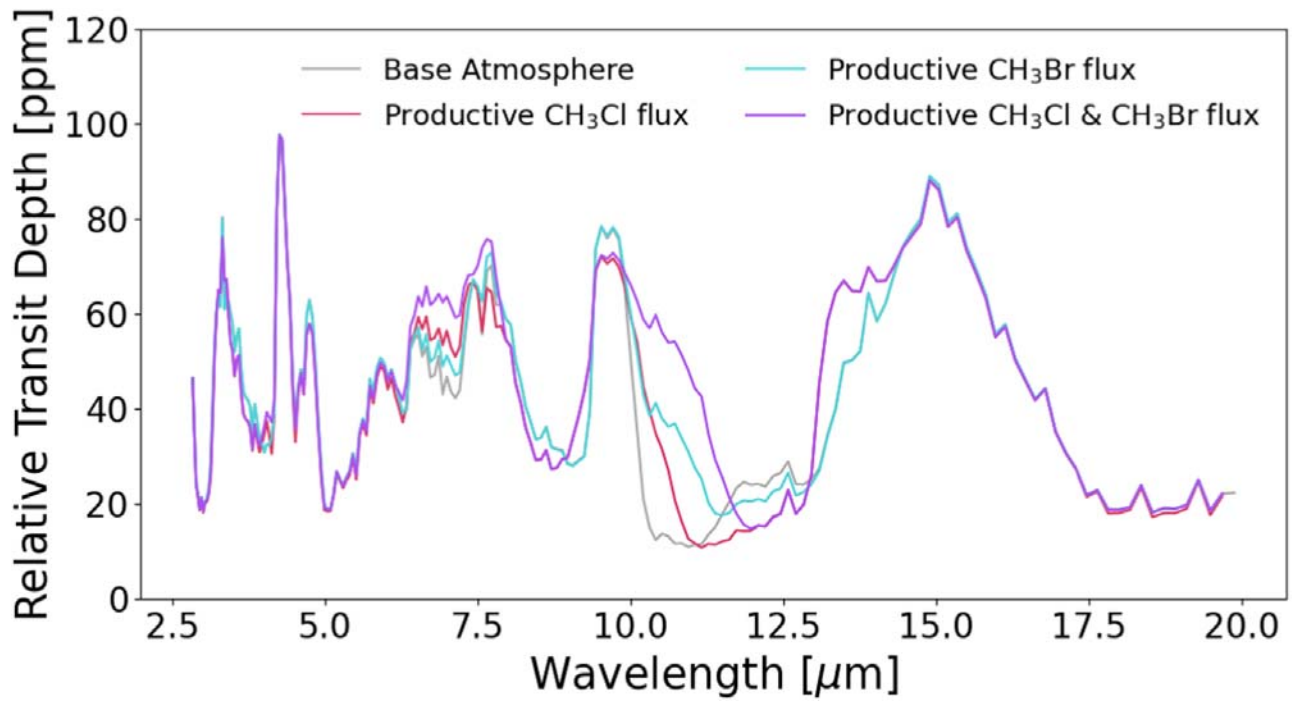

**Figure 9.** Transmission spectra simulated by PSG for 100× Earth-averaged $CH_3Br$, $CH_3Cl$, and both gas fluxes for TRAPPIST-1e. The combined atmosphere typically has an absorption feature the size of or larger than the larger of the two individual gas features, with the major exceptions seen around 3.3 $\mu$m due to decreased $CH_4$ and 9.65 $\mu$m from diminished $O_3$. See Section 4.1 for more information.

Here we examine the detectability of a maximum productivity scenario for $CH_3Cl$ and $CH_3Br$. We consider three test cases: 100× the global Earth average $CH_3Cl$ flux only, 100× the global Earth average $CH_3Br$ fluxes only, and both together. Figure 9 compares transmission spectra for these cases. Generally, the combined $CH_3X$ flux case has a feature equivalent to or greater than the larger of the $CH_3Cl$ or $CH_3Br$ features. Note the feature in the $CH_3Br$ atmosphere is ∼3.3 $\mu$m; this is a $CH_4$ feature resulting from a decrease in the $O_3$ concentration. The $CH_4$ feature is suppressed in the atmospheres containing $CH_3Cl$ because Cl radicals sourced from $CH_3Cl$ destroy $O_3$, which reduces UV shielding of $CH_4$. Consequently, methane abundances are depressed in scenarios with high $CH_3Cl$ fluxes. In addition, Cl radicals directly attack and destroy $CH_4$, further reducing the $CH_4$ mixing ratio. The $O_3$ suppression is significant in atmospheres with $CH_3Cl$, but is modest in those with $CH_3Br$ alone. This gaseous interaction is highly apparent around 9.65 $\mu$m where the usually prominent $O_3$ feature is truncated due to lower high-altitude abundances. Both methylated gases contribute to a broad feature between 9 and 11.5 $\mu$m. The 17.5 $\mu$m $CH_3Br$ feature is not seen because only line-by-line absorbers were used for these simulations, and 17.5 $\mu$m is outside the range of the input line lists. Note that the additive impact of the $CH_3X$ gases on the spectrum is not due to a doubling of the total molecular $CH_3X$ flux. The flux is not doubled, since the Earth's $CH_3Br$ flux is substantially less than the $CH_3Cl$ flux.

Figure 11 compares an Earthlike atmosphere without $CH_3Cl$ or $CH_3Br$ with 100× flux scenario including both gases. It shows the combined error bars for 27 transits, which is the minimum number of transits we determine necessary to detect the presence of methylated gas(es) in a TRAPPIST-1e atmosphere at 3$\sigma$ confidence outside of runaway conditions. A 3$\sigma$ detection would motivate additional follow-up missions, especially for a low false-positive potential detection of a methylated gas. Figure 10

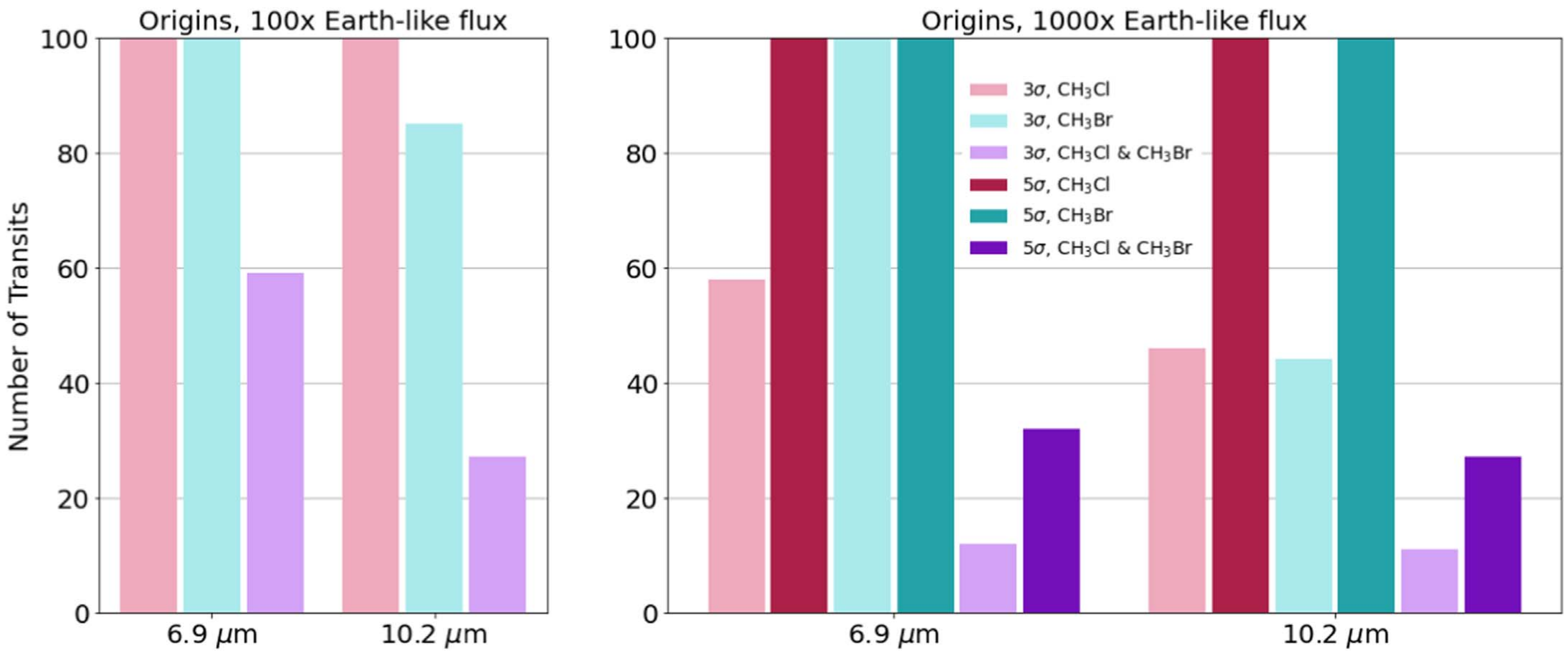

**Figure 10.** Comparison of the number of transits needed to reach $3\sigma$ or $5\sigma$ detection for two different bands where both $CH_3Cl$ and $CH_3Br$ absorb for three scenarios: $CH_3Cl$ only, $CH_3Br$ only, and both gases together. We also consider both the 100× (prerunaway) and 1000× (runaway) flux conditions. All transits are calculated for TRAPPIST-1e, using planetary parameters from Agol et al. (2021) and the MISC-TRA instrument on the Origins Space Telescope concept, simulated using PSG.

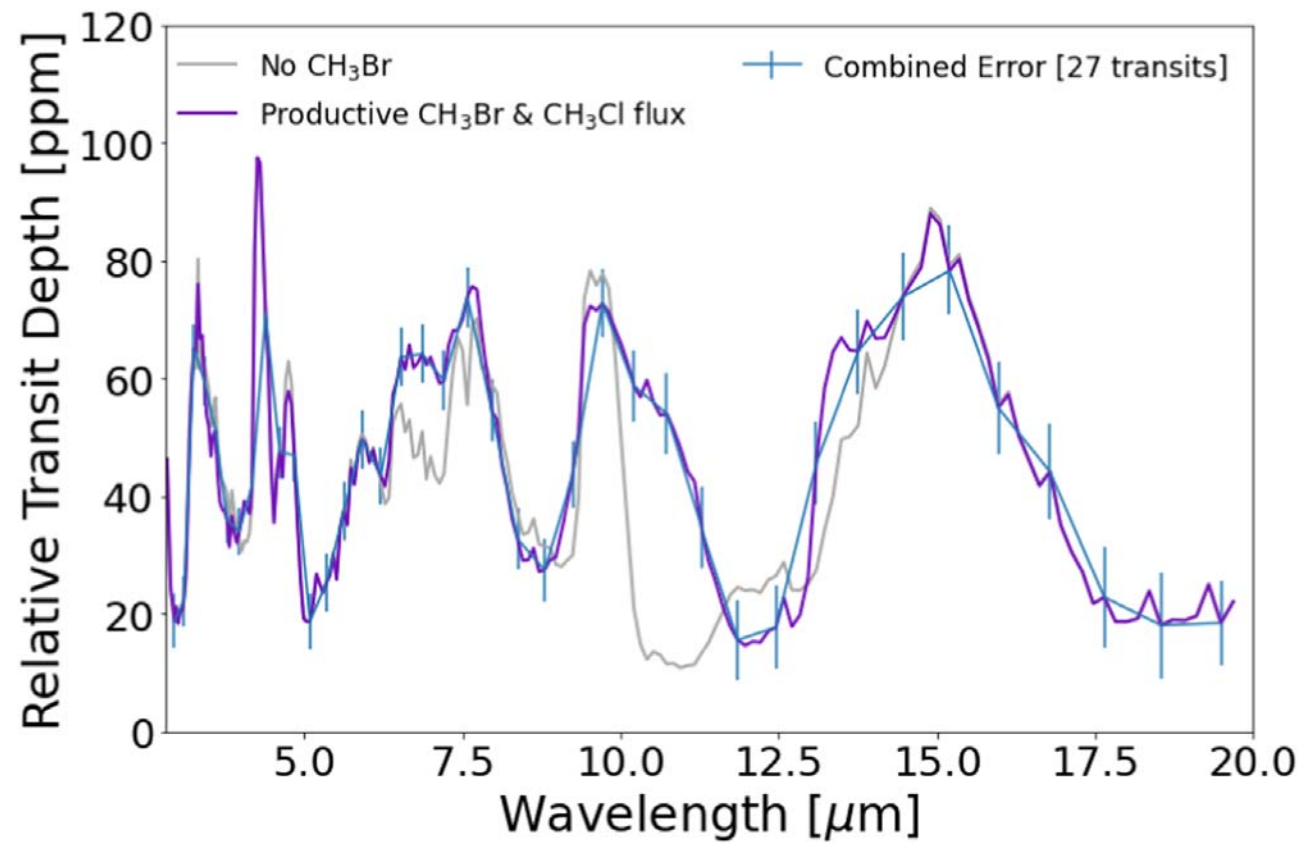

**Figure 11.** Transmission spectra simulated by PSG for 100× Earth-averaged $CH_3Br$ and $CH_3Cl$ flux for TRAPPIST-1e. Assumes Origins space telescope in the MISC-T configuration observation of TRAPPIST-1e. Includes error bars for 27 transits, the minimum number to detect the broad coadded feature around 10 $\mu$m at this flux level.

compares $3\sigma$ and $5\sigma$ detections for pre- and postrunaway scenarios.

Figure 10 shows the number of transits needed to achieve $3\sigma$ or $5\sigma$ confidence at ∼7 $\mu$m and 10 $\mu$m, two wavelength bands where both $CH_3Cl$ and $CH_3Br$ have strong absorption for the two greatest flux regimes considered. The wavelength bands were calculated based on examination of the input opacity data and were kept constant for these calculations. For the atmosphere with both $CH_3Cl$ and $CH_3Br$, calculations of both $CH_3Cl$ and $CH_3Br$ centered features were performed, with the more favorable shown here, although the difference in number of transits is less than two between calculation methods. Although the combined feature is larger than either of the two contributing features, we sample a subsection of it to preserve a baseline comparison with the single gas atmospheres.

For both wavelengths, the most favorable atmosphere to observe would be one with both $CH_3Cl$ and $CH_3Br$. While a scenario with $CH_3Br$ alone is difficult to detect on its own at 7 $\mu$m (>100 transits required) for any flux scenario, the addition of $CH_3Br$ to $CH_3Cl$ significantly reduces the number of transits needed to detect a methylated gas feature to 59 for 100× flux, and 12 for 1000× flux at 7 $\mu$m. For $5\sigma$ confidence at 7 $\mu$m, 27 transits are needed. At ∼10 $\mu$m, $CH_3Br$ is more favorable than $CH_3Cl$, requiring 110 versus 85 transits for 100× flux, but the combined atmosphere still requires the fewest transits to detect (27). For 1000× flux, at $3\sigma$ confidence, the combined atmosphere can be detected in just 11 transits. For $5\sigma$, 27 transits are required for the coadded features. In contrast, the individual gas cases require over 100 transits to detect at $5\sigma$, or 44 at $3\sigma$.

#### 4.1.1. Application of JWST to Methylated Biosignature Gases

We simulated transit observations of TRAPPIST-1e using the NIRSpec and MIRI instruments on JWST. However, in the wavelength range of NIRSpec, there are no distinct $CH_3Cl$ or $CH_3Br$ spectral features that do not significantly overlap with other gases. $CH_3Cl$ absorbs at 3.4 $\mu$m; however, $CH_4$ also has a band at that location, and the nonoverlapping part is ∼0.1 $\mu$m wide. At 4.1 $\mu$m, an additional $CH_3Cl$ absorption feature is confounded by $N_2O$ absorption over the same wavelength range. Our simulations of the MIRI instrument suggest that detecting methylated biosignatures would require a prohibitively large number of transits (>100). This number exceed the number of observable transits (85; Fauchez et al. 2019) during the nominal lifetime of JWST (5.5 yr), even for the largest flux examined in this work, 1000× the Earthlike flux. However, JWST's lifetime may be extended to 10 or 20 yr, theoretically leading to 340 available transits. An extended number of transits would yield results similar to those seen in Figure 11, which is simulated using parameters of the Origins Space Telescope, as described previously.

We also consider the inverse problem of determining what surface molecular flux would be required to detect either gas in a reasonable number of transits. In order to detect these gases in 10 transits with JWST MIRI-LRS, we calculated that it would be necessary to have surface fluxes of approximately $10^{12}$ molecules $cm^{-2}$ $s^{-1}$, at which point the atmosphere would be in an accelerated runaway state similar to that described by Huang et al. (2022), Zhan et al. (2021), and Ranjan et al. (2022). This surface flux, while higher than known production levels, is only 10 times higher than current $CH_4$ production levels.

## 5. Discussion

### 5.1. Additive Effect of Methylated Biosignature Gases

In the previous section, we explored the spectral impact of photochemically self-consistent levels of $CH_3Cl$ and $CH_3Br$ and demonstrated a coadditive effect from overlapping absorption features at high but biologically plausible flux levels. There are two primary contributing factors to this effect. Because of the different atomic masses and electron structures of Br and Cl, and the corresponding difference in the C-Br and C-Cl bond energies, the $CH_3Br$ spectral absorption feature centered at 10.2 $\mu$m is offset from the corresponding 9.9 $\mu$m band of $CH_3Cl$, widening the overall band and enhancing its potential detectability. At modest resolution, the features enhance each other and reduce the number of transits needed to detect the overall $CH_3X$ feature, while at high resolution, the two gas features are distinct. Additionally, both $CH_3Cl$ and $CH_3Br$ compete for the same radical sinks, such as OH, so an added flux of both gases increases their photochemical lifetime relative to a scenario considering each gas alone. Abundances are increased not only at the surface level, but also at higher altitudes containing the needed opacity to create transmission spectral features.

Importantly, this additive effect is not equivalent to simply doubling the flux of a single gas such $CH_3Cl$ because the reaction rates and absorption cross sections vary between the molecules. We demonstrated in Section 3.2 that the "biosignature boost" (mixing ratio increase) is greater in relative terms for $CH_3Br$ than $CH_3Cl$ when altering the host star spectrum from a Sunlike star to that of an M dwarf.

Additionally, both gases' presence may yield additional information. Studies of the Earth reveal that chlorinated metabolic intermediates are more common in terrestrial settings, with bromine-bearing metabolites dominating in marine ecosystems (van Pée & Unversucht 2003). We also maintain the globally averaged Earth's $CH_3Cl$ to $CH_3Br$ ratio, which is known to vary in local environments as much as $2\times$–$20\times$ (Blei et al. 2010). This could result in higher levels of $CH_3Br$ than $CH_3Cl$ prerunaway or for earlier-type stars.

These factors suggest that there may be additional detectability advantages and information yield when considering halomethanes as a general class. For example, in this initial study, we did not include the other major biologically produced methyl halide $CH_3I$, but our results suggest its impact would further enhance the overall $CH_3X$ atmospheric signature.

### 5.2. Methylation of Halogens, Metalloids, and Metals as a General Biosignature Class

Biologically mediated methylation and volatilization of halogens, metals, and metalloids is a common and widespread detoxification process across the domains of life, and across different ecosystems on Earth (e.g., Bentley & Chasteen 2002; Thayer 2002; Chasteen & Bentley 2003; Meyer et al. 2008). A wide variety of elemental substrates are methylated in bacteria, archaea, and eukarya by a variety of enzymes (see Table 1). $CH_3Cl$ and $CH_3Br$ are only two members of a large class of halomethanes, potential methylated biosignature gases. Halomethanes include methyl halides ($CH_3X$; X = F, Cl, Br, I) and polyhalomethanes (e.g., $CH_2Br_2$, $CHBr_3$, and CH2BrI, $CH_2I_2$) with the general formula $CH_{4-a,b,c,d}F_aCl_bBr_cI_d$; X = F, Cl, Br, I. $CH_3F$ is an outlier in this category because it has not been observed to be produced biologically, and has even been shown to arrest methanogenesis in several organisms (Miller et al. 1998; Conrad & Klose 1999). However, some rare organofluorine compounds are formed biologically including fluoroacetate, $\omega$-enriched fatty acids, and complex compounds (Carvalho & Oliveira 2017). These molecules are thought to be relatively scarce due to the high electronegativity of fluorine (inability to form F+ ions), the low solubility of most fluoride-containing minerals, and the poor nucleophilicity of the hydrated fluoride anion (Murphy et al. 2009; Odar et al. 2015). Most of these organofluorine compounds are not volatile, and there is no major biological source of biological fluorinated gases into Earth's atmosphere today. The extent to which these factors would predict correspondingly low organoflourine production by an alien biosphere on an exoplanet is not known and is an area for further astrobiological research.

Halomethanes are produced in marine and terrestrial environments and are widespread across the three domains (Thayer 2002; Paul & Pohnert 2011; St et al. 2016), although the highest production rates today occur in marine micro- and macroalgae (Manley & Dastoor 1988). The fluxes of these gases are distributed such that the methyl halides ($CH_3X$) do not necessarily represent the largest source of the corresponding methylated halogen (Cl, Br, I). For example, the estimated global flux of $CHBr_3$ (boromethane) is actually three times that of $CH_3Br$ (methyl bromide), and the flux of $CH_2Br_2$ (dibromomethane) is equivalent to that of $CH_3Br$ (Yang et al. 2005). Because these gases will compete for the same radical sinks, and because C–X bond energies will have overlapping absorption features, we expect the addition of polyhalomethanes to have a similar effect on the planetary spectra as the addition of $CH_3Br$ to $CH_3Cl$. We did not self-consistently include fluxes of $CHBr_3$ or $CH_2Br_2$ because essential photochemical reactions and cross section data remain unavailable. It may be possible to loosely estimate the potential impact of $CHBr_3$, $CH_2Br_2$, and other polyhalomethanes through comparison of our results to the total flux of these gases. For Br-containing halomethanes only, this is equivalent to adjusting the integrated $CH_{x-y}Br_y$ flux by a factor of five to six; our flux–abundance relationships would be adjusted in flux for Earthlike values by a factor of five to six to account for the missing polymethane fluxes. This overly simplified estimation does not account for the different reaction rates, photolysis, and absorption cross sections for the other polyhalomethane permutations, which are impeded by the lack of data. However, it is demonstrably clear this estimation would result in a significant enhancement of halomethane detectability when estimated from the Earth production rates, particularly for M dwarf hosts. Similarly, in this initial study, we did not include the I-containing halomethanes (e.g., $CH_3I$, $CH_2I_2$, $CHI_3$). The flux of $CH_3I$ is estimated to be 0.3 Tg $yr^{-1}$ or about two times that of $CH_3Br$ (Bell et al. 2002). We expect addition of these gases may further lower the threshold for runaway conditions, which could then be triggered for planets orbiting late-type stars such as TRAPPIST-1 by fluxes closer to those seen on the modern Earth. Future work will self-consistently include updated reaction rate and cross section data for these halomethanes, some of which are yet to be measured or made available. Upon incorporation of this new data, our preliminary photochemical and spectral results suggest the fluxes required to produce observable halomethane signatures may be substantially less than we have estimated here for $CH_3Cl$ and $CH_3Br$.

Halomethanes represent only one subclass of volatile metabolites. Pilcher (2003) first proposed the potential for DMS ($(CH_3)_2S$) and DMDS ($(CH_3)_2S_2$) to serve as biosignatures of early Earthlike planets, which was further explored with self-consistent photochemical models by Domagal-Goldman et al. (2011). Numerous phyla of life on Earth including vascular plants and bacteria generate the corresponding Se-containing organic molecules DMSe ($(CH_3)_2Se$) and DMDSe ($(CH_3)_2Se_2$; Chasteen & Bentley 2003; Bañuelos et al. 2017). Additionally, bacteria and fungi that are observed to produce DMS and DMDS in the environment will also generate DMTe and DMDTe when grown as cultures in Te-rich media in the lab (Basnayake et al. 2001; Chasteen & Bentley 2003), although the extent to which these gases are produced in the natural environment is unknown. Thus, the entire chalcogen family (S, Se, Te) represents a potential methylated biosignature subclass.

The biologically mediated volatilization of metals and metalloids extends even beyond the halogens and chalcogens, including arsenic (As; e.g., $(CH_3)_3As$), antimony (Sb; e.g., $(CH_3)_3Sb$), bismuth (Bi; e.g., $(CH_3)_3Bi$), mercury (Hg; e.g., $(CH_3)_2Hg$), and potentially others (e.g., Bentley & Chasteen 2002; Thayer 2002; Cima et al. 2003; Meyer et al. 2008; Yang et al. 2016). These gases are produced by bacteria, archaea, and eukarya across marine and terrestrial environments from organic-rich sediments to inside the human body (Meyer et al. 2008). None of these gases have yet been as systematically or rigorously explored as exoplanetary biosignatures, likely due to their trace concentrations on Earth, though we qualitatively anticipate enhancements of their photochemical lifetimes and therefore concentrations for Earthlike planets orbiting M dwarf stars and/or those planets with anoxic atmospheres. If such gases could be produced in $H_2$-rich super-Earths, an additional detectability advantage would be conferred in transmission spectroscopy due to increased scale heights. To advance this study requires investment in laboratory or theoretical evaluations of chemical reaction rates, photolysis cross sections, and multiwavelength line or opacity measurements in addition to future photochemical modeling analyses of the type we present here.

In sum, the strengths of methylated gases to serve as potential agnostic biosignatures independent of Earth biology center around (1) the nature of methylation as a basic metabolic process that occurs in all cells, (2) the potential of these gases to be produced in a variety of terrestrial, lacustrine, and marine environments, and (3) their ability to build up photochemically on planets orbiting M dwarfs. Most currently accepted biosignatures are single or pairs of gases that are tied to Earth's biochemistry, such as $O_2$ produced by metabolically complex oxygenic photosynthesis, and are the result of specific evolutionary paths shaped by Earth's environmental selective pressures. In contrast, we suggest that the generic process of methylation (of biomolecules as well as metals) generates a suite of methylated gases, many or all of which may serve as a general class of biosignatures. Even if different biochemistries evolved on other planetary bodies, methylation is such a base cellular process that it is not unreasonable to speculate that it would evolve in different primitive, and possibly complex, life forms.

### 5.3. Ruling Out False Positives

The primary biosignature gases $CH_4$, $NH_3$, and $PH_3$ can be produced at thermodynamic equilibrium at high-temperature atmospheric layers of $H_2$-rich planets (and brown dwarfs; Lodders & Fegley 2002; Visscher et al. 2003) and are detectable in the gas giant planets of the solar system (Dougherty et al. 2009). While the abiotic production of these gases is disfavored on terrestrial worlds with temperate surfaces (e.g., Seager et al. 2013; Sousa-Silva et al. 2020), mass–radius degeneracies (e.g., Guimond & Cowan 2018) and uncertainties concerning the location and thermodynamic state of an unconstrained surface layer may challenge biosignature interpretations for $H_2$-rich super-Earths. Even on temperate terrestrial planets, there are substantial abiotic sources of $CH_4$ such as serpentization reactions (Etiope & Sherwood-Lollar 2013), and the atmospheric mixing ratio of even smaller abiotic fluxes may be enhanced when considering the photochemical impact of M dwarf stars' lower NUV fluxes (Segura et al. 2005; Seager et al. 2013; Schwieterman et al. 2019). Confirming the biogenicity of these potential biosignature gases will rely on additional contextual information, such as characterizing the surface state of the planet or detecting biosignature pairs like $CO_2$-$CH_4$ to confirm a chemical disequilibrium consistent with biology (Krissansen-Totton et al. 2018; Wogan & Catling 2020).

In contrast, there are very few truly abiotic sources of halomethanes, methylated chalcogens, and other methylated metals and metalloids in a planetary context. For example, hydrogen halides (e.g., HF, HCl, HBr, and H I) are the expected form of halogens in $H_2$-rich atmospheres (though reactions with $NH_3$ depletes them below detectable concentrations at observable layers (Showman 2001; Teanby et al. 2014). Terrestrial volcanism also emits halogens primarily in the form of hydrogen halides (Pyle & Mather 2009). Small volcanic sources of methyl halides are known, but these sources are several orders of magnitude smaller than biogenic sources and could not build up to appreciable levels in a planetary atmosphere (Frische et al. 2006). Other known "abiotic" sources of halomethanes on Earth include the decay of organic matter such as dead plants and organic-rich soil (Wishkerman et al. 2008) and biomass burning (Andreae & Merlet 2001). However, in a planetary context these "abiotic" sources rely on the recent presence of life and so would not complicate biosignature interpretations.

$CH_3Cl$ has been detected in protostellar environments and cometary nuclei (Fayolle et al. 2017), likely the result of strong nonequilibrium ion-molecule gas-phase chemistry and/or heterogeneous chemistry on grain surfaces (Acharyya & Herbst 2017). Fayolle et al. (2017) estimated that a maximum of 600 tons per year of $CH_3Cl$ could have been delivered to the young Earth from cometary sources in the first 80 Myr of its history, assuming all such $CH_3Cl$ survives the high temperatures of impact. For comparison, the surface flux of $CH_3Cl$ on Earth today is approximately 330,000 tons per year (300 Gg $yr^{-1}$) or $\sim$500 times greater than this maximal exogenous flux (Engel et al. 2019). To contextualize the impact of a source 500 times smaller than Earth's current flux, our results in Section 3.1 suggest this level of delivery to a late M dwarf terrestrial planet would produce $CH_3Cl$ concentrations consistent with its concentration on Earth today, at the approximately hundreds of parts per trillion level ($10^{-10}$ v/v), assuming the planetary atmosphere is oxygen-rich. We have shown here that this level of $CH_3Cl$ would not be detectable on an exoplanet. All other scenarios, such as planets orbiting G dwarf stars, would lead to less abundant $CH_3Cl$ mixing ratios even at this maximum flux. Additional calculations would need to assess the level of maximal exogenous buildup for an anoxic terrestrial planet. Similarly, the process of meteoric infall can generate abiotic $CH_3Cl$ via pyrolysis of petrologic Cl (Keppler et al. 2014). However, only 1 ppm of Cl is converted to $CH_3Cl$, so extreme fluxes would again be required to produce a false positive. The possibility of erroneously assigning biogenicity to exogenously delivered organohalogens could be avoided by characterization of the target system to determine its age and likely impact rate via dust emission.

Additionally, the ensemble of abiotically produced organohalogen gases is likely to include greater concentrations of hydrogen halides including HF, HCl, HBr, and HI (Acharyya & Herbst 2017; De Keyser et al. 2017) and potentially $CH_3F$- and F-containing polyhalomethane gases in addition to potentially biogenic Cl-, Br-, and I-bearing halomethanes. Because hydrogen halides are known abiotic products, their presence would suggest an abiotic source. On the other hand, the presence of detectable Cl, Br, and I halomethanes without F-bearing fluorinated homologues or corresponding hydrogen halides could be a strong biosignature if F-bearing methylated volatiles can be excluded as biological products. The potential for exogeneous delivery is not a unique challenge for organohalogen biosignature gases, as canonical biosignature gases such as $CH_4$, $NH_3$, and even $O_2$ are contained within or produced in the coma of cometary bodies (Mumma & Charnley 2011; Bieler et al. 2015; Yao & Giapis 2017). However, most analyses have found that external delivery or transfer of these molecules is unlikely to compete with planetary sinks and lead to false-positive biosignatures, except perhaps in the youngest planetary systems (Court & Sephton 2012; Felton et al. 2022).

### 5.4. Alternative Environments and Future Work

In this paper, we have examined methylated biosignatures in $O_2$-rich, Earthlike atmospheres because these gases are demonstrated to be produced in modern environments today. However, other bulk atmospheric composition may confer a more favorable detectability potential, though the plausible generation and fluxes of these gases are less constrained. It is well-known that $H_2$-rich atmospheres confer a detectability advantage in transmission spectroscopy due to extended atmospheric scale heights (Seager et al. 2013; Phillips et al. 2021). The plausibility of biosignature generation in these environments is still uncertain. On a planet with a highly reducing atmosphere, the necessary terminal electron acceptors (oxidants) may not be available or may have a strongly limited abundance. Additional requirements such as a source of carbon that can be biologically fixed to an intermediate redox potential might depend on the specific geologic and atmospheric properties of such a planet. Bains et al. (2014) explored the possibilities for photosynthesis on a rocky planet with a thin $H_2$-dominated atmosphere, finding that oxidation and fixation of $CH_4$ would be a plausible source of carbon for life, and other redox couples such as $N_2$ and $H_2$ may provide catabolic energy on these worlds (Seager et al. 2013; Huang et al. 2022). If the production of halomethanes is possible in such an environment, the overall detectability of the features in transit would increase due to the inflated size of low molecular mass atmospheres, as has been shown for $CH_3Cl$ in Madhusudhan et al. (2021). Searching for alternative methylated biosignature gases on $H_2$-rich super-Earths will be complementary to other biosignature searches, however, a thorough analysis of microbial metabolism under an $H_2$ atmosphere remains understudied in the literature

Anoxic, Archean-like ($N_2$–$CO_2$–$H_2O$–$CH_4$) planetary atmospheres are another potential setting for alternative methylated biosignature gases such as halomethanes and others. Methylated chalcogens are know to be produced in anoxic conditions, and Domagal-Goldman et al. (2011) found that anoxic atmospheres are the optimal environment for observing secondary impacts of DMS and DMDS. The ability of halomethane-producing microbes to grow under anoxic conditions is unknown; further laboratory and/or field studies are needed to address this question. Even if halomethane microbial producers are strictly aerobic, it is also possible that local oxic conditions could exist under a predominantly anoxic atmosphere. There is precedent for this scenario in the late Archean Eon on Earth where "whiffs" of oxygen appeared before the overall oxygenation of the atmosphere (Anbar et al. 2007; Olson et al. 2013; Riding et al. 2014). It is plausible that such a scenario could also occur on an exoplanet with similar prevailing biogeochemical conditions.

In addition, Archean-like atmospheres may have extensive photochemically produced hazes that shield the lower atmosphere and surface from UV penetration (e.g., Arney et al. 2016, 2018). This process may support longer photochemical lifetimes for methylated biosignatures since the main pathways for consumption are created by the photochemical destruction of $H_2O$. Since late-type stars such as M dwarfs already have lowered NUV fluxes, the impact of additional shielding would be less than the impact around a Sunlike star. This effect could result in increased detectability of methylated biosignature features for G and K type stars. Coupled photochemical and detectability studies of halomethanes in anoxic atmospheres is a target for future work.

In this work we used a 1D photochemical model to calculate self-consistent chemical profiles. One-dimensional models are appropriate for initial exploration of a large parameter space, and can more easily accommodate new chemistry such as additional species, but they have limitations including the inability to include effects from atmospheric circulation, local differences in photochemical radical sinks, and possible heterogeneous emission of gases. The emission of halomethanes on Earth today is highly heterogeneous across both marine and terrestrial environments (Yang et al. 2005). Future work using 3D general circulation models coupled with atmospheric chemistry could assess the magnitude of the potential divergence between predictions from 1D and 3D treatments (e.g., Chen et al. 2018, 2019).

The detectability analyses presented here focused on quantifying the detectability of $CH_3X$ gases with transmission spectroscopy, though we also presented synthetic emission spectra showing the spectral impact of $CH_3Cl$ and $CH_3Br$ under a variety of conditions. Future work will provide additional quantitative constraints on the detectability of these gases with future observatories and/or methods. For example, our calculations suggest isolating $CH_3Cl$ features in the near-infrared requires a high spectral resolution (upwards of 100,000) to differentiate individual lines. However, cross-correlation techniques may lower this number (e.g., Brogi & Line 2019), and could be simulated in an additional analysis. Other potential instruments include concepts such as the MIRECLE mission (Staguhn et al. 2019) or LIFE (Quanz et al. 2022), next-generation concepts that may be more advantageous in the mid-infrared than JWST or ground-based observatories.

In future work, we will also expand our analysis of methylated biosignatures, beginning with $CH_3I$ and methylated chalcogens, in addition to polyhalomethanes and other methylated metals and metalloids. Additional studies of this general class will expand its applicability and further develop the exoplanetary biosignature toolkit.

## 6. Conclusion

Methylated biosignatures such as $CH_3Cl$ and $CH_3Br$ are potential capstone biosignatures, serving as corroborative evidence of life on a planet with an ambiguous primary biosignature such as $O_2$. $CH_3Br$, a novel biosignature candidate, is produced by oxygenic phototrophic cyanobacteria (Shibazaki et al. 2016), and can serve as a secondary biosignature to help discriminate the biogenicity of $O_2$ in the atmosphere. We find that $CH_3Br$, which is first extensively analyzed in this work, follows the same buildup trends as $CH_3Cl$ with even greater relative atmospheric enhancements around late-type stars. $CH_3Br$ in combination with $CH_3Cl$, significantly reduces the number of transits necessary to detect combined "$CH_3X$" spectral features in mid-infrared transmission spectroscopy. The coadded features are also enhanced in simulated mid-infrared emission spectra.

The halomethanes explored here are only a small portion of the potential class of methylated biosignature gases, which include additional methyl halides, polyhalomethanes, methylated chalcogens, and other methylated metals and metalloids. We expect that an atmosphere with multiple methylated biogenic gases would have have an even larger coadded feature, and therefore require less observing time and spectral resolving power. Since these gases are nonspecific products of environmental detoxification, they may have a greater likelihood of being produced in an exoplanetary environment, which lends them an advantage over biosignature searches for specific narrow evolutionary outcomes of Earth's biology. In the coming decades, through a combination of ground- and space-based telescopes, it will be possible to detect methylated biosignature gases. Our future work will further analyze additional methylated gases to expand on this new class of biosignatures.

This work was supported by the NASA Interdisciplinary Consortia for Astrobiology Research (ICAR) program via the Alternative Earths team with funding issued under grant No. 80NSSC21K0594 and the CHAMPs (Consortium on Habitability and Atmospheres of M-dwarf Planets) team with funding issued under grant No. 80NSSC21K0905. This work was also supported by the Virtual Planetary Laboratory, which is a member of the NASA Nexus for Exoplanet System Science and funded via NASA Astrobiology Program grant No. 80NSSC18K0829. M.L. gratefully acknowledges additional support from UCR's Provost Research Fellowship. The authors would like to thank UCR graphic artist Sohail Wasif for help in creating the concept figure, Daria Pidhorodetska for support using the Planetary Spectrum Generator, and an anonymous reviewer, whose comments allowed us to improve the manuscript.

*Software:* atmos (Arney et al. 2016; Lincowski et al. 2018), jcamp.py, matplotlib (Hunter 2007), numpy (Harris et al. 2020), Planetary Spectrum Generator (Villanueva et al. 2018), SMART (Meadows & Crisp 1996), spectres (Carnall 2017).

## Appendix A
## Boundary Conditions

Table 2 provides photochemical boundary conditions necessary for reproducing our results.

## Appendix B
## Reactions

Table 3 provides a complete list of reactions used for photochemical modeling.

**Table 2**
Boundary Conditions Used in the Model

| Chemical Species | Deposition Velocity (cm $s^{-1}$) | Flux (molecules $cm^{-2}$ $s^{-1}$) | Mixing Ratio (v/v) |
|---|---|---|---|
| O | 1 | ... | ... |
| $O_2$ | ... | ... | 0.21 |
| $O_3$ | 0.07 | ... | |
| $H_2O$ | ... | ... | fixed |
| H | 1.0 | ... | ... |
| OH | 1.0 | ... | ... |
| $HO_2$ | 1.0 | ... | ... |
| $H_2O_2$ | 1.0 | ... | ... |
| $H_2$ | $2.4 \cdot 10^{-4}$ | ... | ... |
| CO | $1.2 \cdot 10^{-4}$ | $2.0 \cdot 10^{11}$ | ... |
| HCO | 1 | ... | ... |
| $H_2CO$ | 0.2 | ... | ... |
| $CH_4$ | ... | $1.8 \cdot 10^{11}$ | ... |
| $CH_3$ | 1.0 | ... | ... |
| $C_2H_6$ | ... | $9.0 \cdot 10^{8}$ | ... |
| NO | $3.0 \cdot 10^{-3}$ | $1.0 \cdot 10^{9}$ | ... |
| $NO_2$ | $2.0 \cdot 10^{-2}$ | ... | ... |
| $NO_3$ | $2.0 \cdot 10^{-2}$ | ... | ... |
| $N_2O_5$ | 1.0 | ... | ... |
| HNO | 1.0 | ... | ... |
| $H_2S$ | 0.2 | $2.0 \cdot 10^{8}$ | ... |
| $SO_2$ | 1.0 | $9.0 \cdot 10^{9}$ | ... |
| $H_2SO_4$ | ... | $7.0 \cdot 10^{8}$ | ... |
| HSO | 1.0 | ... | ... |
| OCS | ... | $1.5 \cdot 10^{7}$ | ... |
| $SO_4$ [aerosol] | 0.01 | ... | ... |
| $S_8$ [aerosol] | 0.01 | ... | ... |
| $HNO_3$ | **1.0** | ... | ... |
| $N_2O$ | ... | $1.5 \cdot 10^{9}$ | ... |
| $HO_2NO_2$ | 1 | ... | ... |
| $CH_3O$ | 1 | ... | ... |
| $CH_3O_2NO_2$ | 0.2 | ... | ... |
| $CH_3OOH$ | 0.2 | ... | ... |
| $CH_2OH$ | 1 | ... | ... |
| $CO_2$ | ... | ... | 400 ppm |
| $CH_3Cl$ | ... | variable | ... |
| $CCl_4$ | ... | $2.0 \cdot 10^{5}$ | ... |
| ClO | 0.5 | ... | ... |
| HOCl | 0.5 | ... | ... |
| $Cl_2$ | 1.0 | ... | ... |
| $ClONO_2$ | 0.5 | ... | ... |
| $CH_2ClO_2$ | 1.0 | ... | ... |
| HCl | 0.2 | $1 \cdot 10^{8}$ | ... |
| Cl | 1.0 | ... | ... |
| $HClO_4$ | 0.2 | ... | ... |
| $Cl_2O_4$ | 1.0 | ... | ... |
| Br | 1.0 | ... | ... |
| HBr | 0.5 | $1 \cdot 10^{6}$ | ... |
| HOBr | 0.2 | ... | ... |
| $CH_3Br$ | ... | variable | ... |
| $CH_2Br$ | **1.0** | ... | ... |
| $CHBr_3$ | ... | $1.6 \cdot 10^{7}$ | ... |

**Note.** Additional species with zero deposition velocity and zero surface flux: HS, S, SO, $S_2$, $S_4$, $S_8$, $SO_3$, $S_3$, N, $CH_3ONO$, $CH_3ONO_2$, $CH_2ONO_2$, $CH_3O_2$, $CH_3OH$, $CH_2O_2$, $CH_2OOH$, $CH_2ClO$, CHClO, $CCl_3$, $CCl_3O_2$, $COCl_2$, $CCl_3NO_4$, OClO, ClOO, ClONO, ClNO, $ClNO_2$, $CH_2Cl$, $CH_2OCl$, $Cl_2O_2$, $ClO_3$, $Cl_2O$, BrO, $Br_2$, and $BrONO_2$. Closed reaction loops are not available for $CHBr_3$, BrCl, and $CBr_3$.

**Table 3**
Comparison of Reactions in Segura et al. (2005) and This Work

| # | Reaction | Segura et al. (2005) Rate | New Rate | Reference |
|---|---|---|---|---|
| 1 | $H_2O + O^1D \rightarrow OH + OH$ | $2.2 \cdot 10^{-10}$ | $1.63 \cdot 10^{-10} \times e^{60/T}$ | Burkholder et al. (2015) |
| 2 | $H_2 + O^1D \rightarrow OH + H$ | $1.0 \cdot 10^{-10}$ | $1.2 \cdot 10^{-10}$ | Burkholder et al. (2015) |
| 3 | $H_2 + O \rightarrow OH + H$ | $3 \cdot 10^{-14} \times T \times e^{-4480/T}$ | $1.46 \cdot 10^{-9} \times e^{T/9650} + 6.34 \cdot 10^{-12} \times e^{T/4000}$ | National Institute of Standards & Technology (2018) |
| 4 | $H_2 + OH \rightarrow H_2O + H$ | $5.5 \cdot 10^{-12} \times e^{-2000/T}$ | $2.8 \cdot 10^{-12} \times e^{-1800/T}$ | Burkholder et al. (2015) |
| 5 | $H + O_3 \rightarrow OH + O_2$ | $1.4 \cdot 10^{-10} \times e^{-470/T}$ | $1.4 \cdot 10^{-10} \times e^{-470/T}$ | Burkholder et al. (2015) |
| 6 | $H + O_2 \rightarrow HO_2$ | $5.7 \cdot 10^{-32} \times \frac{298}{T}^{1.6}$ | $4.4 \cdot 10^{-32} \times \frac{298}{T}^{1.3}$ | Burkholder et al. (2015) |
| 7 | $H + HO_2 \rightarrow H_2 + O_2$ | $6.48 \cdot 10^{-12}$ | $6.9 \cdot 10^{-12}$ | Burkholder et al. (2015) |
| 8 | $H + HO_2 \rightarrow H_2O + O$ | $1.62 \cdot 10^{-12}$ | $1.6 \cdot 10^{-12}$ | Burkholder et al. (2015) |
| 9 | $H + HO_2 \rightarrow OH + OH$ | $7.29 \cdot 10^{-11}$ | $7.2 \cdot 10^{-11}$ | Burkholder et al. (2015) |
| 10 | $OH + O \rightarrow H + O_2$ | $2.2 \cdot 10^{-11} \times e^{120/T}$ | $1.8 \cdot 10^{-11} \times e^{180/T}$ | Burkholder et al. (2015) |
| 11 | $OH + HO_2 \rightarrow H_2O + O_2$ | $4.8 \cdot 10^{-11} \times e^{250/T}$ | $4.8 \cdot 10^{-11} \times e^{250/T}$ | Burkholder et al. (2015) |
| 12 | $OH + O_3 \rightarrow HO_2 + O_2$ | $1.6 \cdot 10^{-11} \times e^{-940/T}$ | $1.7 \cdot 10^{-12} \times e^{-940/T}$ | Burkholder et al. (2015) |
| 13 | $HO_2 + O \rightarrow OH + O_2$ | $3.0 \cdot 10^{-11} \times e^{200/T}$ | $3.0 \cdot 10^{-11} \times e^{200/T}$ | Burkholder et al. (2015) |
| 14 | $HO_2 + O_3 \rightarrow OH + O_2 + O_2$ | $1.1 \cdot 10^{-14} \times e^{-500/T}$ | $1.0 \cdot 10^{-14} \times e^{-490/T}$ | Sander et al. (2003) |
| 15 | $HO_2 + HO_2 \rightarrow H_2O_2 + O_2$ | $2.3 \cdot 10^{-13} \times e^{600/T} + 1.7 \cdot 10^{-33} \times e^{1000/T} \times \rho$ | $2.3 \cdot 10^{-13} \times e^{590/T} + 1.2 \cdot 10^{-33} \times e^{1000/T} \times \rho$ | Sander et al. (2003) |
| 16 | $H_2O_2 + OH \rightarrow HO_2 + H_2O$ | $2.9 \cdot 10^{-12} \times e^{-160/T}$ | $2.9 \cdot 10^{-12} \times e^{-160/T}$ | Sander et al. (2003) |
| 17 | $O + O \rightarrow O_2$ | $2.76 \cdot 10^{-34} \times e^{710/T} \times \rho$ | $6.0 \cdot 10^{-34} \times 300/T^{2.6} \cdot \rho$ | National Institute of Standards & Technology (2018) |
| 18 | $O + O_2 \rightarrow O_3$ | $6.0 \cdot 10^{-34} \times 298/T^{1.6}$ | $5.12 \cdot 10^{-34} \times 298/T^{2.6} \cdot \rho \times e^{40.5/T}$ | National Institute of Standards & Technology (2018) |
| 19 | $O + O_3 \rightarrow O_2 + O_2$ | $8.0 \cdot 10^{-12} \times e^{-2060/T}$ | $8.0 \cdot 10^{-12} \times e^{-2060/T}$ | Burkholder et al. (2015) |
| 20 | $OH + OH \rightarrow H_2O + O$ | $4.2 \cdot 10^{-12} \times e^{-240/T}$ | $1.8 \cdot 10^{-12}$ | Burkholder et al. (2015) |
| 21 | $O^1D + N_2 \rightarrow O + N_2$ | $1.8 \cdot 10^{-11} \times e^{110/T}$ | $2.15 \cdot 10^{-11} \times e^{110/T}$ | Burkholder et al. (2015) |
| 22 | $O^1D + O_2 \rightarrow O + O_2$ | $3.2 \cdot 10^{-11} \times e^{70/T}$ | $3.3 \cdot 10^{-11} \times e^{55/T}$ | Burkholder et al. (2015) |
| 23 | $O_2 + h\nu \rightarrow O + O^1D$ | | | *photolysis* |

**Table 3**
(Continued)

| # | Reaction | Segura et al. (2005) Rate | New Rate | Reference |
|---|---|---|---|---|
| 24 | $O_2 + h\nu \rightarrow O + O$ | | | *photolysis* |
| 25 | $H_2O + h\nu \rightarrow H + OH$ | | | *photolysis* |
| 26 | $O_3 + h\nu \rightarrow O_2 + O^1D$ | | | *photolysis* |
| 27 | $O_3 + h\nu \rightarrow O_2 + O$ | | | *photolysis* |
| 28 | $O_3 + h\nu \rightarrow O + O + O$ | | | *photolysis* |
| 29 | $H + CO \rightarrow HCO$ | $2.0 \cdot 10^{-33} \times e^{-850/T} \times \rho$ | $1.0 \cdot 10^{-33} \times e^{-629/T} \times \rho$ | National Institute of Standards & Technology (2018) |
| 30 | $H + HCO \rightarrow H_2 + CO$ | $1.2 \cdot 10^{-10}$ | $1.8 \cdot 10^{-10}$ | National Institute of Standards & Technology (2018) |
| 31 | $HCO + HCO \rightarrow H_2CO + CO$ | 0 | $4.5 \cdot 10^{-11}$ | National Institute of Standards & Technology (2018) |
| 32 | $OH + HCO \rightarrow H_2O + CO$ | $5 \cdot 10^{-11}$ | $1.0 \cdot 10^{-10}$ | National Institute of Standards & Technology (2018) |
| 33 | $O + HCO \rightarrow H + CO_2$ | $1.0 \cdot 10^{-10}$ | $5.0 \cdot 10^{-11}$ | National Institute of Standards & Technology (2018) |
| 34 | $O + HCO \rightarrow OH + CO$ | $1.0 \cdot 10^{-10}$ | $5.0 \cdot 10^{-11}$ | National Institute of Standards & Technology (2018) |
| 35 | $H_2O_2 + h\nu \rightarrow OH + OH$ | | | *photolysis* |
| 36 | $CO_2 + h\nu \rightarrow CO + O$ | | | *photolysis* |
| 37 | $CO_2 + h\nu \rightarrow CO + O^1D$ | | | *photolysis* |
| 38 | $CO + OH \rightarrow CO_2 + H$ | $1.5 \cdot 10^{-13}(1 + 0.6 \cdot P_{atm})$ | $\frac{k_0}{1+\frac{k_0}{\frac{k_{inf}}{\rho}}} \times 0.6^{\frac{1+log_{10}(\frac{k_0}{k_{inf}\rho^2})}{-1}}$ $k_0 = 1.5 \cdot 10^{-13}, k_{inf} = 2.1 \cdot 10^9 \times \frac{T}{300}^{6.1}$ | National Institute of Standards & Technology (2018) |
| 39 | $CO + O \rightarrow CO_2$ | $6.5 \cdot 10^{-33} \times e^{-2180/T} \times \rho$ | $1.6 \cdot 10^{-32} \times e^{-2184/T} \times \rho$ | Gao et al. (2015) |
| 40 | $H_2CO + h\nu \rightarrow H + H + CO$ | | | *photolysis* |
| 41 | $H_2CO + h\nu \rightarrow H_2 + CO$ | | | *photolysis* |
| 42 | $H_2CO + h\nu \rightarrow HCO + H$ | | | *photolysis* |
| 43 | $HCO + h\nu \rightarrow H + CO$ | $1.0 \cdot 10^{-2}$ | | *photolysis* |

**Table 3**
(Continued)

| # | Reaction | Segura et al. (2005) Rate | New Rate | Reference |
|---|---|---|---|---|
| 44 | $H_2CO + H \rightarrow H_2 + HCO$ | $2.8 \cdot 10^{-11} \times e^{-1540/T}$ | $2.14 \cdot 10^{-12} \times T/298^{1.62} \times e^{-1090/T}$ | National Institute of Standards & Technology (2018) |
| 45 | $H + H \rightarrow H_2$ | $2.6 \cdot 10^{-33} \times e^{375/T} \times \rho$ | $8.85 \cdot 10^{-33} \times (T/287)^{-0.6\rho}$ | Tsang & Hampson (1986) |
| 46 | $HCO + O_2 \rightarrow HO_2 + CO$ | $3.5 \cdot 10^{-12} \times e^{140/T}$ | $5.2 \cdot 10^{-12}$ | Burkholder et al. 2015 |
| 47 | $H_2CO + OH \rightarrow H_2O + HCO$ | $1.0 \cdot 10^{-11}$ | $5.5 \cdot 10^{-12} \times e^{125/T}$ | Burkholder et al. (2015) |
| 48 | $H + OH \rightarrow H_2O$ | $6.1 \cdot 10^{-26}/T^2 \times \rho$ | $6.9 \cdot 10^{-31} \times \left(\frac{298}{T}\right)^2 \times \rho$ | Baulch et al. (1992) |
| 49 | $OH + OH \rightarrow H_2O_2$ | $6.9 \cdot 10^{-31} \times \left(\frac{298}{T}\right)^{0.8}$<br>$1.5 \cdot 10^{-11}$ | $6.9 \cdot 10^{-31} \times (\frac{298}{T})$<br>$2.6 \cdot 10^{-11}$ | Burkholder et al. (2015) |
| 50 | $H_2CO + O \rightarrow HCO + OH$ | $3.4 \cdot 10^{-11} \times e^{-1600/T}$ | $3.4 \cdot 10^{-11} \times e^{-1600/T}$ | Burkholder et al. (2015) |
| 51 | $H_2CO + O \rightarrow OH + HO_2$ | | $1.4 \cdot 10^{-12} \times e^{-2000/T}$ | Burkholder et al. (2015) |
| 52 | $HO_2 + h\nu \rightarrow OH + O$ | | | *photolysis* |
| 53 | $CH_4 + h\nu \rightarrow CH_{21} + H_2$ | | | *photolysis* |
| 54 | $CH_4 + h\nu \rightarrow CH_3 + H$ | | | *photolysis* |
| 55 | $CH_4 + h\nu \rightarrow CH_{23} + H + H$ | | | *photolysis* |
| 56 | $C_2H_6 + h\nu \rightarrow CH_{23} + CH_{23} + H_2$ | | | *photolysis* |
| 57 | $C_2H_6 + h\nu \rightarrow CH_4 + CH_{21}$ | | | *photolysis* |
| 58 | $HNO_2 + h\nu \rightarrow NO + OH$ | | | *photolysis* |
| 59 | $HNO_3 + h\nu \rightarrow NO_2 + OH$ | | | *photolysis* |
| 60 | $HNO_3 + h\nu \rightarrow HNO_2 + O$ | | | *photolysis* |
| 61 | $HNO_3 + h\nu \rightarrow HNO_2 + O^1D$ | | | *photolysis* |
| 62 | $NO + h\nu \rightarrow N + O$ | | | *photolysis* |
| 63 | $NO_2 + h\nu \rightarrow NO + O^1D$ | | | *photolysis* |
| 64 | $NO_2 + h\nu \rightarrow NO + O$ | | | *photolysis* |

**Table 3**
(Continued)

| # | Reaction | Segura et al. (2005) Rate | New Rate | Reference |
|---|---|---|---|---|
| 65 | $CH_4 + OH \rightarrow CH_3 + H_2O$ | $2.9 \cdot 10^{-12} \times e^{-1820/T}$ | $2.45 \cdot 10^{-12} \times e^{-1775/T}$ | Burkholder et al. (2015) |
| 66 | $CH_4 + O^1D \rightarrow CH_3 + OH$ | $1.4 \cdot 10^{-10}$ | $1.125 \cdot 10^{-10}$ | Sander et al. (2006) |
| 67 | $CH_4 + O^1D \rightarrow H_2CO + H_2$ | $1.4 \cdot 10^{-11}$ | $7.5 \cdot 10^{-12}$ | Sander et al. (2006) |
| 68 | $CH_{21} + CH_4 \rightarrow CH_3 + CH_3$ | $1.9 \cdot 10^{-12}$ | $3.6 \cdot 10^{-11}$ | Böhland et al. (1985) |
| 69 | $CH_{21} + O_2 \rightarrow HCO + OH$ | $3.0 \cdot 10^{-11}$ | $3.0 \cdot 10^{-11}$ | Ashfold et al. (1981) |
| 70 | $CH_{21} + N_2 \rightarrow CH_{23} + N_2$ | $5.0 \cdot 10^{-13}$ | $5.0 \cdot 10^{-14}$ | Harman et al. (2015) |
| 71 | $CH_{23} + H_2 \rightarrow CH_3 + H$ | $5.0 \cdot 10^{-14}$ | $5 \cdot 10^{-14}$ | Harman et al. (2015) |
| 72 | $CH_{23} + CH_4 \rightarrow CH_3 + CH_3$ | $5.0 \cdot 10^{-14}$ | $7.1 \cdot 10^{-12} \times e^{-5051/T}$ | Harman et al. (2015) |
| 73 | $CH_{23} + O_2 \rightarrow HCO + OH$ | $1.5 \cdot 10^{-12}$ | $4.1 \cdot 10^{-11} \times e^{-750/T}$ | National Institute of Standards & Technology (2018) |
| 74 | $CH_3 + O_2 \rightarrow H_2CO + OH$ | | $4.5 \cdot 10^{-31} \times \frac{298}{T}^{3.0}$ | Sander et al. (2003) |
| 75 | $CH_3 + O \rightarrow H_2CO + H$ | $1.0 \cdot 10^{-14}$ | $1.1 \cdot 10^{-10}$ | Burkholder et al. (2015) |
| 76 | $CH_3 + CH_3 \rightarrow C_2H_6$ | | $1.17 \cdot 10^{-25} \times e^{-500/T^{3.75}}$ | National Institute of Standards & Technology (2018) |
| 77 | $CH_3 + h\nu \rightarrow CH_{21} + H$ | | | *photolysis* |
| 78 | $CH_3 + H \rightarrow CH_4$ | | $1.0 \cdot 10^{-28} \times \frac{298}{T}^{1.8}$ | Parkes et al. (1973), Allen et al. (1980) |
| 79 | $CH_3 + HCO \rightarrow CH_4 + CO$ | | $5.0 \cdot 10^{-11}$ | National Institute of Standards & Technology (2018) |
| 80 | $CH_3 + HNO \rightarrow CH_4 + NO$ | | $3.3 \cdot 10^{-12} \times e^{-1000/T}$ | National Institute of Standards & Technology (2018) |
| 81 | $CH_3 + H_2CO \rightarrow CH_4 + HCO$ | | $4.9 \cdot 10^{-15} \times \frac{T}{298}^{4.4} \times e^{-2450/T}$ | National Institute of Standards & Technology (2018) |
| 82 | $H + NO \rightarrow HNO$ | | $(1.2 \cdot 10^{-31} \times e^{-210/T})^{1.17}$ | National Institute of Standards & Technology (2018) |

**Table 3**
(Continued)

| # | Reaction | Segura et al. (2005) Rate | New Rate | Reference |
| --- | --- | --- | --- | --- |
| 83 | $N + N \rightarrow N_2$ | | $(1.25 \cdot 10^{-32} \times \rho$ | National Institute of Standards & Technology (2018) |
| 84 | $N + O_2 \rightarrow NO + O$ | $1.5 \cdot 10^{-11} \times e^{-3600/T}$ | $1.5 \cdot 10^{-11} \times e^{-3600/T}$ | Burkholder et al. (2015) |
| 85 | $N + O_3 \rightarrow NO + O_2$ | $1.0 \cdot 10^{-16}$ | $1.0 \cdot 10^{-16}$ | Burkholder et al. (2015) |
| 86 | $N + OH \rightarrow NO + H$ | $5.3 \cdot 10^{-11}$ | $3.8 \cdot 10^{-11} \times e^{85/T}$ | National Institute of Standards & Technology (2018) |
| 87 | $N + NO \rightarrow N_2 + O$ | $3.4 \cdot 10^{-11}$ | $2.1 \cdot 10^{-11} \times e^{100/T}$ | Burkholder et al. (2015) |
| 88 | $NO + O_3 \rightarrow NO_2 + O_2$ | $2.0 \cdot 10^{-12} \times e^{-1400/T}$ | $3.0 \cdot 10^{-12} \times e^{-1500/T}$ | Burkholder et al. (2015) |
| 89 | $NO + O \rightarrow NO_2$ | $9.0 \cdot 10^{-32} \times \left(\frac{298}{T}\right)^{1.5}$ | $9.0 \cdot 10^{-32} \times \left(\frac{298}{T}\right)^{1.5}$ | Burkholder et al. (2015) |
| 90 | $NO + HO_2 \rightarrow NO_2 + OH$ | $3.7 \cdot 10^{-12} \times e^{250/T}$ | $3.3 \cdot 10^{-12} \times e^{270/T}$ | Burkholder et al. (2015) |
| 91 | $NO + OH \rightarrow HNO_2$ | $7 \cdot 10^{-31} \times \left(\frac{298}{T}\right)^{2.6}$ | $7 \cdot 10^{-31} \times \left(\frac{298}{T}\right)^{2.6}$ | Burkholder et al. (2015) |
| 92 | $NO_2 + O \rightarrow NO + O_2$ | $6.5 \cdot 10^{-12} \times e^{120/T}$ | $5.1 \cdot 10^{-12} \times e^{210/T}$ | Burkholder et al. (2015) |
| 93 | $NO_2 + OH \rightarrow HNO_3$ | $2.6 \cdot 10^{-30} \times \left(\frac{298}{T}\right)^{3.2}$ | $1.8 \cdot 10^{-30} \times \left(\frac{298}{T}\right)^{3}$ | Burkholder et al. (2015) |
| 94 | $NO_2 + H \rightarrow NO + OH$ | $4.8 \cdot 10^{-10} \times e^{-340/T}$ | $4.0 \cdot 10^{-10} \times e^{-340/T}$ | Burkholder et al. (2015) |
| 95 | $HNO_3 + OH \rightarrow H_2O + NO_3$ | $A_{K0} + \frac{A_{K3M}}{1+\frac{A_{K3M}}{A_{K2}}}$, $A_{K0}=7.2 \cdot 10^{-15} \times e^{785/T}$, $A_{K2}=4.1 \cdot 10^{-16} \times e^{1440/T}$, $A_{K3M}=1.9 \cdot 10^{-33} \times e^{725/T} \times \rho$ | $A_{K0} + \frac{A_{K3M}}{1+\frac{A_{K3M}}{A_{K2}}}$, $A_{K0}=2.4 \cdot 10^{-14} \times e^{460/T}$, $A_{K2}=2.7 \cdot 10^{-17} \times e^{2199/T}$, $A_{K3M}=6.5 \cdot 10^{-34} \times e^{1335/T} \times \rho$ | Burkholder et al. (2015) |
| 96 | $HCO + NO \rightarrow HNO + CO$ | | $1.3 \cdot 10^{-11}$ | National Institute of Standards & Technology (2018) |
| 97 | $HNO + hv \rightarrow NO + H$ | | $1.7 \cdot 10^{-3}$ | Cox (1974) |
| 98 | $H + HNO \rightarrow H_2 + NO$ | | $3.0 \cdot 10^{-11} \times e^{-500/T}$ | National Institute of Standards & Technology (2018) |
| 99 | $O + HNO \rightarrow OH + NO$ | | $3.8 \cdot 10^{-11}$ | National Institute of Standards & Technology (2018) |
| 100 | $OH + HNO \rightarrow H_2O + NO$ | | $5.0 \cdot 10^{-11}$ | National Institute of Standards & Technology (2018) |
| 101 | $HNO_2 + OH \rightarrow H_2O + NO_2$ | | $1.8 \cdot 10^{-11} \times e^{-390/T}$ | Burkholder et al. (2015) |
| 102 | $CH_4 + O \rightarrow CH_3 + OH$ | | $5.8 \cdot 10^{-11} \times e^{-4450/T}$ | National Institute of Standards & Technology (2018) |

**Table 3**
(Continued)

| # | Reaction | Segura et al. (2005) Rate | New Rate | Reference |
|---|---|---|---|---|
| 103 | $CH_{21} + H_2 \rightarrow CH_3 + H$ | | $5 \cdot 10^{-15}$ | Tsang & Hampson (1986), Arney et al. (2016) |
| 104 | $CH_{21} + CO_2 \rightarrow H_2CO + CO$ | | $1.0 \cdot 10^{-12}$ | Zahnle (1986), Arney et al. (2016) |
| 105 | $CH_{23} + O \rightarrow HCO + H$ | | $1.0 \cdot 10^{-11}$ | National Institute of Standards & Technology (2018) |
| 106 | $CH_{23} + CO_2 \rightarrow H_2CO + CO$ | | $1.0 \cdot 10^{-14}$ | National Institute of Standards & Technology (2018) |
| 107 | $C_2H_6 + OH \rightarrow C_2H_5 + H_2O$ | | $7.66 \cdot 10^{-12} \times e^{-1020/T}$ | National Institute of Standards & Technology (2018) |
| 108 | $C_2H_6 + O \rightarrow C_2H_5 + OH$ | | $8.54 \cdot 10^{-12} \times \frac{T}{300}^{1.5} \times e^{-2920/T}$ | Baulch et al. (1994), Arney et al. (2016) |
| 109 | $C_2H_6 + O^1D \rightarrow C_2H_5 + OH$ | | $6.29 \cdot 10^{-10}$ | Matsumi et al. (1993), Arney et al. (2016) |
| 110 | $C_2H_5 + H \rightarrow CH_3 + CH_3$ | | $6.0 \cdot 10^{-11}$ | National Institute of Standards & Technology (2018) |
| 111 | $C_2H_5 + O \rightarrow CH_3 + HCO + H$ | | $1.1 \cdot 10^{-10}$ | National Institute of Standards & Technology (2018) |
| 112 | $C_2H_5 + OH \rightarrow CH_3 + HCO + H_2$ | | $4.0 \cdot 10^{-11}$ | National Institute of Standards & Technology (2018) |
| 113 | $C_2H_5 + HCO \rightarrow C_2H_6 + CO$ | | $1.0 \cdot 10^{-10}$ | National Institute of Standards & Technology (2018), Zahnle (1986) |
| 114 | $C_2H_5 + HNO \rightarrow C_2H_6 + NO$ | | $1.6 \cdot 10^{-12} \times e^{-1000/T}$ | National Institute of Standards & Technology (2018), Zahnle (1986) |
| 115 | $C_2H_5 + O_2 \rightarrow CH_3 + HCO + OH$ | | $1.5 \cdot 10^{-28} \times \left(\frac{298}{T}\right)^3$ | Burkholder et al. (2015) |
| 116 | $SO + h\nu \rightarrow S + O$ | | | *photolysis* |
| 117 | $H_2S + h\nu \rightarrow HS + H$ | | | *photolysis* |
| 118 | $SO_2 + h\nu \rightarrow SO + O$ | | | *photolysis* |
| 119 | $SO_2 + h\nu \rightarrow SO_{21}$ | | | *photolysis* |

**Table 3**
(Continued)

| # | Reaction | Segura et al. (2005) Rate | New Rate | Reference |
|---|---|---|---|---|
| 120 | $SO_2 + h\nu \rightarrow$ $SO_{23}$ | | | *photolysis* |
| | 121 | $SO + O_2 \rightarrow O + SO_2$ $2.6 \cdot 10^{-13} \times e^{-2400/T}$ | $2.4 \cdot 10^{-13} \times e^{-2370/T}$ National Institute of Standards & Technology (2018), Zahnle (1986) | |
| 122 | $SO + HO_2 \rightarrow SO_2 + OH$ | $2.8 \cdot 10^{-11}$ | $2.8 \cdot 10^{-11}$ | Kasting (1990), Harman et al. (2015) |
| 123 | $SO + O \rightarrow SO_2$ | $6.0 \cdot 10^{-31} \times \rho$ | $6.0 \cdot 10^{-31} \times \rho$ | Kasting (1990), Harman et al. (2015) |
| 124 | $SO + OH \rightarrow SO_2 + H$ | $8.6 \cdot 10^{-11}$ | $2.7 \cdot 10^{-11} \times e^{335/T}$ | Burkholder et al. (2015) |
| 125 | $SO_2 + OH \rightarrow HSO_3$ | $3 \cdot 10^{-31} \times \left(\frac{298}{T}\right)^{3.3}$ | $3.3 \cdot 10^{-32} \times \left(\frac{298}{T}\right)^{4.3}$ | Burkholder et al. (2015) |
| 126 | $SO_2 + O \rightarrow SO_3$ | $3.4 \cdot 10^{-32} \times e^{-1130/T} \times \rho$ | $1.8 \cdot 10^{-33} \times \left(\frac{298}{T}\right)^{-2.0}$ | Burkholder et al. (2015) |
| 127 | $SO_3 + H_2O \rightarrow H_2SO_4$ | $6.0 \cdot 10^{-15}$ | $2.3 \cdot 10^{-43} \times T \times e^{6540/T} \times \rho \times f_{H_2O}$ | Krasnopolsky (2012) |
| 128 | $HSO_3 + O_2 \rightarrow HO_2 + SO_3$ | $1.3 \cdot 10^{-12} \times e^{-330/T}$ | $1.3 \cdot 10^{-12} \times e^{-330/T}$ | Burkholder et al. (2015) |
| 129 | $HSO_3 + OH \rightarrow H_2O + SO_3$ | $1.0 \cdot 10^{-11}$ | $1.0 \cdot 10^{-11}$ | Kasting (1990), Harman et al. (2015) |
| 130 | $HSO_3 + H \rightarrow H_2 + SO_3$ | $1.0 \cdot 10^{-11}$ | $1.0 \cdot 10^{-11}$ | Kasting (1990), Harman et al. (2015) |
| 131 | $HSO_3 + O \rightarrow OH + SO_3$ | $1.0 \cdot 10^{-11}$ | $1.0 \cdot 10^{-11}$ | Kasting (1990), Harman et al. (2015) |
| 132 | $H_2S + OH \rightarrow H_2O + HS$ | $6.0 \cdot 10^{-12} \times e^{-75/T}$ | $6.1 \cdot 10^{-12} \times e^{-75/T}$ | Burkholder et al. (2015) |
| 133 | $H_2S + H \rightarrow H_2 + HS$ | $1.3 \cdot 10^{-11} \times e^{-860/T}$ | $1.5 \cdot 10^{-11} \times e^{-850/T}$ | Balch & Wolfe (1976) |
| 134 | $H_2S + O \rightarrow OH + HS$ | $9.2 \cdot 10^{-12} \times e^{-1800/T}$ | $9.2 \cdot 10^{-12} \times e^{-1800/T}$ | Burkholder et al. (2015) |
| 135 | $HS + O \rightarrow H + SO$ | $1.6 \cdot 10^{-10}$ | $1.6 \cdot 10^{-10}$ | Burkholder et al. (2015) |
| 136 | $HS + O_2 \rightarrow OH + SO$ | $4.0 \cdot 10^{-19}$ | $4.0 \cdot 10^{-19}$ | Burkholder et al. (2015) |
| 137 | $HS + HO_2 \rightarrow H_2S + O_2$ | $3.0 \cdot 10^{-11}$ | $1.0 \cdot 10^{-11}$ | National Institute of Standards & Technology (2018) |
| 138 | $HS + HS \rightarrow H_2S + S$ | $1.2 \cdot 10^{-11}$ | $2.0 \cdot 10^{-11}$ | National Institute of Standards & Technology (2018) |
| 139 | $HS + HCO \rightarrow H_2S + CO$ | $5.0 \cdot 10^{-11}$ | $2.0 \cdot 10^{-11}$ | Kasting (1990) |
| 140 | $HS + H \rightarrow H_2 + S$ | $1.0 \cdot 10^{-11}$ | $1.0 \cdot 10^{-11}$ | Schofield (1973), Arney et al. (2016) |

**Table 3**
(Continued)

| # | Reaction | Segura et al. (2005) Rate | New Rate | Reference |
|---|---|---|---|---|
| 141 | $HS + S \rightarrow H + S_2$ | $2.2 \cdot 10^{-11} \times e^{120/T}$ | $1.0 \cdot 10^{-11}$ | Zahnle (1986) |
| 142 | $S + O_2 \rightarrow SO + O$ | $2.3 \cdot 10^{-12}$ | $1.6 \cdot 10^{-12} \times e^{100/T}$ | Burkholder et al. (2015) |
| 143 | $S + OH \rightarrow SO + H$ | $6.6 \cdot 10^{-11}$ | $6.6 \cdot 10^{-12}$ | Burkholder et al. (2015) |
| 144 | $S + HCO \rightarrow HS + CO$ | | $1.0 \cdot 10^{-11}$ | Zahnle (1986) |
| 145 | $S + HO_2 \rightarrow HS + O_2$ | $1.5 \cdot 10^{-11}$ | $5.0 \cdot 10^{-12}$ | Zahnle (1986) |
| 146 | $S + HO_2 \rightarrow SO + OH$ | $1.5 \cdot 10^{-11}$ | $5.0 \cdot 10^{-12}$ | Zahnle (1986) |
| 147 | $S + S \rightarrow S_2$ | | $1.87 \cdot 10^{-33} \times e^{-206/T} \times \rho$ | Domagal-Goldman et al. (2011) |
| 148 | $S + S \rightarrow S_2$ | | $1.1 \cdot 10^{-31} \times e^{-206/T} \times \rho$ | Du et al. (2008), Domagal-Goldman et al. (2011) |
| 149 | $S_2 + O \rightarrow S + SO$ | | $1.1 \cdot 10^{-11}$ | Hills et al. (1987), Domagal-Goldman et al. (2011) |
| 150 | $HS + H_2CO \rightarrow H_2S + HCO$ | $1.7 \cdot 10^{-11} \times e^{-800/T}$ | $1.7 \cdot 10^{-11} \times e^{-800/T}$ | Harman et al. (2015) |
| 151 | $S_2 + h\nu \rightarrow S + S$ | | | *photolysis* |
| *ph-* | 152 *otolysis* | $S_2 + h\nu \rightarrow S_2$ | | |
| Ka- | 153 sting (1990), Domagal-Goldman et al. (2011) | $S + S_2 \rightarrow S_3$ | | $2.8 \cdot 10^{-32} \times \rho$ |
| 154 | $S_2 + S_2 \rightarrow S_4$ | | $2.8 \cdot 10^{-31} \times \rho$ | Balch & Wolfe (1976), Domagal-Goldman et al. (2011) |
| 155 | $S + S_3 \rightarrow S_4$ | | $2.8 \cdot 10^{-31} \times \rho$ | Balch & Wolfe (1976), Domagal-Goldman et al. (2011) |
| 156 | $S_4 + S_4 \rightarrow S_8$ | | $2.8 \cdot 10^{-31} \times \rho$ | Balch & Wolfe (1976), Domagal-Goldman et al. (2011) |
| 157 | $S_4 + h\nu \rightarrow S_2 + S_2$ | | | *photolysis* |
| 158 | $S_3 + h\nu \rightarrow S_2 + S$ | | | *photolysis* |
| 159 | $S_8 + h\nu \rightarrow S_4 + S_4$ | | | *photolysis* |
| 160 | $S_8 + h\nu \rightarrow S_4 + S_4$ | | | *photolysis* |
| 161 | | | | *photolysis* |

**Table 3**
(Continued)

| # | Reaction | Segura et al. (2005) Rate | New Rate | Reference |
|---|---|---|---|---|
| | $S_8 + h\nu \rightarrow S_4 + S_4$ | | | |
| 162 | $SO_3 + h\nu \rightarrow SO_2 + O$ | | | *photolysis* |
| 163 | $SO_{21} + M \rightarrow SO_{23} + M$ | $1.0 \cdot 10^{-12}$ | $1.0 \cdot 10^{-12}$ | Turco et al. (1982) |
| 164 | $SO_{21} + M \rightarrow SO_2 + M$ | $1.0 \cdot 10^{-11}$ | $1.0 \cdot 10^{-11}$ | Turco et al. (1982) |
| 165 | $SO_{21} + \mathrm{hv} \rightarrow SO_{23} + M$ | $1.5 \cdot 10^{3}$ | $1.5 \cdot 10^{3}$ | Turco et al. (1982) |
| 166 | $SO_{21} + h\nu \rightarrow SO_2 + h\nu$ | $2.2 \cdot 10^{4}$ | $2.2 \cdot 10^{4}$ | Turco et al. (1982) |
| 167 | $SO_{21} + O_2 \rightarrow SO_3 + O$ | $1.0 \cdot 10^{-16}$ | $1.0 \cdot 10^{-16}$ | Turco et al. (1982) |
| 168 | $SO_{21} + SO_2 \rightarrow SO_3 + SO$ | $4.0 \cdot 10^{-12}$ | $4.0 \cdot 10^{-12}$ | Turco et al. (1982) |
| 169 | $SO_{23} + M \rightarrow SO_2 + M$ | $1.5 \cdot 10^{-13}$ | $1.5 \cdot 10^{-13}$ | Turco et al. (1982) |
| 170 | $SO_{23} + h\nu \rightarrow SO_2 + h\nu$ | $1.1 \cdot 10^{3}$ | $1.1 \cdot 10^{3}$ | Turco et al. (1982) |
| 171 | $SO_{23} + SO_2 \rightarrow SO_3 + SO$ | $7.0 \cdot 10^{-14}$ | $7.0 \cdot 10^{-14}$ | Turco et al. (1982) |
| 172 | $SO + NO_2 \rightarrow SO_2 + NO$ | $1.4 \cdot 10^{-11}$ | $1.4 \cdot 10^{-11}$ | Burkholder et al. (2015) |
| 173 | $SO + O_3 \rightarrow SO_2 + O_2$ | $3.6 \cdot 10^{-12} \times e^{-1100/T}$ | $3.4 \cdot 10^{-12} \times e^{-1100/T}$ | Burkholder et al. (2015) |
| 174 | $SO_2 + HO_2 \rightarrow SO_3 + OH$ | 0 | $1.0 \cdot 10^{-18}$ | Harman et al. (2015) |
| 175 | $HS + O_3 \rightarrow HSO + O_2$ | $9.0 \cdot 10^{-12} \times e^{-280/T}$ | $9.0 \cdot 10^{-12} \times e^{-280/T}$ | Burkholder et al. (2015) |
| 176 | $HS + NO_2 \rightarrow HSO + NO$ | $2.9 \cdot 10^{-11} \times e^{240/T}$ | $2.9 \cdot 10^{-11} \times e^{250/T}$ | Burkholder et al. (2015) |
| 177 | $S + O_3 \rightarrow SO + O_2$ | $1.2 \cdot 10^{-11}$ | $1.2 \cdot 10^{-11}$ | National Institute of Standards & Technology (2018) |
| 178 | $SO + SO \rightarrow SO_2 + S$ | $8.3 \cdot 10^{-15}$ | $2.0 \cdot 10^{-15}$ | National Institute of Standards & Technology (2018) |
| 179 | $SO_3 + SO \rightarrow SO_2 + SO_2$ | $2.0 \cdot 10^{-15}$ | $2.0 \cdot 10^{-15}$ | National Institute of Standards & Technology (2018) |
| 180 | $S + CO_2 \rightarrow SO + CO$ | $1.0 \cdot 10^{-20}$ | $1.0 \cdot 10^{-20}$ | Harman et al. (2015) |
| 181 | $SO + HO_2 \rightarrow HSO + O_2$ | 0 | $2.8 \cdot 10^{-11}$ | Harman et al. (2015) |
| 182 | | $3.5 \cdot 10^{-12} \times e^{140/T}$ | $5.6 \cdot 10^{-12} \times \frac{T}{298}^{-0.4}$ | Kasting (1990) |

**Table 3**
(Continued)

| # | Reaction | Segura et al. (2005) Rate | New Rate | Reference |
|---|---|---|---|---|
| | SO + HCO → HSO + CO | | | |
| 182 | SO + HCO → HSO + CO | $3.5 \cdot 10^{-12} \times e^{T/-140}$ | $5.6 \cdot 10^{-12} \times \frac{T}{298}^{-0.4}$ | Kasting (1990) |
| 183 | H + SO → HSO | $5.7 \cdot 10^{-32} \times \frac{T}{298}^{1.6}$ | $5.7 \cdot 10^{-32} \times \frac{T}{298}^{1.6}$ | Kasting (1990) |
| 184 | HSO + *hv* → HS + O | | | *photolysis* |
| 185 | HSO + NO → HNO + SO | | $1.0 \cdot 10^{-15}$ | Harman et al. (2015) |
| 186 | HSO + OH → $H_2O$ + SO | $4.8 \cdot 10^{-11} \times e^{250/T}$ | $3.0 \cdot 10^{-11}$ | Kasting (1990) |
| 187 | HSO + H → HS + OH | $7.29 \cdot 10^{-11}$ | $2.0 \cdot 10^{-11}$ | Kasting (1990) |
| 188 | HSO + H → $H_2$ + SO | $6.48 \cdot 10^{-12}$ | $6.48 \cdot 10^{-12}$ | Kasting (1990) |
| 189 | HSO + HS → $H_2S$ + SO | $1 \cdot 10^{-12}$ | $3 \cdot 10^{-11}$ | Kasting (1990) |
| 190 | HSO + O → OH + SO | $3 \cdot 10^{-11} \times e^{200/T}$ | $3 \cdot 10^{-11}$ | Kasting (1990) |
| 191 | HSO + S → HS + SO | $1 \cdot 10^{-11}$ | $3 \cdot 10^{-11}$ | Kasting (1990) |
| 192 | $SO_3$ + CO → $SO_2$ + $CO_2$ | | 0 | |
| 193 | H + OCS → CO + HS | | $9.1 \cdot 10^{-12} \times e^{-1940/T}$ | National Institute of Standards & Technology (2018) |
| 194 | HS + CO → OCS + H | | $4.15 \cdot 10^{-14} \times e^{-7650/T}$ | National Institute of Standards & Technology (2018) |
| 195 | O + OCS → CO + SO | | $2.1 \cdot 10^{-11} \times e^{-2200/T}$ | Burkholder et al. (2015) |
| 196 | O + OCS → S + $CO_2$ | | $8.33 \cdot 10^{-11} \times e^{-5530/T}$ | Singleton & Cvetanović (1988), Arney et al. (2016) |
| 197 | OCS + S → CO + $S_2$ | | $1.5 \cdot 10^{-10} \times e^{-1830/T}$ | National Institute of Standards & Technology (2018) |
| 198 | OCS + OH → $CO_2$ + HS | | $7.00 \cdot 10^{-15} \times e^{-1070/T}$ | Burkholder et al. (2015) |
| 199 | S + HCO → OCS + H | | $8.0 \cdot 10^{-11}$ | Loison et al. (2012) |
| 200 | S + CO → OCS | | $4.0 \cdot 10^{-33} \times e^{-1780/T*}\rho$ | Lincowski et al. (2018) |
| 201 | OCS + *hv* → CO + S | | | *photolysis* |

**Table 3**
(Continued)

| # | Reaction | Segura et al. (2005) Rate | New Rate | Reference |
|---|---|---|---|---|
| 202 | $CO + O^1D \rightarrow CO + O$ | | $7.0 \cdot 10^{-11}$ | National Institute of Standards & Technology (2018) |
| 203 | $CO + O^1D \rightarrow CO_2$ | | 0 | |
| 204 | $H_2S + HO_2 \rightarrow H_2O + HSO$ | | $3.0 \cdot 10^{-15}$ | Burkholder et al. (2015) |
| 205 | $OCS + S \rightarrow OCS_2$ | | $8.33 \cdot 10^{-33} \times \rho$ | National Institute of Standards & Technology (2018) |
| 206 | $OCS_2 + S \rightarrow OCS + S_2$ | | $2.0 \cdot 10^{-11}$ | National Institute of Standards & Technology (2018) |
| 207 | $S_3 + O \rightarrow S_2 + SO$ | | 0 | Zahnle (1986) |
| 208 | $S_4 + O \rightarrow S_3 + SO$ | | 0 | Zahnle (1986) |
| 209 | $OCS_2 + CO \rightarrow OCS + OCS$ | | $3.0 \cdot 10^{-12}$ | Zahnle (1986) |
| 210 | $O + NO_3 \rightarrow O_2 + NO_2$ | $1.0 \cdot 10^{-11}$ | $1.0 \cdot 10^{-11}$ | Burkholder et al. (2015) |
| 211 | $HO_2 + NO_3 \rightarrow OH + NO_2 + O_2$ | $4.1 \cdot 10^{-12}$ | $3.5 \cdot 10^{-12}$ | Burkholder et al. (2015) |
| 212 | $NO + NO_3 \rightarrow NO_2 + NO_2$ | $1.5 \cdot 10^{-11} \times e^{170/T}$ | $1.5 \cdot 10^{-11} \times e^{170/T}$ | Burkholder et al. (2015) |
| 213 | $O + NO_2 \rightarrow NO_3$ | $9 \cdot 10^{-32} \times \left(\frac{298}{T}\right)^{2.0}$ | $2.5 \cdot 10^{-31} \times \left(\frac{298}{T}\right)^{1.8}$ | Burkholder et al. (2015) |
| 214 | $OH + NO_3 \rightarrow HO_2 + NO_2$ | $2.3 \cdot 10^{-11}$ | $2.2 \cdot 10^{-11}$ | Burkholder et al. (2015) |
| 215 | $NO_3 + h\nu \rightarrow NO + O_2$ | | | *photolysis* |
| 216 | $NO_3 + h\nu \rightarrow NO_2 + O$ | | | *photolysis* |
| 217 | $N + NO_2 \rightarrow N_2O + O$ | | $5.8 \cdot 10^{-12} \times e^{220/T}$ | Burkholder et al. (2015) |
| 218 | $O^1D + N_2 \rightarrow N_2O$ | $3.5 \cdot 10^{-37} \times \left(\frac{298}{T}\right)^{0.6}$ | $2.8 \cdot 10^{-36} \times \left(\frac{298}{T}\right)^{0.9}$ | Burkholder et al. (2015) |
| 219 | $O^1D + N_2O \rightarrow N_2 + O_2$ | $4.9 \cdot 10^{-11}$ | $4.64 \cdot 10^{-11} \times e^{20/T}$ | Burkholder et al. (2015) |
| 220 | $O^1D + N_2O \rightarrow NO + NO$ | $6.7 \cdot 10^{-11}$ | $7.26 \cdot 10^{-11} \times e^{20/T}$ | Burkholder et al. (2015) |
| 221 | $N_2O + h\nu \rightarrow N_2 + O^1D$ | | | *photolysis* |
| 222 | $N + HO_2 \rightarrow NO + OH$ | | $2.2 \cdot 10^{-11}$ | Brune et al. (1983) |
| 223 | | | $3.0 \cdot 10^{-16}$ | Burkholder et al. (2015) |

**Table 3**
(Continued)

| # | Reaction | Segura et al. (2005) Rate | New Rate | Reference |
|---|---|---|---|---|
| | $O + N_2O_5 \rightarrow NO_2 + NO_2 + O_2$ | | | |
| 224 | $NO_2 + NO_3 \rightarrow N_2O_5$ | $2.2 \cdot 10^{-30} \times \left(\frac{298}{T}\right)^{3.9}$ | $2.4 \cdot 10^{-30} \times \left(\frac{298}{T}\right)^{3.0}$ | Burkholder et al. (2015) |
| 225 | $N_2O_5 + M \rightarrow NO_3 + NO_2$ | | $\frac{\left(\frac{k_0^{*} k_{\rm inf}}{k_0 + k_{\rm inf}} \times 10\left(\frac{\log_{10}(0.35)}{\left(1 + \left(\frac{log_{10}\left(\frac{k_0}{k_{\rm inf}}\right)}{0.75 - 1.27 * log_{10}(0.35)}\right)^2\right)}\right)\right)}{\cdot 10^{14} \times e^{-11080/T} \times \frac{T}{300}^{\rho}{}^{0.1}} k_0 = 1.3 \cdot 10^{-3} \times e^{-11000/T} \times \frac{300}{T}^{3.5} \times \rho\ k_{inf} = 9.7$ | Atkinson et al. (2007) |
| 226 | $NO_2 + NO_3 \rightarrow NO + NO_2 + O_2$ | $4.5 \cdot 10^{-14} \times e^{-1260/T}$ | $4.5 \cdot 10^{-14} \times e^{-1260/T}$ | Burkholder et al. (2015) |
| 227 | $NO_2 + O_3 \rightarrow NO_2 + O_2$ | $1.2 \cdot 10^{-13} \times e^{-2450/T}$ | $1.2 \cdot 10^{-13} \times e^{-2450/T}$ | Burkholder et al. (2015) |
| 228 | $NO_3 + NO_3 \rightarrow NO_2 + NO_2 + O_2$ | | $8.5 \cdot 10^{-13} \times e^{-2450/T}$ | Burkholder et al. (2015) |
| 229 | $HO_2 + NO_2 \rightarrow HO_2NO_2$ | $1.8 \cdot 10^{-31} \times \left(\frac{298}{T}\right)^{3.2}$ | $1.9 \cdot 10^{-32} \times \left(\frac{298}{T}\right)^{3.4}$ | Burkholder et al. (2015) |
| 230 | $O + HO_2NO_2 \rightarrow OH + NO_2 + O_2$ | $7.8 \cdot 10^{-11} \times e^{-3400/T}$ | $7.8 \cdot 10^{-11} \times e^{-3400/T}$ | Burkholder et al. (2015) |
| 231 | $HO_2NO_2 + OH \rightarrow NO_2 + H_2O + O_2$ | $1.3 \cdot 10^{-12} \times e^{380/T}$ | $1.9 \cdot 10^{-12} \times e^{270/T}$ | Atkinson et al. 2004 |
| 232 | $HO_2NO_2 + M \rightarrow HO_2 + NO_2$ | | $\frac{\left(\frac{k_0^{*} k_{\rm inf}}{k_0 + k_{\rm inf}} \times 10(\frac{\log_{10}(0.6)}{\left(1 + \left(\frac{log_{10}(\frac{k_0}{k_{\rm inf}})}{0.75 - 1.27 * log_{10}(0.6)}\right)^2)\right)}\right)}{\rho} k_0 = 4.1 \cdot 10^{-5} \times e^{-10650/T} \times \rho\ k_{\rm inf} = 4.8 \cdot 10^{15} \times e^{-11170/T}$ | Atkinson et al. 2004 |
| 233 | $HO_2NO_2 + h\nu \rightarrow HO_2 + NO_2$ | | | *photolysis* |
| 234 | $HO_2NO_2 + h\nu \rightarrow OH + NO_3$ | | | *photolysis* |
| 235 | $N_2O_5 + h\nu \rightarrow NO_3 + NO_2$ | | | *photolysis* |

**Table 3**
(Continued)

| # | Reaction | Segura et al. (2005) Rate | New Rate | Reference |
|---|---|---|---|---|
| 236 | $N_2O_5 + h\nu \rightarrow NO_3 + NO + O$ | | | *photolysis* |
| 237 | $Cl_2 + h\nu \rightarrow Cl + Cl$ | | | *photolysis* |
| 238 | $ClONO_2 + h\nu \rightarrow Cl + NO_3$ | | | *photolysis* |
| 239 | $ClONO_2 + h\nu \rightarrow ClO + NO_2$ | | | *photolysis* |
| 240 | $O + HNO_3 \rightarrow OH + NO_3$ | | $3.0 \cdot 10^{-17}$ | Burkholder et al. (2015) |
| 241 | $HO_2 + NO_2 \rightarrow HNO_2 + O_2$ | | $5.0 \cdot 10^{-16}$ | Burkholder et al. (2015) |
| 242 | $O + HCl \rightarrow OH + Cl$ | $1.0 \cdot 10^{-11} \times e^{-3300/T}$ | $1.0 \cdot 10^{-11} \times e^{-3300/T}$ | Sander et al. (2006) |
| 243 | $OH + HCl \rightarrow H_2O + Cl$ | $2.6 \cdot 10^{-12} \times e^{-350/T}$ | $2.6 \cdot 10^{-12} \times e^{-350/T}$ | Sander et al. (2006) |
| 244 | $HO_2 + Cl \rightarrow HCl + O_2$ | $1.8 \cdot 10^{-11} \times e^{170/T}$ | $1.8 \cdot 10^{-11} \times e^{170/T}$ | Sander et al. (2006) |
| 245 | $Cl + H_2 \rightarrow HCl + H$ | $3.7 \cdot 10^{-11} \times e^{-2300/T}$ | $3.05 \cdot 10^{-11} \times e^{-2270/T}$ | Sander et al. (2006) |
| 246 | $Cl + H_2O_2 \rightarrow HCl + HO_2$ | $1.1 \cdot 10^{-11} \times e^{-980/T}$ | $1.1 \cdot 10^{-11} \times e^{-980/T}$ | Sander et al. (2006) |
| 247 | $Cl + CH_4 \rightarrow HCl + CH_3$ | $1.1 \cdot 10^{-11} \times e^{-1400/T}$ | $7.3 \cdot 10^{-12} \times e^{-1280/T}$ | Sander et al. (2006) |
| 248 | $Cl + H_2CO \rightarrow HCl + HCO$ | $8.1 \cdot 10^{-11} \times e^{-30/T}$ | $8.1 \cdot 10^{-11} \times e^{-30/T}$ | Sander et al. (2006) |
| 249 | $Cl + C_2H_6 \rightarrow HCl + C_2H_5$ | | $7.2 \cdot 10^{-11} \times e^{-70/T}$ | Sander et al. (2006) |
| 250 | $O^1D + HCl \rightarrow O + HCl$ | | $1.5 \cdot 10^{-11}$ | Sander et al. (2006) |
| 251 | $O^1D + HCl \rightarrow OH + Cl$ | | $9.75 \cdot 10^{-11}$ | Sander et al. (2006) |
| 252 | $Cl + H_2S \rightarrow HCl + HS$ | | $3.7 \cdot 10^{-11} \times e^{210/T}$ | Sander et al. (2006) |
| 253 | $HCl + h\nu \rightarrow H + Cl$ | | | *photolysis* |
| 254 | $O + ClO \rightarrow Cl + O_2$ | $3.0 \cdot 10^{-11} \times e^{70/T}$ | $2.8 \cdot 10^{-11} \times e^{85/T}$ | Sander et al. (2006) |
| 255 | $O + ClO \rightarrow OClO$ | | $8.6 \cdot 10^{-21} \times T^{-4.1} \times e^{-420/T}$ | Sander et al. (2006) |
| 256 | $O + HOCl \rightarrow OH + ClO$ | $1 \cdot 10^{-11} \times e^{-2200/T}$ | $1.7 \cdot 10^{-13}$ | Sander et al. (2006) |
| 257 | $OH + ClO \rightarrow Cl + HO_2$ | $1.1 \cdot 10^{-11} \times e^{120/T}$ | $7.4 \cdot 10^{-12} \times e^{270/T}$ | Sander et al. (2006) |

**Table 3**
(Continued)

| # | Reaction | Segura et al. (2005) Rate | New Rate | Reference |
|---|---|---|---|---|
| 258 | $OH + ClO \rightarrow HCl + O_2$ | | $6.0 \cdot 10^{-13} \times e^{230/T}$ | Sander et al. (2006) |
| 259 | $OH + HOCl \rightarrow H_2O + ClO$ | $3.0 \cdot 10^{-12} \times e^{-500/T}$ | $3.0 \cdot 10^{-12} \times e^{-500/T}$ | Sander et al. (2006) |
| 260 | $HO_2 + Cl \rightarrow OH + ClO$ | | $4.1 \cdot 10^{-11} \times e^{-450/T}$ | Sander et al. (2006) |
| 261 | $HO_2 + ClO \rightarrow HOCl + O_2$ | $4.8 \cdot 10^{-13} \times e^{700/T}$ | $2.7 \cdot 10^{-12} \times e^{220/T}$ | Sander et al. (2006) |
| 262 | $Cl + O_3 \rightarrow ClO + O_2$ | $2.9 \cdot 10^{-11} \times e^{-260/T}$ | $2.3 \cdot 10^{-11} \times e^{-200/T}$ | Sander et al. (2006) |
| 263 | $ClO + NO \rightarrow NO_2 + Cl$ | $6.4 \cdot 10^{-12} \times e^{290/T}$ | $6.4 \cdot 10^{-12} \times e^{290/T}$ | Sander et al. (2006) |
| 264 | $ClO + SO \rightarrow Cl + CO_2$ | | $2.8 \cdot 10^{-11}$ | Sander et al. (2006) |
| 265 | $O^1D + HCl \rightarrow H + ClO$ | | $3.75 \cdot 10^{-11}$ | Sander et al. (2006) |
| 266 | $ClO + h\nu \rightarrow O^1D + Cl$ | | | *photolysis* |
| 267 | $ClO + h\nu \rightarrow O + Cl$ | | | *photolysis* |
| 268 | $HOCl + h\nu \rightarrow OH + Cl$ | | | *photolysis* |
| 269 | $O^1D + Cl_2 \rightarrow ClO + Cl$ | | $2.025 \cdot 10^{-10}$ | Sander et al. (2006) |
| 270 | $O^1D + Cl_2 \rightarrow Cl_2 + O$ | | $6.75 \cdot 10^{-11}$ | Sander et al. (2006) |
| 271 | $OH + Cl_2 \rightarrow HOCl + Cl$ | $1.4 \cdot 10^{-12} \times e^{-900/T}$ | $1.4 \cdot 10^{-12} \times e^{-900/T}$ | Sander et al. (2006) |
| 272 | $Cl + HOCl \rightarrow Cl_2 + OH$ | $3.0 \cdot 10^{-12} \times e^{-130/T}$ | $1.25 \cdot 10^{-12} \times e^{-130/T}$ | Sander et al. (2006) |
| 273 | $Cl + HOCl \rightarrow HCl + ClO$ | | $1.25 \cdot 10^{-12} \times e^{-130/T}$ | Sander et al. (2006) |
| 274 | $ClO + ClO \rightarrow Cl_2 + O_2$ | | $1.00 \cdot 10^{-12} \times e^{-1590/T}$ | Sander et al. (2006) |
| 275 | $O + OClO \rightarrow ClO + O_2$ | | $2.40 \cdot 10^{-12} \times e^{-960/T}$ | Sander et al. (2006) |
| 276 | $NO + OClO \rightarrow NO_2 + ClO$ | | $2.50 \cdot 10^{-12} \times e^{-600/T}$ | Sander et al. (2006) |
| 277 | $OH + OClO \rightarrow HOCl + O_2$ | | $4.50 \cdot 10^{-13} \times e^{800/T}$ | Sander et al. (2006) |
| 278 | $Cl + O_2 \rightarrow ClOO$ | $2.7 \cdot 10^{-33} \times {\frac{T}{298}}^{1.5}$ | $1.4 \cdot 10^{-33} \times \rho \times {\frac{T}{300}}^{-4.1}$ | Atkinson et al. (2007) |
| 279 | $Cl + OClO \rightarrow ClO + ClO$ | | $3.40 \cdot 10^{-11} \times e^{160/T}$ | Sander et al. (2006) |
| 280 | | $2.30 \cdot 10^{-10}$ | $2.30 \cdot 10^{-10}$ | Sander et al. (2006) |

**Table 3**
(Continued)

| # | Reaction | Segura et al. (2005) Rate | New Rate | Reference |
| --- | --- | --- | --- | --- |
| | $Cl + OClO \rightarrow Cl_2 + O_2$ | | | |
| 281 | $Cl + ClOO \rightarrow ClO + ClO$ | $1.2 \cdot 10^{-11}$ | $1.2 \cdot 10^{-11}$ | Sander et al. (2006) |
| 282 | $ClO + O_3 \rightarrow ClOO + O_2$ | | $1.0 \cdot 10^{-12} \times e^{-3600/T}$ | Sander et al. (2006) |
| 283 | $ClO + O_3 \rightarrow OClO + O_2$ | | $1.0 \cdot 10^{-12} \times e^{-4000/T}$ | Sander et al. (2006) |
| 284 | $ClO + ClO \rightarrow ClOO + Cl$ | | $3.0 \cdot 10^{-11} \times e^{-2450/T}$ | Sander et al. (2006) |
| 285 | $ClO + ClO \rightarrow OClO + Cl$ | | $3.5 \cdot 10^{-13} \times e^{-1370/T}$ | Sander et al. (2006) |
| 286 | $SO + OClO \rightarrow SO_2 + ClO$ | | $1.9 \cdot 10^{-12}$ | Sander et al. (2006) |
| 287 | $CLOO + hv \rightarrow CLO + O$ | | | *photolysis* |
| 288 | $OClO + hv \rightarrow ClO + O$ | | | *photolysis* |
| 289 | $Cl + NO_3 \rightarrow ClO + NO_2$ | $2.6 \cdot 10^{-11}$ | $2.4 \cdot 10^{-11}$ | Sander et al. (2006) |
| 290 | $NO_3 + ClO \rightarrow ClOO + NO_2$ | | $4.7 \cdot 10^{-13}$ | Sander et al. (2006) |
| 291 | $Cl + NO_2 \rightarrow ClONO$ | | $1.2 \cdot 10^{-30} \times \left(\frac{298}{T}\right)^{2.0}$ | Sander et al. (2006) |
| 292 | $O + ClONO_2 \rightarrow O_2 + ClONO$ | | $2.9 \cdot 10^{-12} \times e^{-800/T}$ | Sander et al. (2006) |
| 293 | $OH + ClONO_2 \rightarrow HOCl + NO_3$ | | $6.0 \cdot 10^{-13} \times e^{-330/T}$ | Sander et al. (2006) |
| 294 | $OH + ClONO_2 \rightarrow ClO + HNO_3$ | | $6.0 \cdot 10^{-13} \times e^{-330/T}$ | Sander et al. (2006) |
| 295 | $Cl + ClONO_2 \rightarrow Cl_2 + NO_3$ | | $6.5 \cdot 10^{-12} \times e^{135/T}$ | Sander et al. (2006) |
| 296 | $ClO + NO_2 \rightarrow ClONO_2$ | $1.8 \cdot 10^{-31} \times \left(\frac{298}{T}\right)^{3.4}$ | $1.8 \cdot 10^{-31} \times \left(\frac{298}{T}\right)^{3.4}$ | Sander et al. (2006) |
| 297 | $ClONO + hv \rightarrow Cl + NO_2$ | | | *photolysis* |
| 298 | $Cl + NO \rightarrow ClNO$ | $9.0 \cdot 10^{-32} \times \left(\frac{298}{T}\right)^{1.6}$ | $7.6 \cdot 10^{-32} \times \left(\frac{300}{T}\right)^{1.8} \times \rho$ | Sander et al. (2006) |
| 299 | $Cl + ClNO \rightarrow NO + Cl_2$ | $5.8 \cdot 10^{-11} \times e^{100/T}$ | $5.8 \cdot 10^{-11} \times e^{100/T}$ | Sander et al. (2006) |

**Table 3**
(Continued)

| # | Reaction | Segura et al. (2005) Rate | New Rate | Reference |
|---|---|---|---|---|
| 300 | $Cl + NO_2 \rightarrow ClNO_2$ | $1.3 \cdot 10^{-30} \times \left(\frac{298}{T}\right)^{2.0}$ | $1.8 \cdot 10^{-31} \times \left(\frac{298}{T}\right)^{2.0}$ | Sander et al. (2006) |
| 301 | $OH + ClNO_2 \rightarrow HOCl + NO_2$ | | $2.4 \cdot 10^{-12} \times e^{-1250/T}$ | Sander et al. (2006) |
| 302 | $ClONO_2 + h\nu \rightarrow Cl + NO_2$ | | | *photolysis* |
| 303 | $ClNO + h\nu \rightarrow Cl + NO_2$ | | | *photolysis* |
| 304 | $OH + CH_3Cl \rightarrow CH_2Cl + H_2O$ | $2.1 \cdot 10^{-12} \times e^{-1150/T}$ | $1.96 \cdot 10^{-12} \times e^{-1250/T}$ | Burkholder et al. (2020) |
| 305 | $Cl + CH_3Cl \rightarrow CH_2Cl + HCl$ | | $2.17 \cdot 10^{-11} e^{-1130/T}$ | Sander et al. (2006) |
| 306 | $CH_2Cl + O_2 \rightarrow CH_2ClO_2$ | | $1.9 \cdot 10^{-30} \times \left(\frac{298}{T}\right)^{3.2}$ | Sander et al. (2006) |
| 307 | $NO + CH_2ClO_2 \rightarrow CH_2ClO + NO_2$ | | $7.0 \cdot 10^{-12} e^{300/T}$ | Sander et al. (2006) |
| 308 | $CH_2ClO + O_2 \rightarrow CHClO + HO_2$ | | $6.0 \cdot 10^{-14}$ | Sander et al. (2006) |
| 309 | $CHClO + h\nu \rightarrow HCO + Cl$ | | | *photolysis* |
| 310 | $CH_3Cl + h\nu \rightarrow Cl + CH_3$ | | | *photolysis* |
| 311 | $O^1D + CCl_4 \rightarrow ClO + CCl_3$ | | $2.838 \cdot 10^{-10}$ | Sander et al. (2006) |
| 312 | $CCl_3 + O_2 \rightarrow CCl_3O_2$ | | $8.0 \cdot 10^{-31} \times \left(\frac{298}{T}\right)^{6.0}$ | Sander et al. (2006) |
| 313 | $NO + CCl_3O_2 \rightarrow COCl_2 + Cl + NO_2$ | | $7.3 \cdot 10^{-12} \times e^{270/T}$ | Sander et al. (2006) |
| 314 | $O^1D + COCl_2 \rightarrow Cl + ClO + CO$ | | $2.2 \cdot 10^{-10} \times e^{30/T}$ | Sander et al. (2006) |
| 315 | $CCl_4 + h\nu \rightarrow CCl_3 + Cl$ | | | *photolysis* |
| 316 | $COCl_2 + h\nu \rightarrow Cl + Cl + CO$ | | | *photolysis* |
| 317 | $NO_2 + CCl_3O_2 \rightarrow CC_3NO_4$ | | $2.9 \cdot 10^{-29} \times \left(\frac{298}{T}\right)^{6.8}$ | Sander et al. (2006) |

**Table 3**
(Continued)

| # | Reaction | Segura et al. (2005) Rate | New Rate | Reference |
|---|---|---|---|---|
| 318 | $CCl_3NO_4$ + M → $CCl_3O_2$ + $NO_2$ | | $\frac{\frac{k_0^* k_{\rm inf}}{k_0+k_{\rm inf}} \times 10^{\left(\frac{\log_{10}(0.32)}{\left(1+\left(\frac{log_{10}\left(\frac{k_0}{k_{\rm inf}}\right)}{0.75-1.27*log_{10}(0.32)}\right)^2\right)}\right)}}{\rho}$ $k_0 = 4.3 \cdot 10^{-3} \times e^{-10235/T} \times \rho$ $k_{inf}$=$4.8 \cdot 10^{16} \times e^{-11820/T}$ | Sander et al. (2006) |
| 319 | ClOO + M → Cl + $O_2$ | | $2.8 \cdot 10^{-10} \times e^{-1820/T}$ | Atkinson et al. (2007) |
| 320 | ClO + $CH_2O_2$ → ClOO + $CH_3O$ | | $3.3 \cdot 10^{-12} \times e^{-115/T}$ | Sander et al. (2006) |
| 321 | Cl + $CH_3ONO_2$ → $CH_2ONO_2$ + HCl | | $1.3 \cdot 10^{-11} \times e^{-1200/T}$ | Sander et al. (2006) |
| 322 | Cl + $CH_2O_2$ → HCl + $CH_2O_2$ | | $8.0 \cdot 10^{-11}$ | Sander et al. (2006) |
| 323 | Cl + $CH_3O_2$ → ClO + $CH_3O$ | | $8.0 \cdot 10^{-11}$ | Sander et al. (2006) |
| 324 | Cl + $CH_3OH$ → $CH_2OH$ + HCl | | $5.5 \cdot 10^{-11}$ | Sander et al. (2006) |
| 325 | ClO + $CH_3O_2$ →$CH_3OCl$ + $O_2$ | | $1.65 \cdot 10^{-12} \times e^{-115/T}$ | Sander et al. (2006) |
| 326 | ClO + $CH_3O_2$ → ClOO + ClOO | | $1.65 \cdot 10^{-12} \times e^{-115/T}$ | Sander et al. (2006) |
| 327 | OH + $CH_3OCl$ → HOCl + $CH_3O$ | | $2.5 \cdot 10^{-12} \times e^{-370/T}$ | Sander et al. (2006) |
| 328 | $CH_3OCl$ + *hν* → $CH_3O$ + Cl | | | *photolysis* |
| 329 | Cl + $HO_2NO_2$ → ClO + $HNO_3$ | | $1.0 \cdot 10^{-13}$ | Sander et al. (2006) |
| 330 | O + $Cl_2O$ → ClO + ClO | | $2.7 \cdot 10^{-11} \times e^{-530/T}$ | Atkinson et al. (2007) |

**Table 3**
(Continued)

| # | Reaction | Segura et al. (2005) Rate | New Rate | Reference |
|---|---|---|---|---|
| 331 | $Cl + Cl_2O \rightarrow Cl_2 + ClO$ | | $6.2 \cdot 10^{-11} \times e^{130/T}$ | Atkinson et al. (2007) |
| 332 | $NO + Cl_2O_2 \rightarrow Cl_2O + NO_2$ | | $1.0 \cdot 10^{-15}$ | Ingham et al. (2005), Catling et al. (2010) |
| 333 | $CLO2 + hv \rightarrow Cl + ClO$ | | | *photolysis* |
| 334 | $ClO + ClO \rightarrow Cl_2O_2$ | $1.9 \cdot 10^{-32} \times \left(\frac{298}{T}\right)^{3.9}$ | $1.9 \cdot 10^{-32} \times \left(\frac{298}{T}\right)^{3.6}$ | Burkholder et al. (2020) |
| 335 | $Cl + Cl_2O_2 \rightarrow Cl_2 + ClOO$ | | $1.0 \cdot 10^{-10}$ | Sander et al. (2006) |
| 336 | $Cl_2O_2 + M \rightarrow ClO + ClO$ | | $\frac{\left(\frac{k_0^* k_{\rm inf}}{k_0 + k_{\rm inf}} \times 10\left(\frac{\log_{10}(0.45)}{\left(1 + \left(\frac{log_{10}(\frac{k_0}{k_{\rm inf}})}{0.75 - 1.27 * log_{10}(0.45)} 0.32\right)^2\right)}\right)\right)}{\rho}$ $k_0 = 3.7 \cdot 10^{-7} \times e^{-7690/T} \times \rho$ $k_{inf} = 7.9 \cdot 10^{15} \times e^{-882-/T}$ | Sander et al. (2006) |
| 337 | $CL_2O_2 + hv \rightarrow Cl + ClOO$ | | | *photolysis* |
| 338 | $CL_2O_2 + hv \rightarrow ClO + ClO$ | | | *photolysis* |
| 339 | $O + OClO \rightarrow ClO_3$ | | $2.9 \cdot 10^{-31} \times \left(\frac{298}{T}\right)^{3.1})$ | Sander et al. (2006) |
| 340 | $O_3 + OClO \rightarrow ClO_3 + O_2$ | | $2.1 \cdot 10^{-12} \times e^{-4700/T}$ | Sander et al. (2006) |
| 341 | $ClO_3 + hv \rightarrow ClO + O_2$ | | | *photolysis* |
| 342 | $OH + ClO_3 \rightarrow HClO_4$ | | $2.336 \cdot 10^{-9}$ | Simonaitis & Heicklen (1975), Catling et al. (2010) |
| 343 | $OH + ClO_3 \rightarrow HO_2 + OCLO$ | | $1.94 \cdot 10^{36} \times T^{-15.3} \times e^{-5542/T})$ | Zhu & Lin (2001) |
| 344 | $ClO + O_2 \rightarrow ClO_3$ | | $9.00 \cdot 10^{-28} \times T^{-2}$ | |
| 345 | $ClO + CLO_3 \rightarrow Cl_2O_4$ | | $8.62 \cdot 10^{15} \times e^{-1826/T} \times T^{-9.75}$ | Xu & Lin (2003) |
| 346 | $ClO + CLO_3 \rightarrow CLOO + OClO$ | | $1.82 \cdot 10^{-18} \times e^{-2417/T} \times T^{2.28}$ | Xu & Lin (2003) |

**Table 3**
(Continued)

| # | Reaction | Segura et al. (2005) Rate | New Rate | Reference |
|---|---|---|---|---|
| 347 | $ClO + CLO_3 \rightarrow Cl_2O_4$ | | $1.42 \cdot 10^{-18} \times e^{-2870/T} \times T^{2.11}$ | Xu & Lin (2003) |
| 348 | $Cl_2O_4 + h\nu \rightarrow ClOO + OClO$ | | | *photolysis* |
| 349 | $OH + ClO_3 \rightarrow HClO_4$ | | $3.0 \cdot 10^{-30} \times \rho$ | Simonaitis & Heicklen (1975), Catling et al. (2010) |
| 350 | $O + BrO \rightarrow Br + O_2$ | | $1.9 \cdot 10^{-11} \times e^{230/T}$ | Burkholder et al. (2020) |
| 351 | $BrO + h\nu \rightarrow Br + O$ | | | *photolysis* |
| 352 | $Br + NO_3 \rightarrow BrO + NO_2$ | | $1.6 \cdot 10^{-11}$ | Atkinson et al. (2007) |
| 353 | $Br + OClO \rightarrow BrO + ClO$ | | $2.6 \cdot 10^{-11} \times e^{-1300/T}$ | Burkholder et al. (2020) |
| 354 | $BrO + NO \rightarrow NO_2 + Br$ | | $8.8 \times 10^{-12} \times e^{260/T}$ | Burkholder et al. (2020) |
| 355 | $BrO + BrO \rightarrow Br + Br + O_2$ | | $2.7 \cdot 10^{-12}$ | Atkinson et al. (2000) |
| 356 | $BrO + ClO \rightarrow Br + OClO$ | | $9.5 \cdot 10^{-13} \times e^{550/T}$ | Burkholder et al. (2020) |
| 357 | $BrO + ClO \rightarrow Br + ClOO$ | | $2.3 \cdot 10^{-12} \times e^{260/T}$ | Burkholder et al. (2020) |
| 358 | $O^1D + HBr \rightarrow H + BrO$ | | $3.0 \cdot 10^{-11}$ | Burkholder et al. (2020) |
| 359 | $O + HBr \rightarrow OH + Br$ | | $5.8 \cdot 10^{-12} \times e^{-1500/T}$ | Burkholder et al. (2020) |
| 360 | $OH + HBr \rightarrow H_2O + Br$ | | $5.5 \cdot 10^{-12} \times e^{200/T}$ | Burkholder et al. (2020) |
| 361 | $HO_2 + Br \rightarrow O_2 + HBr$ | | $4.8 \cdot 10^{-12} \times e^{-310/T}$ | Burkholder et al. (2020) |
| 362 | $Br + H_2CO \rightarrow HCO + HBr$ | | $1.17 \cdot 10^{-11}$ | Burkholder et al. (2020) |
| 363 | $Br + H_2O_2 \rightarrow HBr + HO_2$ | | $1.0 \cdot 10^{-11} \times e^{-3000/T}$ | Burkholder et al. (2020) |
| 364 | $HBr + OH \rightarrow Br + H_2O$ | | $1.1 \cdot 10^{-11}$ | Atkinson et al. (2000) |
| 365 | $O^1D + HBr \rightarrow Br + OH$ | | $9.0 \cdot 10^{-11}$ | Burkholder et al. (2020) |
| 366 | $O^1D + HBr \rightarrow O + HBr$ | | $3.0 \cdot 10^{-11}$ | Burkholder et al. (2020) |
| 367 | $Br_2 + h\nu \rightarrow Br + Br$ | | | *photolysis* |

**Table 3**
(Continued)

| # | Reaction | Segura et al. (2005) Rate | New Rate | Reference |
|---|---|---|---|---|
| 368 | $O + HOBr \rightarrow OH + BrO$ | | $1.2 \cdot 10^{-10} \times e^{-430/T}$ | Burkholder et al. (2020) |
| 369 | $BrO + HO_2 \rightarrow HOBr + O_2$ | | $3.7 \cdot 10^{-12} \times e^{-545/T}$ | Atkinson et al. (2000) |
| 370 | $Br_2 + OH \rightarrow HOBr + Br$ | | $1.9 \cdot 10^{-11} \times e^{-240/T}$ | Atkinson et al. (2000) |
| 371 | $HOBr + h\nu \rightarrow Br + OH$ | | | *photolysis* |
| 372 | $OH + Br_2 \rightarrow HOBr + Br$ | | $2.1 \cdot 10^{-11} \times e^{240/T}$ | Burkholder et al. (2020) |
| 373 | $O + BrONO_2 \rightarrow NO_3 + BrO$ | | $1.9 \cdot 10^{-11} \times e^{215/T}$ | Burkholder et al. (2020) |
| 374 | $Br + BrONO_2 \rightarrow Br_2 + NO_3$ | | $4.9 \cdot 10^{-11}$ | Orlando & Tyndall (1996) |
| 375 | $BrO + NO_2 \rightarrow M + BrONO_2$ | | $5.5 \cdot 10^{-31} \times \frac{T}{300}^{-3.1}$ | Burkholder et al. (2020) |
| 376 | $BrONO_2 + h\nu \rightarrow BrO + NO_2$ | | | *photolysis* |
| 377 | $BrONO_2 + h\nu \rightarrow Br + NO_3$ | | | *photolysis* |
| 378 | $BrO + NO_2 \rightarrow M + BrONO_2$ | | | *photolysis* |
| 379 | $CH_3Br + h\nu \rightarrow CH_3 + Br$ | | | *photolysis* |
| 380 | $O^1D + CH_3Br \rightarrow O + CH_3Br$ | | $9.1 \cdot 10^{-13}$ | Burkholder et al. (2020) |
| 381 | $CH_3Br + h\nu \rightarrow CH_2Br + H$ | | | *photolysis* |
| 382 | $CH_2Br + O_2 \rightarrow H_2CO + BrO$ | | $1.2 \cdot 10^{-30} \times e^{-300/T}$ | Eskola et al. (2006) |
| 383 | $NO_2 + CH_2O_2 \rightarrow CH_2O + NO_3$ | | $6.3 \cdot 10^{-11}$ | Stone et al. (2014) |
| 384 | $OH + CH_3Br \rightarrow CH_2Br + H_2O$ | | $1.42 \cdot 10^{-12} \times e^{-1150/T}$ | Burkholder et al. (2020) |
| 385 | $Cl + CH_3Br \rightarrow CH_2Br + HCl$ | | $1.46 \cdot 10^{-11} \times e^{-1040/T}$ | Burkholder et al. (2020) |

**Table 3**
(Continued)

| # | Reaction | Segura et al. (2005) Rate | New Rate | Reference |
|---|---|---|---|---|
| 386 | $HBr + CH_2Br \rightarrow CH_3Br + Br$ | | $2.7 \cdot 10^{-13}$ | Seetula (2003) |
| 387 | $HCl + CH_2Br \rightarrow CH_3Br + Cl$ | | $1.6 \cdot 10^{-17}$ | Brudnik et al. (2013) |
| 388 | $CH_3 + Br \rightarrow CH_3Br$ | | $1.1 \cdot 10^{-10}$ | Krasnoperov & Mehta (1999) |
| 389 | $CH_3 + Br_2 \rightarrow CH_3Br + Br$ | | $4.2 \cdot 10^{-11}$ | Khamaganov & Crowley (2010) |
| 390 | $CH_3Cl + Br \rightarrow CH_3Br + Cl$ | | $2.9 \cdot 10^{-35}$ | Irikura & Francisco (2007) |
| 391 | $Cl + CHBr_3 \rightarrow CBr_3 + HCl$ | | $4.85 \cdot 10^{-12} \times e^{-850/T}$ | Burkholder et al. (2020) |
| 392 | $BrO + BrO \rightarrow Br_2 + O_2$ | | $2.9 \cdot 10^{-14} \times e^{-840/T}$ | Atkinson et al. (2000) |
| 393 | $Br + O_3 \rightarrow BrO + O_2$ | | $1.7 \cdot 10^{-11} \times e^{800/T}$ | Atkinson et al. (2000) |
| 394 | $Br + BrONO_2 \rightarrow Br_2 + NO_3$ | | $4.9 \cdot 10^{-11}$ | Atkinson et al. (2000) |
| 395 | $BrO + OH \rightarrow Br + HO_2$ | | $7.5 \cdot 10^{-11}$ | Atkinson et al. (2000) |
| 396 | $OH + CHBr_3 \rightarrow CBr_3 + H_2O$ | | $9.0 \cdot 10^{-13} \times e^{-360/T}$ | Burkholder et al. (2020) |
| 397 | $BrO + ClO \rightarrow BrCl + O_2$ | | $4.1 \cdot 10^{-13} \times e^{290/T}$ | Burkholder et al. (2020) |
| **398** | $ClO + OH \rightarrow Cl + HO_2$ | $1.1 \cdot 10^{-11} \times e^{T/-120}$ | $1.9 \cdot 10^{-11} \times e^{T/-120}$ | Atkinson et al. (2007) |
| **399** | $N_2O_5 + H_2O \rightarrow HNO_3 + HNO_3$ | $2.0 \cdot 10^{-19}$ | $2.0 \cdot 10^{-22}$ | Atkinson et al. (2004) |
| **400** | $CH_3O_2 + HO_2 \rightarrow CH_3OOH + O_2$ | $3.8 \cdot 10^{-13} \times e^{800/T}$ | $5.21 \cdot 10^{-12}$ | Atkinson et al. (2001) |
| **401** | $ClONO_2 + OH \rightarrow Cl + HO_2 + NO_2$ | $1.2 \cdot 10^{-12} \times e^{330/T}$ | $1.2 \cdot 10^{-12} \times e^{330/T}$ | Burkholder et al. (2020) |
| **402** | $ClONO_2 + O \rightarrow Cl + O_2 + NO_2$ | $2.9 \cdot 10^{-12} \times e^{800/T}$ | $3.6 \cdot 10^{-12} \times e^{840/T}$ | Burkholder et al. (2020) |
| **403** | $HO_2 + NO_3 \rightarrow HNO_3 + O_2$ | $4.1 \cdot 10^{-12}$ | | Burkholder et al. (2020) |
| **404** | $ClONO + OH \rightarrow HOCl + NO_2$ | $3.5 \cdot 10^{-14}$ | $2.4 \cdot 10^{-12} timese^{1250/T}$ | Burkholder et al. (2020) |

**Table 3**
(Continued)

| # | Reaction | Segura et al. (2005) Rate | New Rate | Reference |
|---|---|---|---|---|
| **405** | $Cl + ClONO_2 \rightarrow Cl + Cl + NO_2$ | $6.8 \cdot 10^{-11} \times e^{-160/T}$ | $6.8 \cdot 10^{-12} \times e^{-135/T}$ | Burkholder et al. (2020) |
| **406** | $CH_3O_2 + NO \rightarrow H_3CO + NO_2$ | $4.2 \cdot 10^{-12} \times e^{T/-180}$ | $7.29 \cdot 10^{-12}$ | Atkinson et al. (2001) |
| **407** | $H_3CO + O_2 \rightarrow H_2CO + HO_2$ | $3.9 \cdot 10^{-14} \times e^{T/900}$ | $9.6 \cdot 10^{-12}$ | Atkinson et al. (2001) |
| **408** | $H_3CO + O \rightarrow H_2CO + OH$ | $1.0 \cdot 10^{-14}$ | $1.0 \cdot 10^{-14}$ | Tsang & Hampson (1986) |
| **409** | $CH_2Cl + HBr \rightarrow CH_3Cl + Br$ | | $6.59 \cdot 10^{-13}$ | Seetula (1996) |
| **410** | $CH_2Cl + HCl \rightarrow CH_3Cl + Cl$ | | $2.39 \cdot 10^{-13}$ | Brudnik et al. (2013) |
| **411** | $CH_3 + Cl \rightarrow CH_3Cl$ | | $6.12 \cdot 10^{-11}$ | Parker et al. (2007) |
| **412** | $CH_3 + Cl_2 \rightarrow CH_3Cl + Cl$ | | $1.67 \cdot 10^{-13}$ | Eskola et al. (2008) |
| **413** | $CH_3Br + Cl \rightarrow CH_3Cl + Br$ | | $1.3 \cdot 10^{-24}$ | Irikura & Francisco (2007) |
| ** | $CH_3 + OH \rightarrow H_2CO + H_2$ | $1 \cdot 10^{10}$ | | |
| ** | $H_3CO + OH \rightarrow H_2CO + H_2O$ | $5.0 \cdot 10^{-11}$ | | |
| ** | $ClO_2 + O \rightarrow ClO + O_2$ | | | |
| ** | $CH_3 + O_2 \rightarrow CH_3O_2$ | | | |
| ** | $CH_3O_2 + CH_3O_2 \rightarrow H_3CO + H_3CO + O_2$ | $2.5 \cdot 10^{-13} \times e^{190/T}$ | | |
| ** | $ClO + NO_3 \rightarrow ClONO + O_2$ | $4.0 \cdot 10^{-13}$ | | |
| ** | $N_2O + H \rightarrow NO + NO + H$ | $5.03 \cdot 10^{-7} \times \left(\frac{298}{T}\right)^{2.16} \times e^{-18701/T}$ | | |
| ** | $N_2O + NO \rightarrow NO_2 + N_2$ | $2.92 \cdot 10^{-13} \times \left(\frac{298}{T}\right)^{2.23}) \times e^{-23292/T}$ | | |

**Note.** Our code has ∼100 more reactions than the previous version, and many of the overlapping reactions have had rates updated. Reactions indicated with "**" were included in Segura et al. (2005), but are not included in recent development of the model due to a lack of updated rates in the literature. A sensitivity test confirms that these reactions would not significantly change the results presented here.

## ORCID iDs

Michaela Leung https://orcid.org/0000-0003-1906-5093
Edward W. Schwieterman https://orcid.org/0000-0002-2949-2163
Mary N. Parenteau https://orcid.org/0000-0003-1225-6727
Thomas J. Fauchez https://orcid.org/0000-0002-5967-9631

## References

Acharyya, K., & Herbst, E. 2017, ApJ, 850, 105
Agol, E., Dorn, C., Grimm, S. L., et al. 2021, PSJ, 2, 1
Allen, M., Pinto, J. P., & Yung, Y. L. 1980, ApJL, 242, L125
Amouroux, D., & Donard, O. F. X. 1996, GeoRL, 23, 1777
Anbar, A. D., Duan, Y., Lyons, T. W., et al. 2007, Sci, 317, 1903
Andreae, M. O., & Merlet, P. 2001, GBioC, 15, 955
Arney, G., Domagal-Goldman, S. D., Meadows, V. S., et al. 2016, AsBio, 16, 873
Arney, G., Domagal-Goldman, S. D., & Meadows, V. S. 2018, AsBio, 18, 311
Arney, G., Meadows, V., Crisp, D., et al. 2014, JGRE, 119, 1860
Arney, G. N. 2019, ApJL, 873, L7
Ashfold, M. N. R., Fullstone, M. A., Hancock, G., & Ketley, G. W. 1981, CP, 55, 245
Atkinson, R., Baulch, D., Cox, R., et al. 2000, JPCRD, 29, 167
Atkinson, R., Baulch, D., Cox, R., et al. 2000, IUPAC Subcommittee on Gas Kinetic Data Evaluation for Atmospheric Chemistry, http://rpw.chem.ox.ac.uk/IUPACsumm_web_latest.pdf
Atkinson, R., Baulch, D., Cox, R., et al. 2007, ACP, 7, 981
Atkinson, R., Baulch, D., Cox, R. A., et al. 2004, ACP, 4, 1461
Bains, W., Seager, S., & Zsom, A. 2014, Life, 4, 716
Balch, W. E., & Wolfe, R. S. 1976, ApEnM, 32, 781
Bañuelos, G. S., Lin, Z.-Q., & Broadley, M. 2017, Selenium in Plants: Molecular, Physiological, Ecological and Evolutionary Aspects (Cham: Springer), 231
Basnayake, R. S., Bius, J. H., Akpolat, O. M., & Chasteen, T. G. 2001, Appl. Organometal. Chem., 15, 499
Battersby, C., Armus, L., Bergin, E., et al. 2018, NatAs, 2, 596
Baulch, D., Cobos, J., Cox, A., et al. 1994, CoFl, 98, 59
Baulch, D. L., Cobos, C. J., Cox, R. A., et al. 1992, JPCRD, 21, 411
Bell, N., Hsu, L., Jacob, D. J., et al. 2002, JGRD, 8, 4340
Bentley, R., & Chasteen, T. G. 2002, MMBR, 66, 250
Bieler, A., Altwegg, K., Balsiger, H., et al. 2015, Natur, 526, 678
Blackman, J. W., Beaulieu, J. P., Bennett, D. P., et al. 2021, Natur, 598, 272
Blei, E., Heal, M. R., & Heal, K. V. 2010, BGeo, 7, 3657
Böhland, T., Dõbe, S., Temps, F., & Wagner, H. G. 1985, Ber. Bunsenges. Phys. Chem, 89, 1110
Brogi, M., & Line, M. R. 2019, AJ, 157, 114
Brudnik, K., Twarda, M., Sarzyński, D., & Jodkowski, J. T. 2013, J. Mol. Model., 19, 1489
Brune, W. H., Schwab, J. J., & Anderson, J. G. 1983, JPhCh, 87, 4503
Burkholder, J., Sander, S. P., Abbatt, A., et al. 2015, Chemical Kinetics and Photochemical Data for Use in Atmospheric Studies Evaluation Number 18, (Pasadena, CA: Jet Propulsion Laboratory), https://jpldataeval.jpl.nasa.gov/pdf/JPL_Publication_15-10.pdf
Burkholder, J. B., Sander, S. P., Abbatt, J. P. D., et al. 2020, Chemical Kinetics and Photochemical Data for Use in Atmospheric Studies, Evaluation Number 19, (Pasadena, CA: Jet Propulsion Laboratory), doi:10.1002/kin.550171010
Carnall, A. 2017, arXiv:1705.05165
Carpenter, L. J., Liss, P. S., & Penkett, S. A. 2003, JGRD, 108, 4256
Carpenter, L. J., Reimann, S., Burkholder, J. B., et al. 2014, Update on Ozone-Depleting Substances (ODSs) and Other Gases of Interest to the Montreal Protocol, (Geneva: World Meteorological Organization), https://orbi.uliege.be/handle/2268/175647
Carrión, O., Curson, A. R., Kumaresan, D., et al. 2015, NatCo, 6, 6579
Carvalho, M. F., & Oliveira, R. S. 2017, Crit. Rev. Biotechnol., 37, 880
Catling, D. C., Claire, M. W., Zahnle, K. J., et al. 2010, JGRE, 115, E00E11
Chasteen, T. G., & Bentley, R. 2003, ChRv, 103, 1
Chau, Y. K., Wong, P. T. S., Silverberg, B. A., Luxon, P. L., & Bengert, G. A. 1976, Sci, 192, 1130
Chen, H., Wolf, E. T., Kopparapu, R., Domagal-Goldman, S., & Horton, D. E. 2018, ApJL, 868, L6
Chen, H., Wolf, E. T., Zhan, Z., & Horton, D. E. 2019, ApJ, 886, 16
Cima, F., Craig, P., & Harrington, C. 2003, Organotin Compounds in the Environment (New York: Wiley), 101
Comes, F. J. 1994, AngCh, 33, 1816
Conrad, R., & Klose, M. 1999, FEMS Microbiol. Ecol., 30, 47
Court, R. W., & Sephton, M. A. 2012, P&SS, 73, 233
Cox, R. A. 1974, J. Photochem., 3, 175
Crisp, D. 1997, GeoRL, 24, 571
De Keyser, J., Dhooghe, F., Altwegg, K., et al. 2017, MNRAS, 469, S695
Defrère, D., Léger, A., Absil, O., et al. 2018, ExA, 46, 543
Domagal-Goldman, S. D., Meadows, V. S., Claire, M. W., & Kasting, J. F. 2011, AsBio, 11, 419
Dougherty, M. K., Esposito, L. W., & Krimigis, S. M. (ed.) 2009, Saturn from Cassini-Huygens (Dordrecht: Springer Netherlands), 1
Du, S., Francisco, J. S., Shepler, B. C., & Peterson, K. A. 2008, JChPh, 128, 204306
Dungan, R. S., Yates, S. R., & Frankenberger, W. T. 2003, Environ. Microbiol., 5, 287
Engel, A., Rigby, M., Burkholder, J. B., et al. 2019, Update on Ozone-depleting Substances (ODSs) and Other Gases of Interest to the Montreal Protocol, (Geneva: World Meteorological Organization), https://library.wmo.int/doc_num.php?explnum_id=5705
Eskola, A. J., Timonen, R. S., Marshall, P., Chesnokov, E. N., & Krasnoperov, L. N. 2008, JPCA, 112, 7391
Eskola, A. J., Wojcik-Pastuszka, D., Ratajczak, E., & Timonen, R. S. 2006, PCCP, 8, 1416
Etiope, G., & Sherwood-Lollar, B. S. 2013, RvGeo, 51, 276
Fan, T. W., Lane, A. N., & Higashi, R. M. 1997, EnST, 31, 569
Farhan Ul Haque, M., Besaury, L., Nadalig, T., et al. 2017, NatSR, 7, 17589
Fauchez, T. J., Turbet, M., Villanueva, G. L., et al. 2019, ApJ, 887, 194
Fauchez, T. J., Villanueva, G. L., Schwieterman, E. W., et al. 2020, NatAs, 4, 372
Fayolle, E. C., Öberg, K. I., Jørgensen, J. K., et al. 2017, NatAs, 1, 703
Felton, R. C., Bastelberger, S. T., Mandt, K. E., et al. 2022, JGRE, 127, e06853
Frische, M., Garofalo, K., Hansteen, T. H., & Borchers, R. 2006, GGG, 7, Q05020
Fujii, Y., Angerhausen, D., Deitrick, R., et al. 2018, AsBio, 18, 739
Fujimori, T., Yoneyama, Y., Taniai, G., et al. 2012, LimOc, 57, 154
Gao, P., Hu, R., Robinson, T. D., Li, C., & Yung, Y. L. 2015, ApJ, 806, 249
Gialluca, M. T., Robinson, T. D., Rugheimer, S., & Wunderlich, F. 2021, PASP, 133, 054401
Gillon, M., Triaud, A. H. M. J., Demory, B.-O., et al. 2017, Natur, 542, 456
Gordon, I., Rothman, L., Hargreaves, R., et al. 2022, JQSRT, 277, 107949
Grenfell, J. L. 2017, PhR, 713, 1
Guimond, C. M., & Cowan, N. B. 2018, AJ, 155, 230
Harman, C. E., Schwieterman, E. W., Schottelkotte, J. C., & Kasting, J. F. 2015, ApJ, 812, 137
Harper, D. B. 1995, in Naturally-Produced Organohalogens, ed. A. Grimvall & E. W. B. de Leer (Dordrecht: Springer Netherlands), 21
Harris, C. R., Millman, K. J., Van Der Walt, S. J., et al. 2020, Natur, 585, 357
Hills, A. J., Cicerone, R. J., Calvert, J. G., & Birks, J. W. 1987, JPhCh, 91, 1199
Hu, H., Mylon, S. E., & Benoit, G. 2007, Chmsp, 67, 911
Huang, J., Seager, S., Petkowski, J. J., Ranjan, S., & Zhan, Z. 2022, AsBio, 22, 171
Hunter, J. D. 2007, CSE, 9, 90
Ingham, T., Sander, S. P., & Friedl, R. R. 2005, FaDi, 130, 89
Irikura, K. K., & Francisco, J. S. 2007, JPCA, 111, 6852
Jacob, D. J. 1999, Introduction to Atmospheric Chemistry (Princeton, NJ: Princeton Univ. Press)
Jia, Y., Huang, H., Zhong, M., et al. 2013, EnST, 47, 3141
Kaltenegger, L. 2017, ARA&A, 55, 433
Kaltenegger, L., MacDonald, R. J., Kozakis, T., et al. 2020, ApJL, 901, L1
Kasting, J. F. 1990, Orig. Life Evol. Biosph., 20, 199
Kasting, J. F., Liu, S., & Donahue, T. 1979, JGR, 84, 3097
Keller-Rudek, H., Moortgat, G. K., Sander, R., & Sörensen, R. 2013, ESSD, 5, 365
Keppler, F., Harper, D. B., Greule, M., et al. 2014, NatSR, 4, 7010
Khamaganov, V., & Crowley, J. N. 2010, Int. J. Chem. Kinet., 42, 575
Klick, S., & Abrahamsson, K. 1992, JGR, 97, 12683
Krasnoperov, L. N., & Mehta, K. 1999, JPCA, 103, 8008
Krasnopolsky, V. A. 2012, Icar, 218, 230
Krissansen-Totton, J., Garland, R., Irwin, P., & Catling, D. C. 2018, AJ, 156, 114
Lewis, N., Clampin, M., Mountain, M., et al. 2017, Transit Spectroscopy of TRAPPIST-1e JWST Cycle 1 Proposal (GTO 1331), https://www.stsci.edu/cgi-bin/get-proposal-info?observatory=JWST&id=1331
Lin, Z., Seager, S., Ranjan, S., Kozakis, T., & Kaltenegger, L. 2022, ApJL, 925, L10

Lincowski, A. P., Meadows, V. S., Crisp, D., et al. 2018, ApJ, 867, 76
Liuzzi, G., Villanueva, G. L., Crismani, M. M. J., et al. 2020, JGRE, 125, e2019JE006250
Lodders, K., & Fegley, B. 2002, Icar, 155, 393
Loison, J.-C., Halvick, P., Bergeat, A., Hickson, K. M., & Wakelam, V. 2012, MNRAS, 421, 1476
López-Morales, M., Currie, T., Teske, J., et al. 2019, arXiv:1903.09523
Loyd, R. O. P., France, K, Youngblood, A., et al. 2016, ApJ, 824, 102
Lustig-Yaeger, J., Meadows, V. S., & Lincowski, A. P. 2019, AJ, 158, 27
Macdonald, M. L., Wadham, J. L., Young, D., et al. 2020, ACP, 20, 7243
Madhusudhan, N., Piette, A. A. A., & Constantinou, S. 2021, ApJ, 918, 1
Manley, S. L. 2002, Biogeochemistry, 60, 163
Manley, S. L., & Dastoor, M. N. 1988, Marine Biol., 98, 477
Manley, S. L., Goodwin, K., & North, W. J. 1992, LimOc, 37, 1652
Manley, S. L., Wang, N. Y., Waiser, M. L., & Cicerone, R. J. 2006, GBioC, 20, GB3015
Matsumi, Y., Tonokura, K., Inagaki, Y., & Kawasaki, M. 1993, JPhCh, 97, 6816
Meadows, V., & Crisp, D. 1996, JGR, 101, 4595
Meadows, V. S. 2017, AsBio, 17, 1022
Meadows, V. S., Arney, G. N., Schwieterman, E. W., et al. 2018b, AsBio, 18, 133
Meadows, V. S., Reinhard, C. T., Arney, G. N., et al. 2018a, AsBio, 18, 630
Meixner, M., Cooray, A., Leisawitz, D., et al. 2019, arXiv:1912.06213
Messenger, S. J. 2013, PhD thesis, Massachusetts Institute of Technology
Meyer, J., Michalke, K., Kouril, T., & Hensel, R. 2008, Syst. Appl. Microbiol., 31, 81
Miller, L. G., Sasson, C., & Oremland, R. S. 1998, ApEnM, 64, 4357
Montzka, S. A., Reimann, S., O'Doherty, S., et al. 2011, Scientific Assessment of Ozone Depletion: 2010 (Geneva: World Meteorological Organization)
Morley, C. V., Kreidberg, L., Rustamkulov, Z., Robinson, T., & Fortney, J. J. 2017, ApJ, 850, 121
Mumma, M. J., & Charnley, S. B. 2011, ARA&A, 49, 471
Murphy, C. D., Clark, B. R., & Amadio, J. 2009, Appl. Microbiol. Biotechnol., 84, 617
National Institute of Standards & Technology 2018, NIST Chemistry WebBook, SRD 69, https://webbook.nist.gov/cgi/inchi/InChI%3D1S/CH3Br/c1-2/h1H3
Neilson, A. H. 2003, in Organic Bromine and Iodine Compounds, ed. A. H. Neilson (Berlin: Springer), 75
Odar, C., Winkler, M., & Wiltschi, B. 2015, Biotechnol. J, 10, 427
Olson, S. L., Kump, L. R., & Kasting, J. F. 2013, ChGeo, 362, 35
Ooki, A., & Yokouchi, Y. 2011, Mar. Chem., 124, 119
Orlando, J. J., & Tyndall, G. S. 1996, JPhCh, 100, 19398
Parker, J. K., Payne, W. A., Cody, R. J., et al. 2007, JPCA, 111, 1015
Parkes, D. A., Paul, D. M., Quinn, C. P., & Robson, R. C. 1973, CPL, 23, 425
Parmentier, V., Line, M. R., Bean, J. L., et al. 2018, A&A, 617, A110
Paul, C., & Pohnert, G. 2011, Nat. Prod. Rep., 28, 186
Peacock, S., Barman, T., Shkolnik, E. L., Hauschildt, P. H., & Baron, E. 2019, ApJ, 871, 235
Phillips, C. L., Wang, J., Kendrew, S., et al. 2021, ApJ, 923, 144
Pidhorodetska, D., Fauchez, T., Villanueva, G., Domagal-Goldman, S., & Kopparapu, R. K. 2020, ApJL, 898, L33
Pilcher, C. B. 2003, AsBio, 3, 471
Pyle, D., & Mather, T. 2009, ChGeo, 263, 110
Quanz, S. P., Kammerer, J., Defrère, D., et al. 2018, Proc. SPIE, 10701, 107011I
Quanz, S. P., Ottiger, M., Fontanet, E., et al. 2022, A&A, 664, A21
Ranjan, S., Schwieterman, E. W., Harman, C., et al. 2020, ApJ, 896, 148
Ranjan, S., Seager, S., Zhan, Z., et al. 2022, ApJ, 930, 131
Rhew, R. C., Chen, C., Teh, Y. A., & Baldocchi, D. 2010, AtmEn, 44, 2054
Rhew, R. C., Miller, B. R., Vollmer, M. K., & Weiss, R. F. 2001, JGR, 106, 20875
Rhew, R. C., Miller, B. R., & Welss, R. F. 2000, Natur, 403, 292
Riding, R., Fralick, P., & Liang, L. 2014, PreR, 251, 232
Robinson, T. D., Ennico, K., Meadows, V. S., et al. 2014, ApJ, 787, 171
Robinson, T. D., Meadows, V. S., Crisp, D., et al. 2011, AsBio, 11, 393
Rugheimer, S., Kaltenegger, L., Segura, A., Linsky, J., & Mohanty, S. 2015, ApJL, 809, 57
Sander, S., Friedl, R. R., Ravishankara, A. R., et al. 2003, Chemical Kinetics and Photochemical Data for Use in Atmospheric Studies Evaluation Number 14, (Pasadena, CA: Jet Propulsion Laboratory), https://jpldataeval.jpl.nasa.gov/pdf/JPL_02-25_rev02.pdf
Sander, S. P., Abbatt, J., Barker, J. R., et al. 2006, Chemical Kinetics and Photochemical Data for Use in Atmospheric Studies Evaluation Number 15, (Pasadena, CA: Jet Propulsion Laboratory), https://jpldataeval.jpl.nasa.gov/pdf/JPL_15_AllInOne.pdf
Schaefer, L., & Fegley, B. 2008, Chemical Evolution across Space & Time (Washington, DC: American Chemical Society), 187
Schall, C., Heumann, K. G., & Kirst, G. O. 1997, Fresenius J. Anal. Chem., 359, 298
Schall, C., Laturnus, F., & Heumann, K. G. 1994, Chmsp, 28, 1315
Scheucher, M., Grenfell, J. L., Wunderlich, F., et al. 2018, ApJ, 863, 6
Schofield, K. 1973, JPCRD, 2, 25
Schwieterman, E. W., Kiang, N. Y., Parenteau, M. N., et al. 2018, AsBio, 18, 663
Schwieterman, E. W., Reinhard, C. T., Olson, S. L., et al. 2019, ApJ, 874, 9
Schwieterman, E. W., Robinson, T. D., Meadows, V. S., Misra, A., & Domagal-Goldman, S. 2015, ApJ, 810, 57
Seager, S., Bains, W., & Hu, R. 2013, ApJ, 777, 95
Seager, S., Bains, W., & Petkowski, J. J. 2016, AsBio, 16, 465
Seetula, J. A. 1996, J. Chem. Soc. Faraday Trans., 92, 3069
Seetula, J. A. 2003, PCCP, 5, 849
Segura, A., Kasting, J. F., Meadows, V., et al. 2005, AsBio, 5, 706
Seinfeld, J. H., & Pandis, S. N. 2016, Atmospheric Chemistry and Physics: From Air Pollution to Climate Change (Hoboken, NJ: Wiley)
Shibazaki, A., Ambiru, K., Kurihara, M., Tamegai, H., & Hashimoto, S. 2016, Mar. Chem., 181, 44
Showman, A. P. 2001, Icar, 152, 140
Simmonds, P. G., Derwent, R. G., Manning, A. J., O'Doherty, S., & Spain, G. 2010, AtmEn, 44, 1284
Simonaitis, R., & Heicklen, J. 1975, P&SS, 23, 1567
Singleton, D. L., & Cvetanović, R. J. 1988, JPCRD, 17, 1377
Sousa-Silva, C., Seager, S., Ranjan, S., et al. 2020, AsBio, 20, 235
St, F., Laturnus, F., Doležalová, J., Holík, J., & Wimmer, Z. 2016, Plant Soil Environ., 61, 103
Staguhn, J., Mandell, A., Stevenson, K., et al. 2019, arXiv:1908.02356
Stefels, J., Steinke, M., Turner, S., Malin, G., & Belviso, S. 2007, Biogeochemistry, 83, 245
Stone, D., Blitz, M., Daubney, L., Howes, N. U. M., & Seakins, P. 2014, PCCP, 16, 1139
Suissa, G., Wolf, E. T., Kopparapu, R. K., et al. 2020, AJ, 160, 118
Tait, V. K., & Moore, R. M. 1995, LimOc, 40, 189
Teal, D. J., Kempton, E. M.-R., Bastelberger, S., Youngblood, A., & Arney, G. 2022, arXiv:2201.08805
Teanby, N., Showman, A., Fletcher, L., & Irwin, P. 2014, P&SS, 103, 250
Thayer, J. S. 2002, Appl. Organometal. Chem., 16, 677
Tinetti, G., Meadows, V. S., Crisp, D., et al. 2005, AsBio, 5, 461
Tsang, W., & Hampson, R. 1986, JPCRD, 15, 1087
Turco, R. P., Whitten, R. C., & Toon, O. B. 1982, RvGeo, 20, 233
van Pée, K. H., & Unversucht, S. 2003, Chmsp, 52, 299
Vanderburg, A., Rappaport, S. A., Xu, S., et al. 2020, Natur, 585, 363
Villanueva, G. L., Liuzzi, G., Faggi, S., et al. 2022, Fundamentals of the Planetary Spectrum Generator (Greenbelt, MD: NASA Goddard Space Flight Center)
Villanueva, G. L., Smith, M. D., Protopapa, S., Faggi, S., & Mandell, A. M. 2018, JQSRT, 217, 86
Visscher, P. T., Baumgartner, L. K., Buckley, D. H., et al. 2003, Environ. Microbiol., 5, 296
Wishkerman, A., Gebhardt, S., McRoberts, C. W., et al. 2008, EnST, 42, 6837
Wogan, N. F., & Catling, D. C. 2020, ApJ, 892, 127
Wunderlich, F., Scheucher, M., Grenfell, J. L., et al. 2021, A&A, 647, A48
Xu, Z. F., & Lin, M. C. 2003, JChPh, 119, 8897
Yang, X., Cox, R. A., Warwick, N. J., et al. 2005, JGRD, 110, D23311
Yang, Z., Fang, W., Lu, X., et al. 2016, Environ. Pollut., 214, 504
Yao, Y., & Giapis, K. P. 2017, NatCo, 8, 15298
Yokouchi, Y., Hasebe, F., Fujiwara, M., et al. 2005, JGRD, 110, D23309
Zahnle, K., Claire, M., & Catling, D. 2006, Geobiology, 4, 271
Zahnle, K. J. 1986, JGR, 91, 2819
Zhan, Z., Seager, S., Petkowski, J. J., et al. 2021, AsBio, 2, 765
Zhu, R. S., & Lin, M.-C. 2001, in The 2001 Technical Meeting of the Eastern States Section of the Combustion Institute, 104